\documentclass[aps,amsmath,amssymb,showpacs,showkeys]{revtex4}
\usepackage[dvips]{graphicx,color}

\DeclareMathOperator{\arccsch}{arcCsch}
\usepackage[utf8]{inputenc}
\usepackage{multirow}
\usepackage{times}
\usepackage{xcolor}
\usepackage[%
  colorlinks=true,
  urlcolor=blue,
  linkcolor=red,
  citecolor=blue
]{hyperref}

\newcommand{\orcid}[1]{\href{https://orcid.org/#1}{\includegraphics[width=8pt]{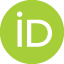}}}

\begin{document}
\title{Quasinormal Modes of Non-Linearly Charged Black Holes surrounded by a Cloud of Strings in Rastall Gravity}

\author{Dhruba Jyoti Gogoi \orcid{0000-0002-4776-8506}}
\email[Email: ]{moloydhruba@yahoo.in}

\affiliation{Department of Physics, Dibrugarh University,
Dibrugarh 786004, Assam, India}

\author{Ronit Karmakar}
\email[Email: ]{ronit.karmakar622@gmail.com}

\affiliation{Department of Physics, Dibrugarh University,
Dibrugarh 786004, Assam, India}

\author{Umananda Dev Goswami  \orcid{0000-0003-0012-7549}}
\email[Email: ]{umananda2@gmail.com}

\affiliation{Department of Physics, Dibrugarh University,
Dibrugarh 786004, Assam, India}

\begin{abstract}
We obtain the black hole solution for the Ay\'on - Beato - Garc\'ia (ABG) type 
black hole surrounded by a cloud of strings in Rastall gravity and calculate 
the scalar quasinormal modes of it for a massless scalar field. To have a 
better visualization of the results, we also introduce a new non-linear 
electrodynamic source and obtain a black hole solution surrounded by a cloud 
of strings in the Rastall gravity. We see that the quasinormal modes are 
affected by the type of non-linear electrodynamic sources in case of higher 
magnitudes of charge $q$. With increase in the magnitude of charge $q$, 
gravitational waves decay rapidly for the new black hole solution, while for 
the ABG black hole, the decay rate increases initially and finally starts to 
decrease near to $q=1$. We also study the impact of the cloud of strings and 
other model parameters, including the Rastall parameter on the quasinormal 
modes for both of the black holes. The gravitational waves decay slowly with 
increase in the cloud of strings parameter for both of the black holes. 
Dependency of quasinormal modes on the Rastall parameter is different from 
a surrounding dark energy field and the decay of gravitational waves may be 
slow or rapid depending on the value of this parameter. 

\end{abstract}


\maketitle
\section{Introduction}
The theory of General Relativity (GR) is widely regarded as the most elegant 
theory ever conceived. This beautiful theory has been put to test for many 
times in the past but it has emerged out to be consistent with experimental 
observations to a fair extent in most of the cases \cite{will2014,hulse1975,damour1992}. The most recent outstanding evidence in favour of GR came in the form 
of observations from the Laser Interferometer Gravitational Wave Observatory 
(LIGO) and the Variability of Solar Irradiance and Gravity Oscillations 
(Virgo), when they reported the first ever detection of 
Gravitational Waves (GWs) from the merger of two black holes in 2015 \cite{abbott2016}. After that, many other events of GWs were detected by these 
observatories in the following years \cite{abbott2016_2,bpabbott2016,bpabbott2017,abbott2017,burns2019,abbott2020,rabbott2021}. These detections have opened 
up a new research avenue to study the universe via GW 
astronomy. They also provide a testing ground for various theories of gravity 
including GR in the extreme gravity regime. Overall, GR has drastically 
uplifted our level of understanding and way of thinking about the universe -- 
its origin, existence and future. However, GR was already known to have many drawbacks. Such as, in GR there is no scope for explaining the cosmic accelerated expansion \cite{riess1998,perlmutter1999}. 
At a highly relativistic regime, GR is non-renormalizable \cite{stelle1977} and 
at some stage, we have to encounter infinities in the theory. GR also does not 
provide any prediction regarding the dark components of the universe, whose 
presence has been indirectly shown by observations \cite{bachal1999,perez2020}. 
Realizing these drawbacks, there have been attempts to modify GR. 

The simplest modification of GR that includes the dark components of the 
observable universe { in the form of some supplementary fields}, which is 
most favorable in terms of cosmological observations at present, is the 
$\Lambda$CDM ($\Lambda$ Cold Dark Matter) model.
According to this model the universe is spatially flat, homogeneous and 
isotropic over the large scales, and is filled with a constant energy density
fluid throughout, known as the cosmological constant $\Lambda$ \cite{bull2016,amendola2010}. This simple model fits well with almost all observational data 
that we have. But, it too has some drawbacks, e.g., $\Lambda$CDM model is not 
able to justify the existence of $\Lambda$. Similarly, $\Lambda$CDM model fails to address the small value of $\Lambda$ and hence the fine-tuning problem with 
the theoretical prediction. This represents the largest 
discrepancy between observed and theoretical predictions in the history of 
theoretical physics \cite{amendola2010}.  Moreover, there is no explanation 
for the constant energy density of the $\Lambda$, in contrast to the variation 
of energy density of the other components of the universe with the scale 
factor \cite{copeland2006, amendola2010}. { To deal with these serious 
issues, different Modified Theories of Gravity
(MTGs) were introduced, so that without introducing the dark components of the 
universe, this class of theories can predict observational data like 
accelerated expansion of the universe and the galactic rotation curves.} Another
key aspect to study in MTGs is the GWs, which has been studied vigorously in 
the past few decades and has been benefited from the endeavours to develop new 
MTGs \cite{Gogoi2020, Miller_2019}. Various novel properties of GWs have 
emerged from the application of various MTGs. As an example, the presence of 
three polarization modes of GWs in metric formalism of $f(R)$ theory 
opens up new possibilities in GW observations. Here, two modes are the GR modes
and the third mode is the scalar polarization mode, which consist of a 
combination of a massive longitudinal mode and a massless breathing mode 
\cite{Gogoi2020,Gogoi2021}. This has inspired a surge in research in this 
field, especially, the applications of new MTGs to black holes and stellar 
dynamics, and that lead to results predicting deviations from GR 
\cite{Gogoi2020,Gogoi2021,Gogoi2021_1}. { One may note that the extra 
polarization modes of GWs have been constrained by recent observations 
\cite{Ezquiaga2018}. Particularly, the LIGO-Virgo collaboration gives 
the current bound on the mass of the quanta of massive mode, the graviton 
as $ m_g \le 7.7\times10^{-23}$ $eV/c^2$ \cite{abbot_bound}.}

Among the modified theories, a special modification to GR was given by P.\ 
Rastall in 1972, where he proposed that the covariant derivative of the 
energy-momentum tensor does not vanish in curved spacetime, or in presence of 
massive objects, but only vanish in flat spacetime. This new model of 
gravity was called the Rastall gravity \cite{rastall1972}. Recently, it has 
drawn the attention of researchers mainly due to the unique behaviour of 
violation of energy conservation in presence of background curvature. In 
this model, Rastall proposed the covariant derivative of the energy-momentum 
tensor to be equal to divergence of the Ricci scalar or curvature scalar $R$. 
However, the Bianchi identity of the vanishing divergence of the Einstein 
tensor was maintained \cite{moraes2019}. It is to be noted that although the 
violation of conservation principle is shown exclusively by the Rastall theory 
only, but the same feature also includes inherently by other theories like 
$f(R,T)$ theory and $f(R,L_M)$ theory \cite{moraes2019,shabani2020}. This 
feature of Rastall gravity caused significant deviations from GR, specially 
near the high curvature regimes, like in the regions near the Black Holes and 
Neutron stars. Although Rastall gravity may be equivalent to GR in some aspects \cite{visser2018}, one can see that in presence of non-zero curvature or some dark energy fields the theory can show significant deviations from GR \cite{darabi2018}.
Recently, a lot of important applications of Rastall theory can be seen in the 
literature, some of which includes gravitational collapse of a homogeneous 
perfect fluid \cite{ziaie2019}, some rotating and non-rotating Black Hole 
solutions \cite{heydarzade2017,heydarzade2017_2,kumar2018} and phenomenon of 
particle creation in Rastall cosmology \cite{batista2012}. Another important 
feature of Rastall gravity is that it can explain the accelerated expansion of 
the universe also \cite{moradpour2017}. Moreover, in recent times, many 
observational constraints were satisfied by Rastall gravity theory, like 
calculation of the age of the universe and the Hubble parameter from this theory
agree well with the observed values \cite{rawaf1996}. It also provides a 
better explanation for the presence of the matter-dominated era than GR 
\cite{rawat1996_2}. Also it is free from entropy and age problems of standard 
cosmology \cite{fabris2000} and can predict gravitational lensing accurately 
\cite{abdel2005,abdel2001}. Another scope of this theory includes the positive 
possibility of occurrence of traversable wormhole solutions 
\cite{moradpour2017_2}. Further, a work highlighted the use of this theory in 
the Chaplygin model \cite{fabris2011}, attempting to combine dark matter and 
dark energy into a single theory, and then studying the structure formation 
constraints. Ref.\ \cite{carames2015, carames_2014} discusses the combined treatment of 
Rastall gravity and scalar-tensor theory, and formed a new Brans-Dicke-Rastall 
(BDR) gravity, where PPN \cite{peri_ppn} constraint was satisfied and it led to an 
accelerated-decelerated cycle in matter-dominated era of the universe. In other words, Rastall gravity has provided us with a wide range of application scope 
in various domains. It converges to GR in a weak field regime and thus 
satisfies the solar system tests along with other constraints satisfied by GR. 
{ It is an important feature of this theory that the field equations are 
obtained directly from the violation condition of energy conservation law and 
there is no form of Lagrangian associated with the actual 
theory \cite{Gogoi2021, rastall1972}.} 
This makes Rastall theory different and unique from the other MTGs including 
$f(R)$ theories.

The impacts of the cloud of strings on the properties of black holes have been 
studied in detail in the last a few years for different modified theories. In 
a recent study, the cloud of strings and the quintessence have been considered 
together in Lovelock gravity \cite{toledo18}. The magnetic Lovelock black 
holes have been considered in Ref. \cite{mag_string} in presence of a cloud of 
strings. The black hole solution in Rastall gravity surrounded by a cloud of 
strings has been obtained in Ref. \cite{Cai_2020} for the first time where the 
authors studied the Quasinormal Modes (QNMs) explicitly. Being motivated by 
these studies, in this work, we have considered the non-linearly charged black 
hole solutions surrounded by a cloud of strings for the first time in Rastall 
gravity. This study will shed light on the impacts of clouds of strings on the 
properties of the black hole in presence of non-linear charge distributions. 
To have a clear picture on the scenario, we have considered two types of 
non-linear charged distribution in the presence of a cloud of strings in 
Rastall gravity.
 
The primary focus of our work shall be on the study of the scalar 
QNMs of Ay\'on - Beato - Garc\'ia (ABG) black hole \cite{ayon1999} and a new 
black hole in Rastall gravity with a cloud of strings as the surrounding field. 
QNMs are basically some complex numbers arising out of emissions of GWs from 
the compact, massive objects that are undergoing perturbations in spacetime 
\cite{Kokkotas}. The real part of the QNM signifies the frequency of GW 
emission from the source, while the imaginary coefficient represents the 
damping of GW during its propagation. The idea of QNMs was first proposed by 
Vishveshwara in 1970 \cite{vishveshwara1970} and it was confirmed by 
Press in his work in 1971 \cite{press1971}. Later, quasinormal eigenfrequency 
calculations were done by Chandrasekhar and Detweiler in 1975 
\cite{chandrasekhar1975}. These QNMs and eigenfrequency calculations were 
utilized to test theories of gravity including GR in 
\cite{berti2015,dreyer2004}.

Recently, the study of black holes and neutron stars in Rastall gravity has 
been frequently carried out with the hope of results of notable deviations 
from GR. One such work was carried out in Ref.\ \cite{oliveira2015}, where 
neutron stars were studied in Rastall theory. On the same line, another work 
was done in Ref.\ \cite{heydarzade2017_2}, where black holes were studied 
in Rastall theory. Black holes surrounded by a perfect fluid were also studied 
in Rastall gravity in the Ref.\ \cite{heydarzade2017, kumar21}. Similar study on 
QNMs of black holes surrounded by quintessence scalar fields in Rastall gravity 
was done extensively in \cite{chen2005,zhang2006,zhang2007}. QNMs of black hole in Rastall gravity have been studied in Ref.\ \cite{Liang_2018}, where the 
variation of QNMs with respect to the Rastall parameter $\lambda$ has been 
studied for different perturbations using WKB approximation method. The author 
showed that for $k\lambda < 0$ there is a rapid damping of QNMs in 
gravitational, scalar and electromagnetic fields, whereas for $k\lambda > 0$ 
the damping occurs slowly. Further, black holes with non-linear 
electrodynamic sources were studied in Ref.s\ \cite{balart2014, nojiri_2017, jusufi20}. In this context, more recently QNMs of black holes with different
non-linear electrodynamic sources in Rastall gravity have been studied
in detail in \cite{Gogoi2021}. Here it was found that the behaviour of the 
black holes and their QNMs depend on the considered surrounding field type.  
 
In this work, we study the QNMs of regular black holes surrounded by a 
cloud of strings together with a non-linear electrodynamic source in Rastall 
gravity as mentioned above. Here our aim is to investigate any dependency of 
the quasinormal frequencies with the charge of black holes, the Rastall 
parameter and the characteristics of the cloud of strings. The work is 
organized as follows. In Section \ref{section2}, we derive the charged black 
hole solutions in Rastall gravity surrounded by a cloud of strings. In Section 
\ref{section3}, behaviour of potential and QNMs of the black hole solutions 
are studied explicitly. The properties of the Hawking temperature for the 
black hole solutions are studied in Section \ref{section4}. Finally, we 
conclude the work with a brief summary of the results in Section 
\ref{conclusion}. Throughout the paper, we have used the natural unit system 
($G = c = 1$).

\section{Charged black hole Solutions surrounded by a cloud of strings in 
Rastall Gravity} \label{section2}
The Rastall gravity is a type of modification of GR in which the general 
covariant conservation condition  $T^{\mu \nu}_{\ ; \, \nu} = 0$ is changed 
to the following form:
\begin{equation}\label{r1}
\nabla_\nu T^{\mu\nu} = \lambda \nabla^\mu R,  
\end{equation}
where $\lambda$ is known as the Rastall parameter. This condition modifies 
the general energy momentum conservation in presence of non-vanishing 
curvature. When the curvature tends to zero, i.e. in absence of any matter 
energy content the theory tends to GR. The consequent 
Rastall field equations are given by \cite{rastall1972,Gogoi2021}
\begin{eqnarray}
&& R_{\mu\nu}-\frac{1}{2}\left(\, 1 -2\beta\,\right) g_{\mu\nu}R=\kappa T_{\mu\nu}\ .\label{E1}
\end{eqnarray}
Taking trace of the above equation, we obtain
\begin{equation}
R=\frac{\kappa}{\left(4\,\beta-1\right)}\,T\ ,\quad \beta\ne 1/4, \label{E2}
\end{equation}
where we have used $\beta = \kappa \lambda$ and from here onwards we shall 
denote $\beta$ as the Rastall parameter. Now, to obtain the black hole 
solutions in this framework, we consider the following ansatz: 
\begin{equation} \label{metric}
ds^2 = f(r) dt^2 - \dfrac{dr^2}{f(r)} - r^2 d\Omega^2,
\end{equation}
which represents a spherically symmetric general spacetime metric in 
Schwarzschild coordinate. Here $f(r)$ is the metric function and 
$d\Omega^2 = d\theta^2 + \sin^2\theta d\phi^2$. Next, we define the Rastall 
tensor as $\mathcal{S}_{\mu\nu} = \mathcal{G}_{\mu\nu} + \beta g_{\mu\nu} R$, where $\mathcal{G}_{\mu\nu}$ is the Einstein tensor. The Rastall tensor has the 
following components:

\begin{eqnarray}\label{H}
&&{\mathcal{S}^{0}}_{0}={\mathcal{G}^{0}}_{0}+\beta R=\frac{1}{f(r)}\mathcal{G}_{00}+\beta R=-r^{-2}\Big \lbrace f^{\prime}(r)r-1+f(r)  \Big \rbrace +\beta R,\nonumber\\
&&{\mathcal{S}^{1}}_{1}={\mathcal{G}^{1}}_{1}+\beta R=-f(r) \mathcal{G}_{11}+\beta R=-r^{-2}\Big \lbrace f^{\prime}(r)r-1+f(r)  \Big \rbrace +\beta R,\nonumber\\
&&{\mathcal{S}^{2}}_{2}={\mathcal{G}^{2}}_{2}+\beta R=-r^{-2}\mathcal{G}_{22}+\beta R=-r^{-2}\Big \lbrace rf^{\prime}(r)+\frac{1}{2}r^2 f^{\prime\prime}(r)\Big \rbrace +\beta R,\nonumber\\
&&{\mathcal{S}^{3}}_{3}={\mathcal{G}^{3}}_{3}+\beta R=-\frac{1}{r^2 \sin^2 \theta}\mathcal{G}_{33}+\beta R=-r^{-2}\Big \lbrace rf^{\prime}(r)+\frac{1}{2}r^2 f^{\prime\prime}(r)\Big \rbrace +\beta R,
\end{eqnarray}
 where the Ricci scalar reads as
\begin{equation}\label{R}
R=r^{-2}\Big \lbrace {r}^{2}f^{\prime\prime}(r)  +4r f^{\prime}(r)-2+2\,f(r) \Big \rbrace
\end{equation}
and the prime denotes the derivative with respect to the radial coordinate 
$r$. We define a general total energy momentum tensor $T^\mu_\nu$ as
\begin{equation} \label{gen_T}
{{T}^{\mu}}_{\nu}={E^{\mu}}_{\nu}+{\mathcal{T}^{\mu}_{\nu}},
\end{equation}
where $\mathcal{T}^{\mu}_{\nu}$ is the energy-momentum tensor for the 
surrounding field, which we will define later and  ${E^{\mu}}_{\nu}$ is the 
trace-free Maxwell tensor expressed as
\begin{equation}\label{E*}
E_{\mu\nu}={\frac{2}{\kappa}}\left(F_{\mu\alpha}{F_{\nu}}^{\alpha}-
\frac{1}{4}g_{\mu\nu}F^{\alpha\beta}F_{\alpha\beta}\right),
\end{equation}
where $F_{\mu\nu}$ is the antisymmetric Faraday tensor, which satisfies the
conditions:
\begin{eqnarray}\label{max}
&&{F^{\mu\nu}}_{;\mu}=0,\nonumber\\
&&\partial_{[\sigma} F_{\mu\nu]}=0,
\end{eqnarray}
For spherical symmetry it gives,
\begin{equation}
F^{01}=\frac{q}{r^2}.
\end{equation}
This parameter $q$ plays the role of electrostatic charge in the theory. Hence, the Maxwell tensor takes the form:
\begin{equation}\label{E**}
{E^{\mu}}_{\nu}={\frac{q^2}{\kappa r^4}}~\begin{pmatrix}1 & 0 & 0 & 0 \\
0& 1 & 0 & 0 \\
0 & 0 & -1 & 0 \\
0 & 0 & 0 & -1\\
\end{pmatrix}.
\end{equation}
In GR, the first static spherically symmetric black hole solution surrounded 
by a cloud of string was introduced by Letelier \cite{Letelier}. Considering a 
cloud of strings as world sheets, one can write the energy-momentum tensor of 
a cloud of strings characterized by a proper density $\rho_c $ as given by

\begin{equation}
\mathcal{T}^{\mu \nu }=\frac{\rho_c \mathit{\Sigma} ^{\mu \beta }{\mathit{\Sigma}_{\beta }}^{\nu }}{\sqrt{-\gamma }},
\end{equation}
where $\gamma =\frac{1}{2}\mathit{\Sigma} ^{\mu \nu }\mathit{\Sigma} _{\mu \nu }$. The string bivector $\mathit{\Sigma} ^{\mu \nu }$ is expressed as
\begin{equation}
\mathit{\Sigma} ^{\mu \nu }=\epsilon ^{ab}\frac{\partial x^{\mu }}{\partial\xi^{a}  }\frac{\partial x^{\nu }}{\partial \xi^{b}},
\end{equation}
where $\epsilon ^{ab}$ is the two-dimensional Levi-Civita symbol with 
$\epsilon ^{01}=-\,\epsilon ^{10}=1$. Here, one can see that the cloud of 
strings is spherically symmetric and hence it is a function of the radial 
coordinate $r$ only. As a result, one can have the density $\rho_c$ and the 
string bivector $\mathit{\Sigma} ^{\mu \nu }$ as functions of the radial 
coordinate $r$ only. Thus, the non-zero components of antisymmetric 
$\mathit{\Sigma} ^{\mu \nu }$ are found to be $\mathit{\Sigma} ^{01}$ and 
$\mathit{\Sigma} ^{10}$. They are related as $\mathit{\Sigma} ^{01} = 
-\,\mathit{\Sigma} ^{10}$. Hence, the energy-momentum tensor of cloud of 
strings takes the following form:
\begin{equation}\label{string_EM_tensor}
{\mathcal{T}^{ \mu}}_{\nu }=\begin{pmatrix}
\rho _{c}(r) & 0 & 0 & 0 \\ 
0 &  \rho _{c}(r) & 0 & 0 \\ 
0 & 0 &  0 & 0\\ 
0 & 0 & 0 & 0 \\
\end{pmatrix}.
\end{equation}

From Eq.s \eqref{r1}, \eqref{E2} and \eqref{string_EM_tensor} we may write,
\begin{equation}
\frac{\mathrm{d} \rho _{c}}{\mathrm{d} r}+\frac{2\rho _{c}}{r}=\frac{2\beta }{4\beta -1}\frac{\mathrm{d} \rho _{c}}{\mathrm{d} r}.
\end{equation}
The solution of the above equation reads,
\begin{equation}\label{cloud_density}
\rho _{c}(r)=b\, r^{-\frac{2(4\beta -1)}{2\beta -1}},
\end{equation}
where $b$ is an integration constant. It is associated with the density of the 
cloud of strings. The weak energy condition demands that $b\ge0.$  Now the 
${\mathcal{S}^{0}}_{0}={T^{0}}_{0}$ and ${\mathcal{S}^{1}}_{1}={T^{1}}_{1}$ components 
of Rastall field equations give,
 \begin{equation}\label{e00}
-r^{-2}\left(rf^{\prime}-1+f  \right)+\frac {\beta
}{{r}^{2}}\left({r}^{2}f^{\prime\prime}  +4r f^{\prime}-2+2\,f \right)
=\kappa\rho_c+\frac{q^2}{ r^4},
\end{equation} 
Similarly, the ${\mathcal{S}^{2}}_{2}={T^{2}}_{2}$ and 
${\mathcal{S}^{3}}_{3}={T^{3}}_{3}$ components take the form:
\begin{equation}\label{e22}
-r^{-2}\left(rf^{\prime}+\frac{1}{2}r^2 f^{\prime\prime}\right)+\frac {\beta
}{{r}^{2}}\left({r}^{2}f^{\prime\prime}  +4r f^{\prime}-2+2\,f \right)
=-\frac{q^2}{ r^4}.
\end{equation}
Solving Eq.s \eqref{e00} and \eqref{e22}, we get the general solution for the 
metric function,
\begin{equation}\label{f1}
f(r)=1-\frac{2 M}{r}+\frac{q^2}{r^2}+\frac{a (2 \beta -1)^2 r^{\frac{4 \beta }{1-2 \beta }}}{8 \beta ^2+2 \beta -1},
\end{equation}
where $a = \kappa b$. We shall call this term $a$ as cloud of string parameter 
as from \eqref{cloud_density}, one can see that this parameter or the 
integration constant $b$ directly accounts for the cloud of string density. 
This parameter is again restricted by weak energy condition $a \ge 0$. 

With Eq.\ (\ref{f1}), the metric (\ref{metric}) takes the form: 
\begin{equation}\label{metric01}
ds^2=\left(1-\frac{2 M}{r}+\frac{q^2}{r^2}+\frac{a (2 \beta -1)^2 r^{\frac{4 \beta }{1-2 \beta }}}{8 \beta ^2+2 \beta -1}\right)dt^2
-\frac{dr^2}{1-\frac{2 M}{r}+\frac{q^2}{r^2}+\frac{a (2 \beta -1)^2 r^{\frac{4 \beta }{1-2 \beta }}}{8 \beta ^2+2 \beta -1}}
-r^2 d\Omega^2.
\end{equation}
The above metric \eqref{metric01} represents Reissner-Nordstr\"om black hole 
surrounded by a cloud of strings in Rastall gravity. The Ricci scalar for the 
above metric is given by
\begin{equation} \label{ricci_m01}
R = \frac{2 a r^{\frac{4 \beta }{1-2 \beta }-2}}{4 \beta -1}.
\end{equation}
One can see from this equation that the Ricci scalar depends on the Rastall 
parameter $\beta$ as well as the cloud of string parameter $a$, but not the 
charge of the black hole $q$. The Ricci squared is given by
\begin{align} \label{riccisq_m01}
R_{\mu\nu}R^{\mu\nu} &= 2\,r^{-8} \left(\frac{a^2 \left(8 \beta ^2-4 \beta +1\right) r^{\frac{4}{1-2 \beta }}}{(1-4 \beta )^2}+2 a q^2 r^{\frac{2}{1-2 \beta }}+2 q^4\right),
\end{align}
and the Kretschmann scalar is
\begin{multline} \label{kret_m01}
R_{\alpha\beta\mu\nu}R^{\alpha\beta\mu\nu} = 4\,r^{-8} \bigg(\frac{a^2 (4 \beta  (\beta  (4 \beta  (14 \beta -9)+11)-2)+1) r^{\frac{4}{1-2 \beta }}}{\left(8 \beta ^2+2 \beta -1\right)^2}\\+\frac{2 a r^{\frac{2}{1-2 \beta }} \left(2 (1-6 \beta ) M r+(14 \beta -1) q^2\right)}{2 \beta +1}+ 2 \left(6 M^2 r^2-12 M q^2 r+7 q^4\right)\bigg).
\end{multline}
From the expressions \eqref{ricci_m01}, \eqref{riccisq_m01} and 
\eqref{kret_m01}, one can see that the black hole is a singular black hole 
for any values of $a$ and $\beta$. The Kretschmann scalar as well as the 
Ricci squared depends on the electrodynamic source. In the limit of 
$\beta\rightarrow0$ and $q\rightarrow0$ in the metric \eqref{metric01}, one 
can recover the black hole surrounded by a cloud of strings in GR 
\cite{Letelier} as 
\begin{equation}\label{metric02}
ds^2=\left(1-\frac{2M}{r}-a \right)dt^2
-\frac{dr^2}{
1-\frac{2M}{r}
-a}-r^2 d\Omega^2.
\end{equation} 
{ One may note that the spacetime represented by this  metric 
\eqref{metric02} is asymptotically flat as the third term $a$ in the metric 
function is a constant term. One may realise this from the Ricci curvature 
from Eq.~\eqref{R}, which gives for this metric as
\begin{equation}
R = -\frac{2 a}{r^2}.
\end{equation}
At asymptotic limit i.e. $r \rightarrow \infty$,
\begin{equation}
R = 0.
\end{equation}
Which represents flat spacetime for any values of $a$. Hence, in GR, the term $a$ can mimic the deficit angle \cite{new01,new02}. However, in Rastall gravity, due to the presence of $\beta$, the term can not mimic the deficit angle. Hence, in Rastall gravity, $a$ is expected to provide different signatures on QNMs depending on the value of $\beta$.
}

The QNMs of non-linearly charged black holes surrounded by dark energy type 
fields were studied recently in \cite{Gogoi2021}. In this study it was seen 
that the QNMs of non-linearly charged black holes vary significantly from that 
of linearly charged black holes. However, till now to the best of our 
knowledge, no study has been done using non-linearly charged black holes 
surrounded by a cloud of strings in Rastall gravity. Thus this study will shed 
some light on the behaviour of black holes in Rastall gravity with non-linear 
electrodynamic sources and also surrounded by a cloud of strings. In this 
study, we shall study two types of non-linearly charged black holes viz., 
ABG black hole and a new black hole in Rastall gravity surrounded by a cloud 
of strings.

\subsection{Ay\'on - Beato - Garc\'ia Black Hole surrounded by a Cloud of 
Strings}

Here we use a non-linear electrodynamic distribution function 
$\sigma(r)$ and with this function we can modify the metric \eqref{metric01} as 
\cite{Gogoi2021}:
\begin{equation}\label{gen_metric}
ds^2=\left(1-\frac{2m(r)}{r}
+\frac{a (2 \beta -1)^2 r^{\frac{4 \beta }{1-2 \beta }}}{8 \beta ^2+2 \beta -1}\right)dt^2
-\frac{dr^2}{1-\frac{2m(r)}{r}
+\frac{a (2 \beta -1)^2 r^{\frac{4 \beta }{1-2 \beta }}}{8 \beta ^2+2 \beta -1}}
-r^2 d\Omega^2,
\end{equation}
where $$m(r) = \dfrac{\sigma(r)}{\sigma_\infty}M.$$ The function $\sigma(r)>0$ 
and $\sigma'(r)>0$ for $r \geq 0$. Again, $\sigma(r)/r \rightarrow 0$ as $r 
\rightarrow 0$. $\sigma_\infty$ is a normalisation constant at infinity, i.e. 
$\sigma_\infty = \sigma(r \rightarrow \infty)$. For the ABG black hole the
function $m(r)$ takes the form \cite{ayon1999}:
\begin{equation} \label{mabg}
m_{ABG}(r) = M \left\lbrace 1 - \tanh \left( \dfrac{q^2}{2 M r} \right) \right\rbrace.
\end{equation} 
Using this ABG function the metric \eqref{gen_metric} can be written as
\begin{equation}\label{ABG_metric}
ds^2=\left(1-\frac{2M \lbrace 1 - \tanh( \frac{q^2}{2 M r}) \rbrace}{r}
+\frac{a (2 \beta -1)^2 r^{\frac{4 \beta }{1-2 \beta }}}{8 \beta ^2+2 \beta -1}\right)dt^2
-\frac{dr^2}{1-\frac{2M \lbrace 1 - \tanh( \frac{q^2}{2 M r}) \rbrace}{r}
+\frac{a (2 \beta -1)^2 r^{\frac{4 \beta }{1-2 \beta }}}{8 \beta ^2+2 \beta -1}}
-r^2 d\Omega^2.
\end{equation}
This is the ABG black hole solution surrounded by a cloud of strings in Rastall
gravity. The ABG black hole is a regular black hole \cite{ayon1999}, which 
removes the singularity issue arising from the terms containing the mass and 
charge of the black hole. However, in this case, we expect three black hole 
horizons due to the mass, charge and surrounding field terms containing the hairs of the black hole. Thus, if we do not have any singularity issue arising 
from the surrounding field terms, the black hole will be a regular one.

\begin{figure}[htb]
\centerline{
   \includegraphics[scale = 0.3]{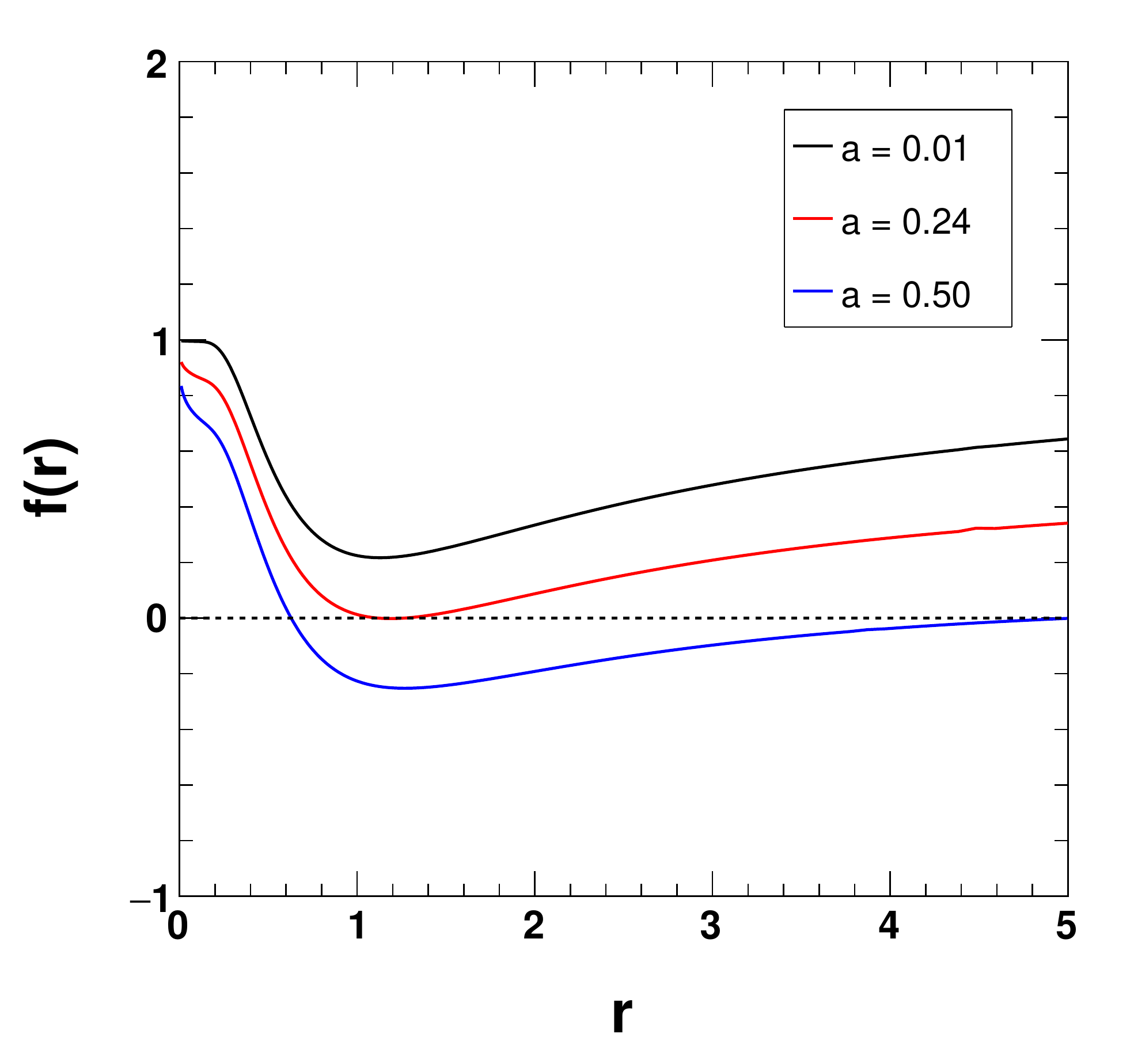}\hspace{1cm}
   \includegraphics[scale = 0.3]{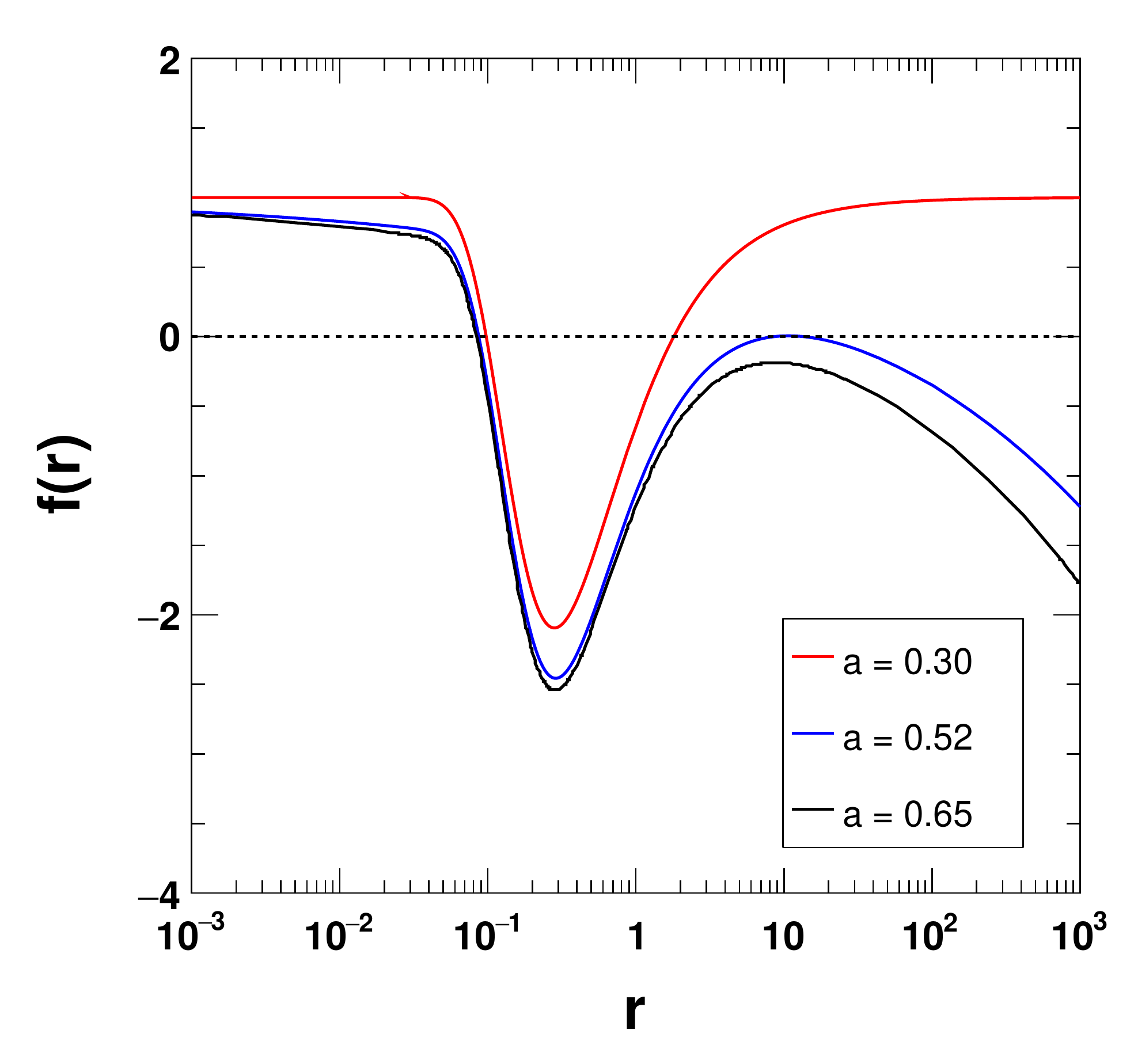}}
\vspace{-0.2cm}
\caption{Variation of the metric function $f(r)$ with respect to $r$ for the 
ABG black hole metric \eqref{ABG_metric} in Rastall gravity surrounded 
by a cloud of strings with $q = 1.2$, $\beta=0.05$ and $M = 1$ (on left panel), 
and $q = 0.6$, $\beta=0.05$ and $M = 1$ (on right panel) for different $a$ 
values.}
\label{fig01}
\end{figure} 

To see the properties of this black hole properly, we have plotted the metric 
function of the black hole with respect to $r$. In Fig.\ \ref{fig01} the black 
hole metric $f(r)$ is plotted with respect to $r$ taking $q = 1.2$, 
$\beta=0.05$ and $M = 1$ (on left panel), and $q = 0.6$, $\beta=0.05$ and 
$M = 1$ (on right panel) for different $a$ values. One can see that for the 
plot on the left panel, the critical value of $a$ is around $0.24$. At this 
value of $a$, in the small $r$ regime, we have one horizon of the black hole. 
If the value of $a$ is increased beyond this, two horizons are obtained within 
the small $r$ range. Basically, the charge and the mass of the black hole are
responsible for the first dip in the curves, but the cloud of strings parameter 
$a$ has very small impact on it unless the charge is high. This can be properly 
visualized from the plot on the right panel of Fig.\ \ref{fig01}. Here the 
charge $q=0.6$ and the variation of $a$ imposes very less impact on the first 
dip of the curves. Thus for small charges, the first horizon of the black hole 
is affected slightly for different values of $a$. However, the second horizon 
varies significantly with respect to $a$. The last horizon or the third horizon
which is appearing due to the cloud of strings field has the maximum $a$ 
dependency as expected. Hence, we will call the third horizon as the cloud of 
string field horizon, $r_c$. From the plot on the right panel of this figure, 
one can also see that for the second and the third horizon the critical value 
of $a$ is around $0.52$ and above this value, the second and the third horizon 
disappear completely. Below this critical value three horizons are obtained. 
But if $a$ is lowered further, the third horizon moves towards $+\infty$ and 
at $a=0$ the third horizon vanishes.

\begin{figure}[htb]
\centerline{
   \includegraphics[scale = 0.3]{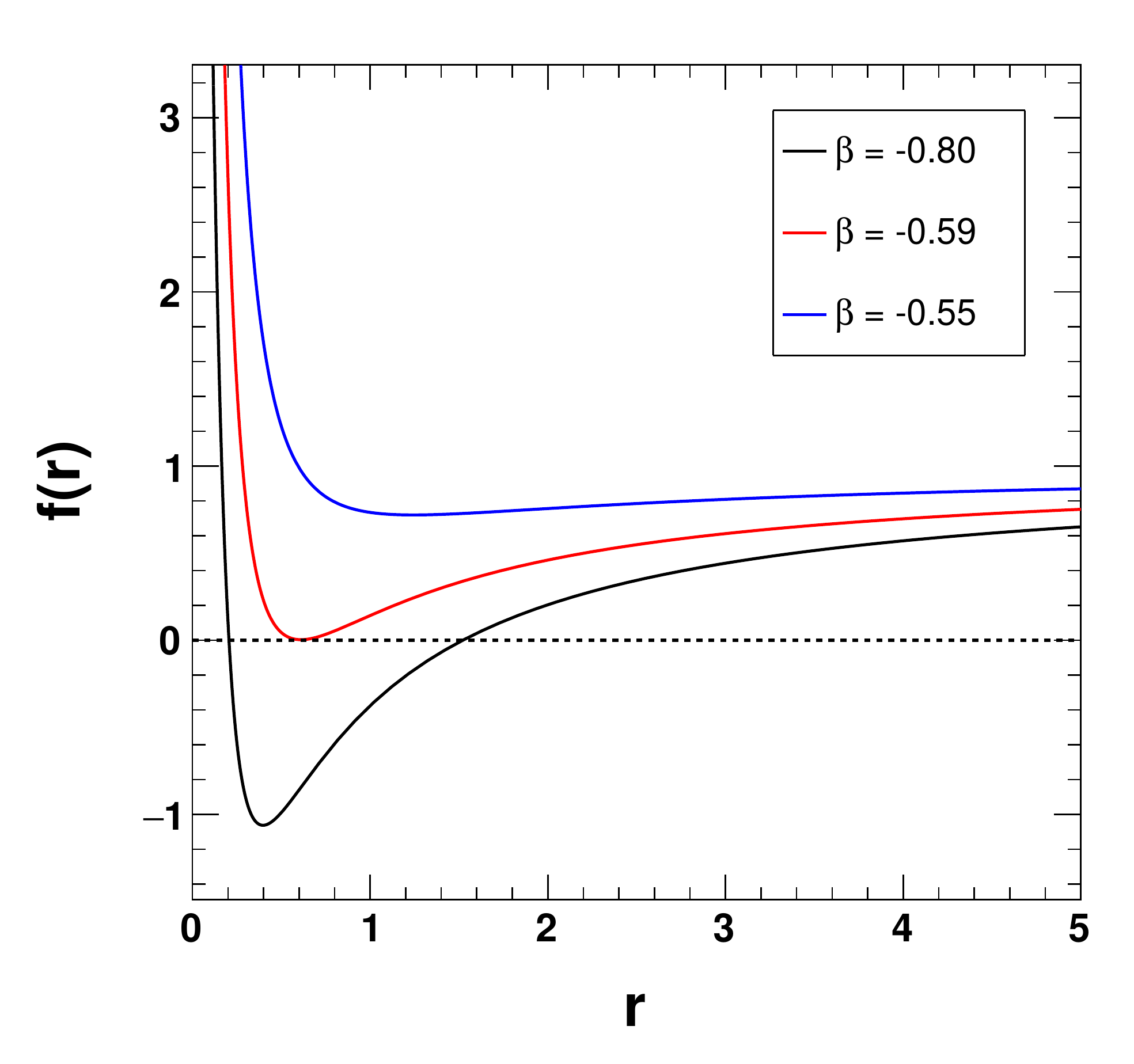}\hspace{1cm}
   \includegraphics[scale = 0.3]{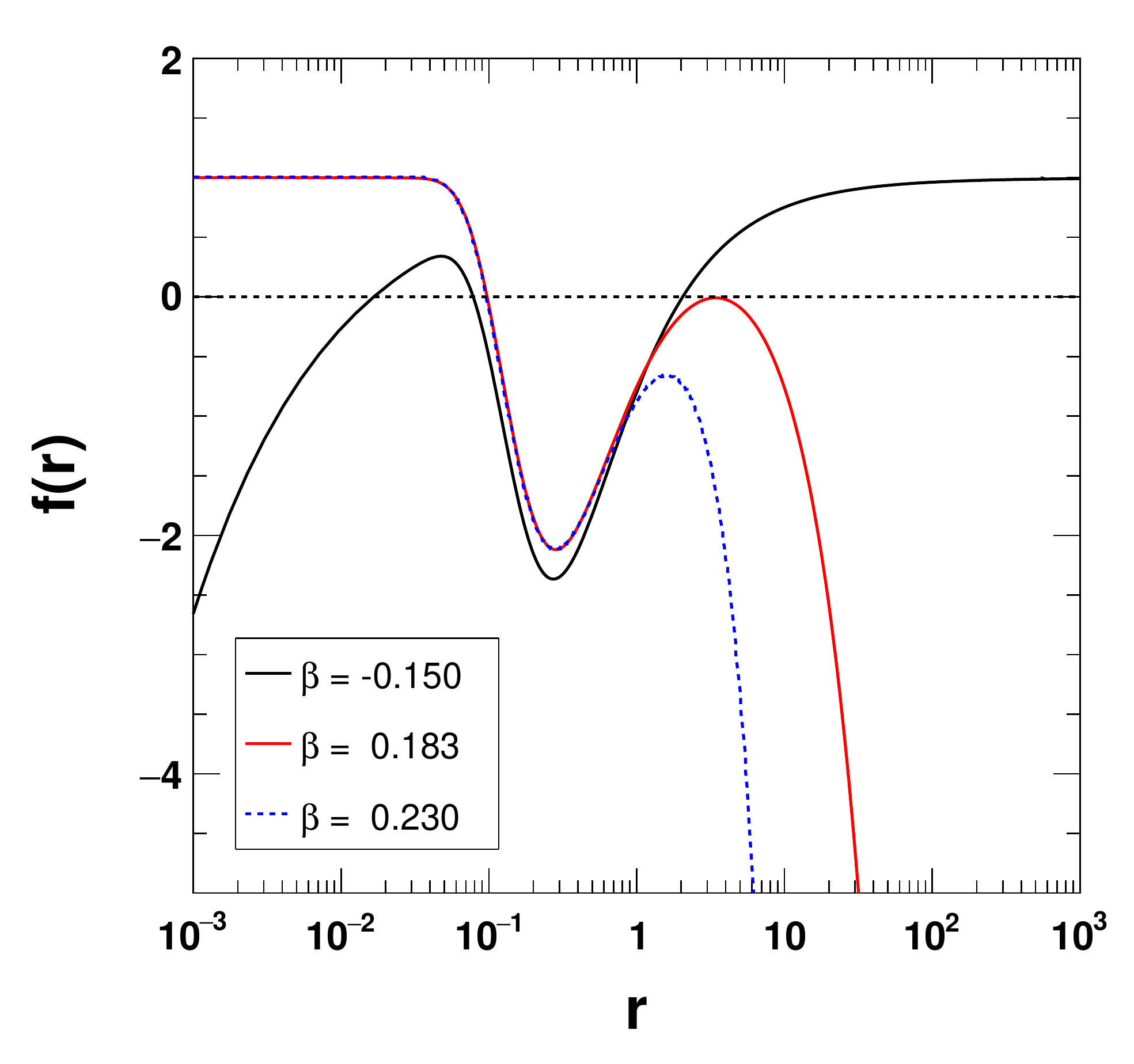}}
\vspace{-0.2cm}
\caption{Variation of the metric function $f(r)$ with respect to $r$ for the 
ABG black hole metric \eqref{ABG_metric} in Rastall gravity surrounded by 
cloud of strings with $q = 0.6$, $a=0.1$ and $M = 1$ for different $\beta$ 
values.}
\label{fig02}
\end{figure}

In Fig.\ \ref{fig02} the metric function $f(r)$ is plotted with respect to 
$r$ for different values of $\beta$. Figure shows that the Rastall parameter 
$\beta$ can influence all the three horizons of the black hole. The plot on 
the left panel shows that for $\beta$ around $-0.59$, we have only one horizon 
of the black hole. For any values greater than this, naked singularity is 
formed. For $\beta$ around $0.183$ (see the plot on the right panel of 
Fig.\ \ref{fig02}), the black hole has two horizons. In this case, the cloud 
of string horizon i.e. $r_c$ and the second horizon coincide to give a single 
horizon. If one increases $\beta$ beyond this value, the black hole shows one 
horizon only. Thus, the Rastall parameter can introduce significant changes on 
all the horizons of the black hole. Positive values of $\beta$ influence the 
cloud of string horizon $r_c$ most, while the negative $\beta$ has more 
impacts on the first and second horizons of the black hole.  

\begin{figure}[htb]
\centerline{
   \includegraphics[scale = 0.3]{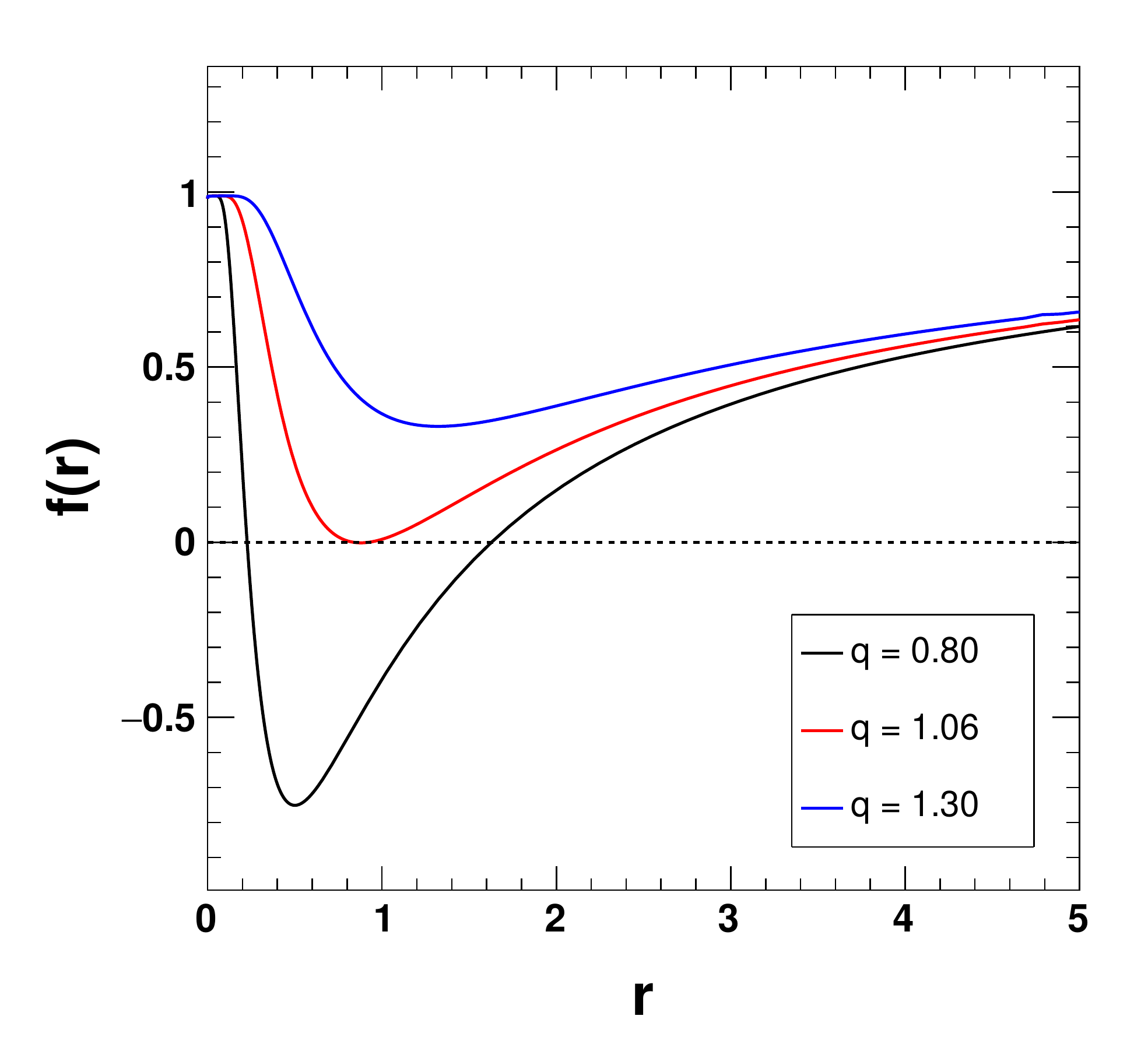}\hspace{1cm}
   \includegraphics[scale = 0.3]{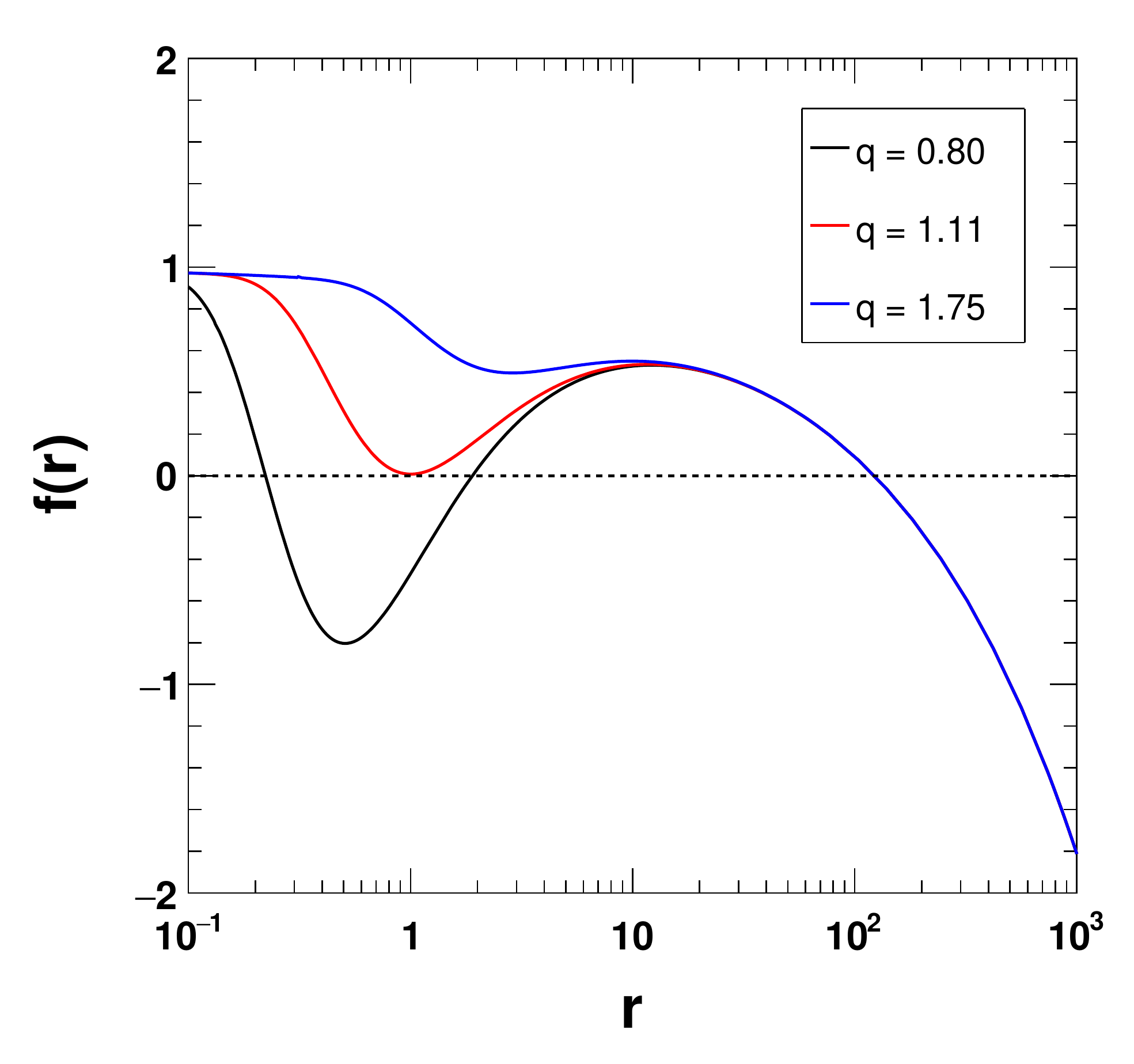}}
\vspace{-0.2cm}
\caption{Variation of metric function $f(r)$ with respect to $r$ for ABG
black hole metric \eqref{ABG_metric} in Rastall gravity surrounded by a cloud 
of strings with $\beta = -0.01$, $a=0.01$ and $M = 1$ (on left panel), and 
$\beta = 0.1$, $a=0.1$ and $M = 1$ (on right panel) for different $q$ values.}
\label{fig03}
\end{figure}

As already seen, another important parameter of the black hole is its charge 
$q$. To clearly see the impact of this parameter on the black hole structure, 
the plots of the metric function $f(r)$ with respect to $r$ for different 
values of charge $q$ of the black hole are shown in Fig. \ref{fig03}. Here, 
for the plot on the left panel we have chosen $\beta = -0.01$, $a = 0.01$ and 
$M = 1$. We see that for the $q$ around $1.06$, the first and the second 
horizons combine to give a single horizon. For this set of parameters, the 
charge $q > 1.06$ can give naked singularity if there is no cloud of string 
horizon, i.e. $r_c$ is present. For the plot on the right panel, we have 
considered $\beta = 0.1$, $a = 0.1$ and $M = 1$. This plot clearly shows that 
a variation in charge does not have any significant influences over the cloud 
of string horizon $r_c$. In this case, for the $q$ around $1.11$, we obtain 
only two black hole horizons and for $q > 1.11$, we have only one horizon 
of the black hole which is the cloud of string horizon $r_c$.

It is possible to obtain the expression for the associated electric field from 
the field equations. In presence of non-linear electrodynamic sources, the 
effective energy momentum tensor can be written as
\begin{equation} \label{add}
T_{\mu\nu} = L(F) g_{\mu\nu} - L_F F_{\mu \lambda} F^\lambda_\nu + \mathcal{T}_{\mu \nu},
\end{equation}
where $F = \dfrac{1}{4} F^{\mu\nu}F_{\mu\nu}$ (Lorentz invariant) and 
$L_F = \dfrac{dL}{dF}.$ Restricting the electric field by considering 
$F_{\mu\nu} = E(r)(\delta^0_\mu \delta^1_\nu - \delta^1_\mu \delta^0_\nu)$, 
we can express the field equations as
\begin{eqnarray} \label{nonlinearfield}
&&{\mathcal{S}^{0}}_{0}={\mathcal{G}^{0}}_{0}+\beta R=\frac{1}{f(r)}\mathcal{G}_{00}+\beta R=-r^{-2}\Big \lbrace f^{\prime}(r)r-1+f(r)  \Big \rbrace +\beta R= -\,\kappa \left[ L(F) + E^2 L_F - \rho_c(r) \right],\nonumber\\
&&{\mathcal{S}^{1}}_{1}={\mathcal{G}^{1}}_{1}+\beta R=-f(r) \mathcal{G}_{11}+\beta R=-r^{-2}\Big \lbrace f^{\prime}(r)r-1+f(r)  \Big \rbrace +\beta R= -\,\kappa \left[ L(F) + E^2 L_F - \rho_c(r) \right],\nonumber\\
&&{\mathcal{S}^{2}}_{2}={\mathcal{G}^{2}}_{2}+\beta R=-r^{-2}\mathcal{G}_{22}+\beta R=-r^{-2}\Big \lbrace rf^{\prime}(r)+\frac{1}{2}r^2 f^{\prime\prime}(r)\Big \rbrace +\beta R= -\,\kappa L(F),\nonumber\\
&&{\mathcal{S}^{3}}_{3}={\mathcal{G}^{3}}_{3}+\beta R=-\frac{1}{r^2 \sin^2 \theta}\mathcal{G}_{33}+\beta R=-r^{-2}\Big \lbrace rf^{\prime}(r)+\frac{1}{2}r^2 f^{\prime\prime}(r)\Big \rbrace +\beta R= -\,\kappa L(F).
\end{eqnarray}
Again the electromagnetic field equations, $\nabla_\mu (F^{\mu \nu} L_F) = 0$ 
give,
 \begin{equation}\label{ELF}
 E(r) L_F = - \dfrac{q}{4 \pi r^2}.
 \end{equation}
Using the above result in the field Eq.s \eqref{nonlinearfield}, we can 
have
 \begin{equation}\label{abgE}
 E(r) =\frac{q \left(4 M r-q^2 \tanh \left(\frac{q^2}{2 M r}\right)\right) \text{sech}^2\left(\frac{q^2}{2 M r}\right)}{4 M r^3},
 \end{equation}
which is the expression of the electric field for the ABG type non-linear 
charge distribution. It is clear from the above expression that the 
 electric field is regular everywhere and asymptotically it results,
 \begin{equation}
 E(r) \approx \dfrac{q}{r^2}.
 \end{equation}
This result ascertains that the ABG type black hole becomes a 
Reissner-Nordstr\"om black hole in the asymptotic regime or for a very very
small magnitude of charge.
 Using Eq.s \eqref{abgE} and \eqref{ELF} in the first field equation of Eq.s \eqref{nonlinearfield}
 we can obtain an explicit the form of $L(F)$ as,
 \begin{equation}
L(F) = \frac{q^2 sech^2\left(\frac{q^2}{2 M r}\right) \left((2 \beta -1) q^2 \tanh \left(\frac{q^2}{2 M r}\right)+2 M r\right)}{16 \pi  M r^5}.
\end{equation}
 It may be noted that one may also obtain this expression directly from third or fourth field equation of Eq.s \eqref{nonlinearfield}.

Now to see the black hole nature, we calculate Ricci scalar, Ricci squared and 
Kretschmann scalar. These scalars are helpful to predict the singularity of 
the black hole, if there exists any. Ricci scalar for ABG black hole is
\begin{equation}
R = \frac{2 a r^{\frac{4 \beta }{1-2 \beta }-2}}{4 \beta -1}-\frac{8 q^4 \sinh ^4\left(\frac{q^2}{2 M r}\right) \text{csch}^3\left(\frac{q^2}{M r}\right)}{M r^5}.
\end{equation}
Ricci squared is found to be, 

\begin{align} \label{ricci_sq_ABG}
R_{\mu\nu}R^{\mu\nu} &= \frac{4 M (2 \beta  M+M) \left(a (1-2 \beta ) r^{\frac{2}{1-2 \beta }}+(1-4 \beta ) q^2 \text{sech}^2\left(\frac{q^2}{2 M r}\right)\right)^2+X}{2 (1-4 \beta )^2 M r^8 (2 \beta  M+M)}
\end{align}
where,
$$X = (2 \beta +1)\,r^{-2} \left((4 \beta -1) q^2 \left(q^2 \tanh \left(\frac{q^2}{2 M r}\right)-2 M r\right) \text{sech}^2\left(\frac{q^2}{2 M r}\right)-4 a \beta  M r^{\frac{4 \beta }{1-2 \beta }+3}\right)^2.$$
And finally the Kretschmann scalar is
\begin{align} \label{kret_ABG}
R_{\alpha\beta\mu\nu}R^{\alpha\beta\mu\nu} &= 4\,r^{-10} \left(r^6 \left(\frac{a (1-2 \beta )^2 r^{\frac{4 \beta }{1-2 \beta }}}{8 \beta ^2+2 \beta -1}+\frac{2 M \left(\tanh \left(\frac{q^2}{2 M r}\right)-1\right)}{r}\right)^2+r^2 Y^2\right)
\end{align}
where,
$$Y = y_1^2+4 \left(\frac{a \beta  (6 \beta -1) r^{\frac{4 \beta }{1-2 \beta }+3}}{8 \beta ^2+2 \beta -1}+M r^2 y_2+q^2 r \text{sech}^2\left(\frac{q^2}{2 M r}\right)-\frac{2 q^4 \sinh ^4\left(\frac{q^2}{2 M r}\right) \text{csch}^3\left(\frac{q^2}{M r}\right)}{M}\right),$$

$$y_1 = \frac{4 a \beta  (2 \beta -1) r^{\frac{2}{1-2 \beta }}}{8 \beta ^2+2 \beta -1}+2 M r y_2+q^2 \text{sech}^2\left(\frac{q^2}{2 M r}\right),$$
$$ y_2 = \left(\tanh \left(\frac{q^2}{2 M r}\right)-1\right).$$
From the above expressions it is seen that the Ricci scalar for this black 
hole differs completely from the Ricci scalar of Reissner-Nordstr\"om black 
hole. In case of the Reissner-Nordstr\"om black hole surrounded by a cloud of 
strings in Rastall gravity defined by the metric \eqref{metric01}, the Ricci 
scalar is independent of the charge of the black hole. But in this case, Ricci 
scalar depends on the charge of the black apart from the Rastall parameter. 
Thus, the non-linear charge distribution function used in  Eq.\ \eqref{add} 
has a significant impact over the Ricci scalar of the ABG black hole. The Ricci squared and the Kretschmann scalar are also different from the 
Reissner-Nordstr\"om black hole surrounded by a cloud of strings in Rastall 
gravity. However, one important point worth to mention is that for this case, 
at $\beta = 0$, i.e. at the GR limit, we have a regular black hole. In case of 
Rastall gravity, we see that for $0.25 < \beta < 0.50$, the singularity 
issue arising from the terms containing $\beta$ vanish resulting in a regular 
black hole. This constraint on $\beta$ can be obtained from the Kretschmann 
scalar \eqref{kret_ABG}. It is seen from Eq. \eqref{kret_ABG} that for the Kretschmann scalar to be finite at the limit $r\rightarrow 0$, the $\beta$ dependent terms should not diverge. One may note that due to the consideration of the ABG non-linear charge distribution, one does not get singularity issues from the mass and charge dependent terms in the metric and the scalars. Now, taking account of the mass and charge independent terms in the Eq. \eqref{kret_ABG}, we have obtained that $0.25 < \beta < 0.50$ gives a regular black hole solution. 

\subsection{A new Black Hole solution surrounded by a Cloud of Strings}
\begin{figure}[htb]
\centerline{
   \includegraphics[scale = 0.3]{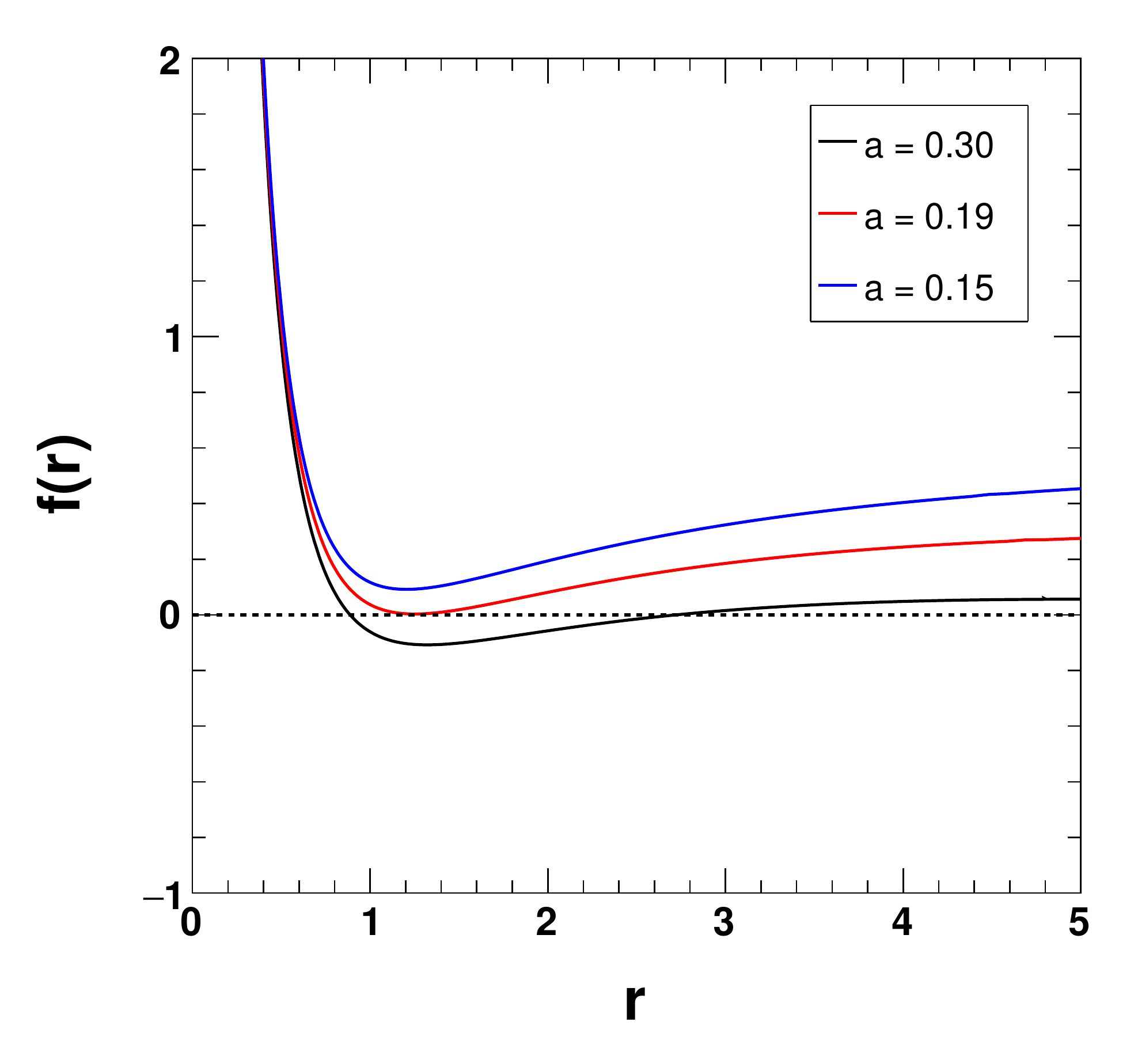}\hspace{1cm}
   \includegraphics[scale = 0.3]{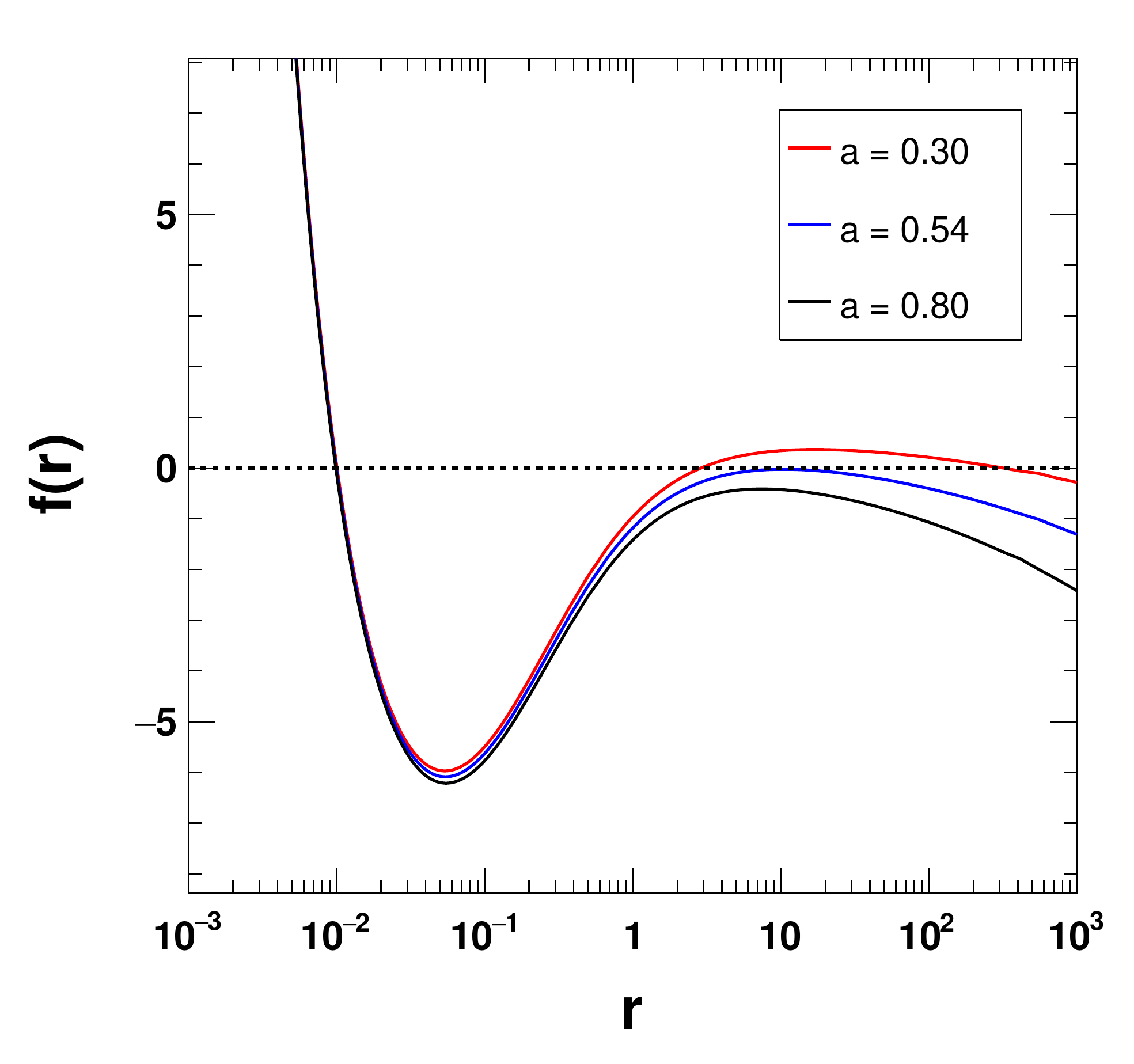}}
\vspace{-0.2cm}
\caption{Variation of metric function $f(r)$ with respect to $r$ for the black 
hole defined by metric \eqref{our_metric} with $q = 1.15$, $\beta = 0.1$, 
$\zeta = -0.05$ and $M = 1$ (on left panel), and $q = 0.60$, $\beta = 0.05$, 
$\zeta = -0.80$ and $M = 1$ (on right panel) for different $a$ values. }
\label{fig04}
\end{figure}

At this stage as a possible alternative to the function \eqref{mabg}, we 
introduce a new distribution function as given by
\begin{equation}\label{our_model}
m(r) = M \left\lbrace 1 + \arccsch\left(\zeta - \frac{2 M r}{q^2} \right)\right\rbrace,
\end{equation}
for which the black hole metric \eqref{gen_metric} can be expressed as
\begin{align}\label{our_metric}
ds^2 &=\left(1-\frac{2M \lbrace 1 + \arccsch(\zeta - \frac{2 M r}{q^2} )\rbrace}{r}
+\frac{a (2 \beta -1)^2 r^{\frac{4 \beta }{1-2 \beta }}}{8 \beta ^2+2 \beta -1}\right)dt^2 \\ \notag 
&-\frac{dr^2}{1-\frac{2M \lbrace 1 + \arccsch(\zeta - \frac{2 M r}{q^2} )\rbrace}{r}
+\frac{a (2 \beta -1)^2 r^{\frac{4 \beta }{1-2 \beta }}}{8 \beta ^2+2 \beta -1}}
-r^2 d\Omega^2.
\end{align}
Here $\zeta$ is a free model parameter. To study the behaviour of this metric 
of the black hole, we have plotted the metric function $f(r)$ with respect to 
$r$ for different values of $a$ in Fig.\ \ref{fig04}. On the left panel of 
Fig.\ \ref{fig04}, we have chosen $q = 1.15$, $\beta = 0.1$, $\zeta = -0.05$ 
and $M = 1$. For this case, we show that when the value of $a$ is around 
$0.19$, the first and second horizons of the black hole are coincided to 
give a single horizon and for $a<0.19$, only the third horizon $r_c$ is left. 
For the plot on the right panel, we have considered $q = 0.6$, $\beta = 0.05$, 
$\zeta = -0.8$ and $M = 1$. In this case, we see that variation of $a$ has 
comparatively less influence over the first horizon and for $a$ around $0.54$, 
the second and the third horizons are combined to give a single horizon. For 
this set of parameters, if one increases $a$ beyond $0.54$, the second and 
the third horizon of the black hole vanish completely.

\begin{figure}[htb]
\centerline{
   \includegraphics[scale = 0.3]{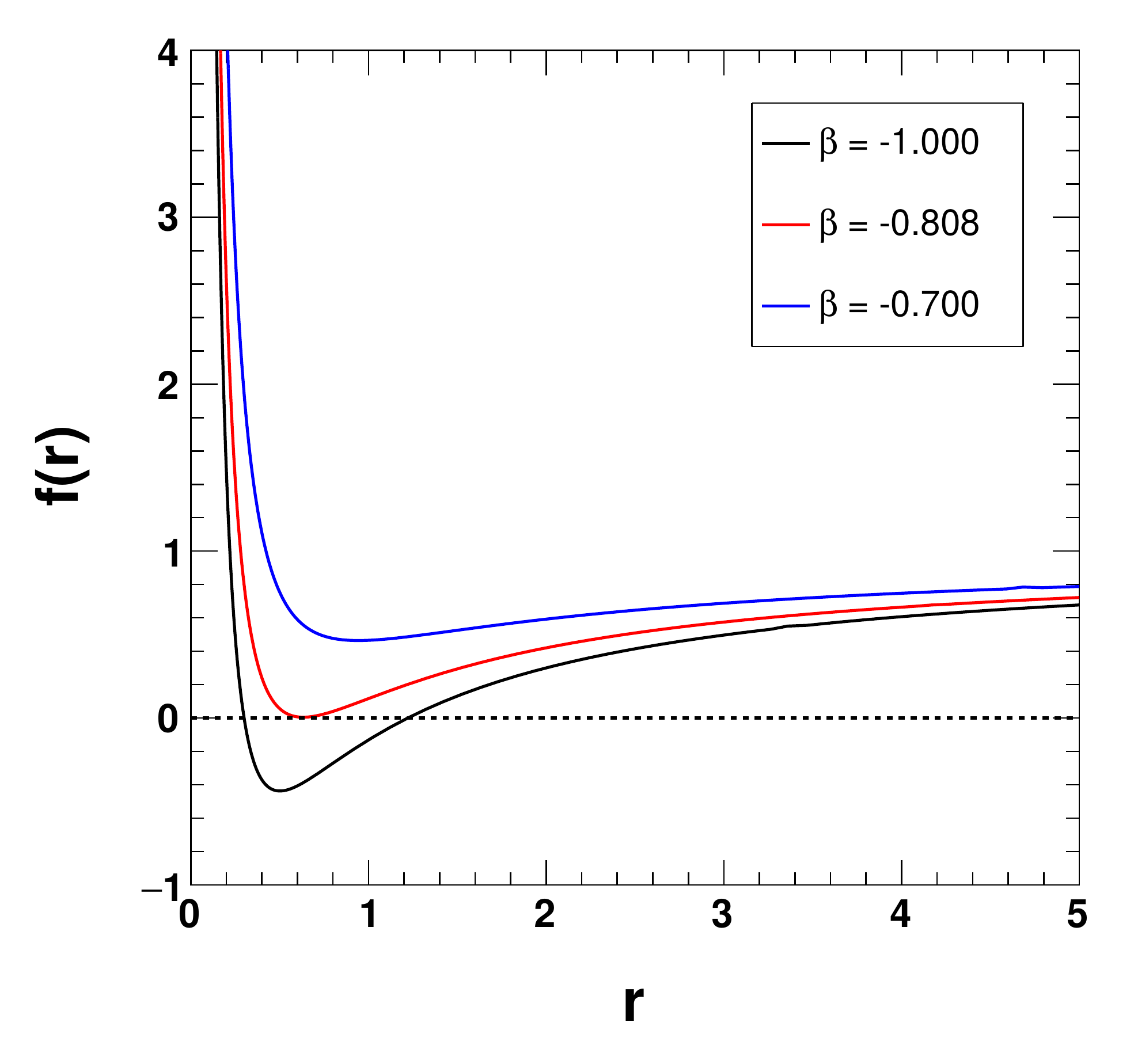}\hspace{1cm}
   \includegraphics[scale = 0.3]{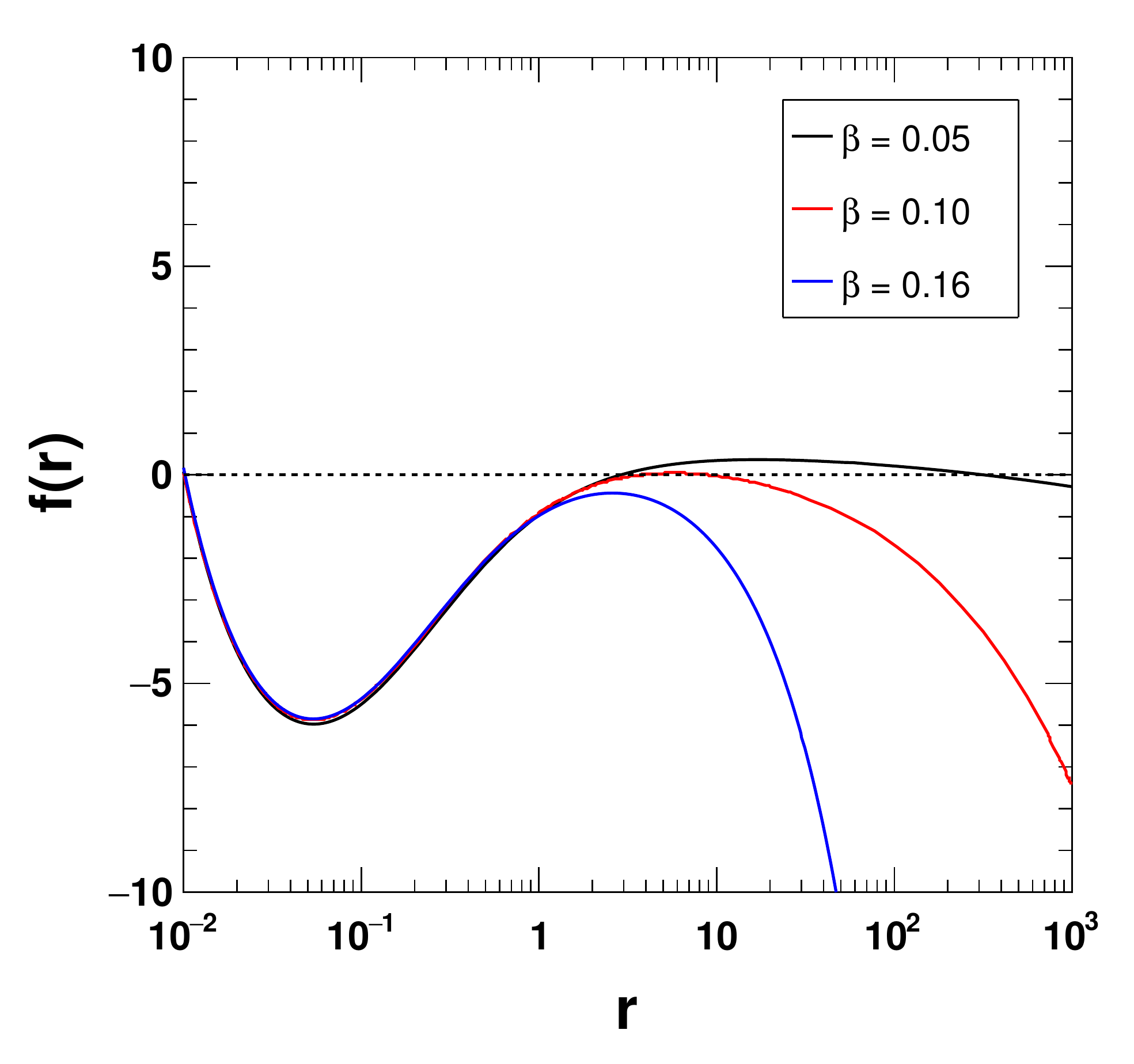}}
\vspace{-0.2cm}
\caption{Variation of metric function $f(r)$ with respect to $r$ for the black 
hole defined by metric \eqref{our_metric} with $q = 0.6$, $a=0.3$, 
$\zeta = -0.5$ and $M = 1$ (on left panel), and $q = 0.6$, $a=0.3$, 
$\zeta = -0.8$ and $M = 1$ (on right panel) for different $\beta$ values.}
\label{fig05}
\end{figure}

In Fig.\ \ref{fig05} the metric function $f(r)$ is plotted with respect to 
$r$ for different $\beta$ values. Similar to the previous black hole metric, 
here also we have seen that the Rastall parameter $\beta $ can influence all 
the three horizons of the black hole. For the plot on the left panel of this 
figure it is seen that for $\beta$ around $-0.808$ the first and second 
horizons are combined to a single horizon. For $\beta > -0.808$, first and 
second horizons vanish and there can be a naked horizon for small values of 
$a$. While on the right panel one can see that for the $\beta$ around $0.10$, 
the second and third horizons are coincided to result in a single black hole 
horizon. For $\beta > 0.10$, second and third horizons vanish completely.

\begin{figure}[htb]
\centerline{
   \includegraphics[scale = 0.3]{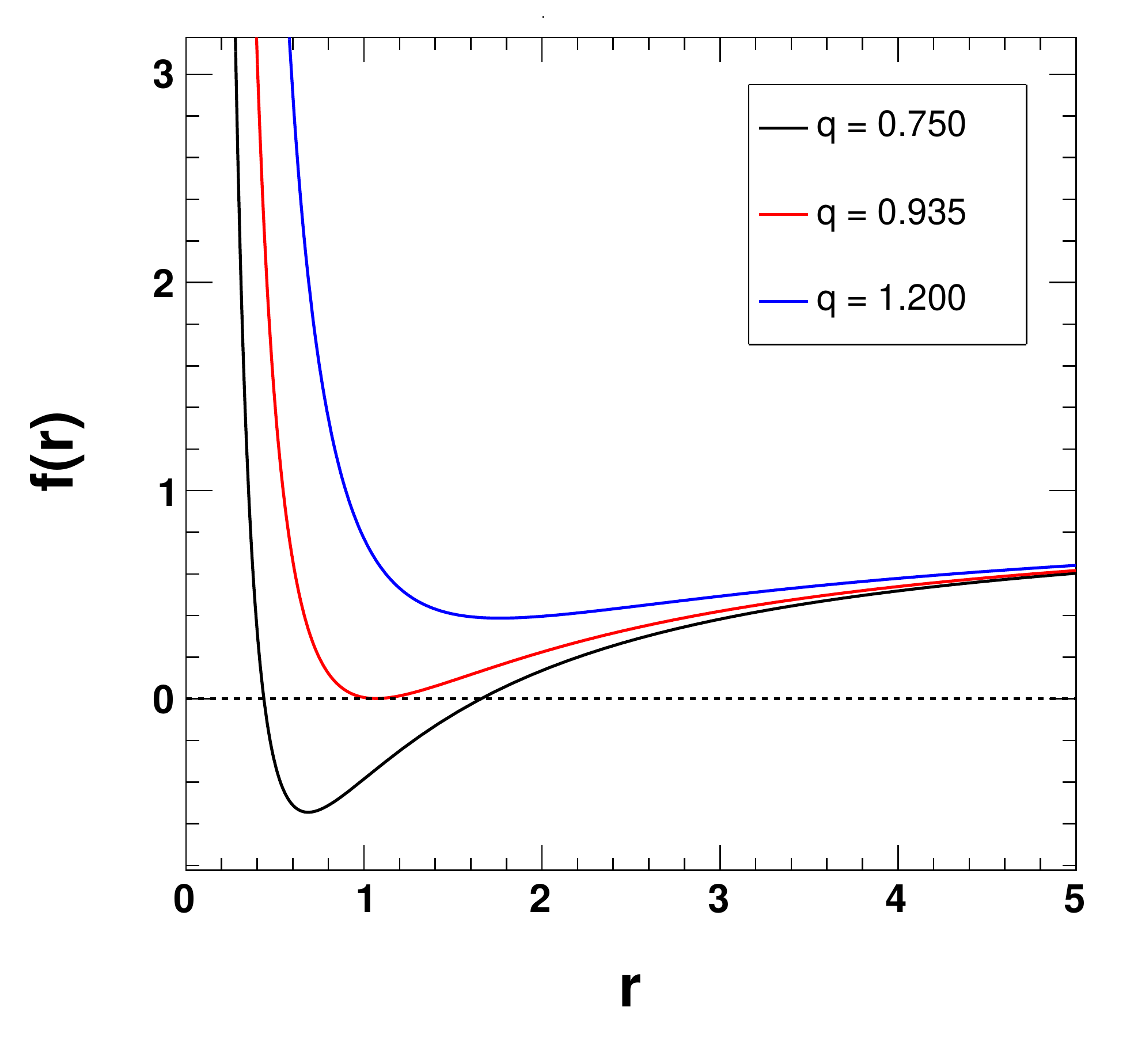}\hspace{1cm}
   \includegraphics[scale = 0.3]{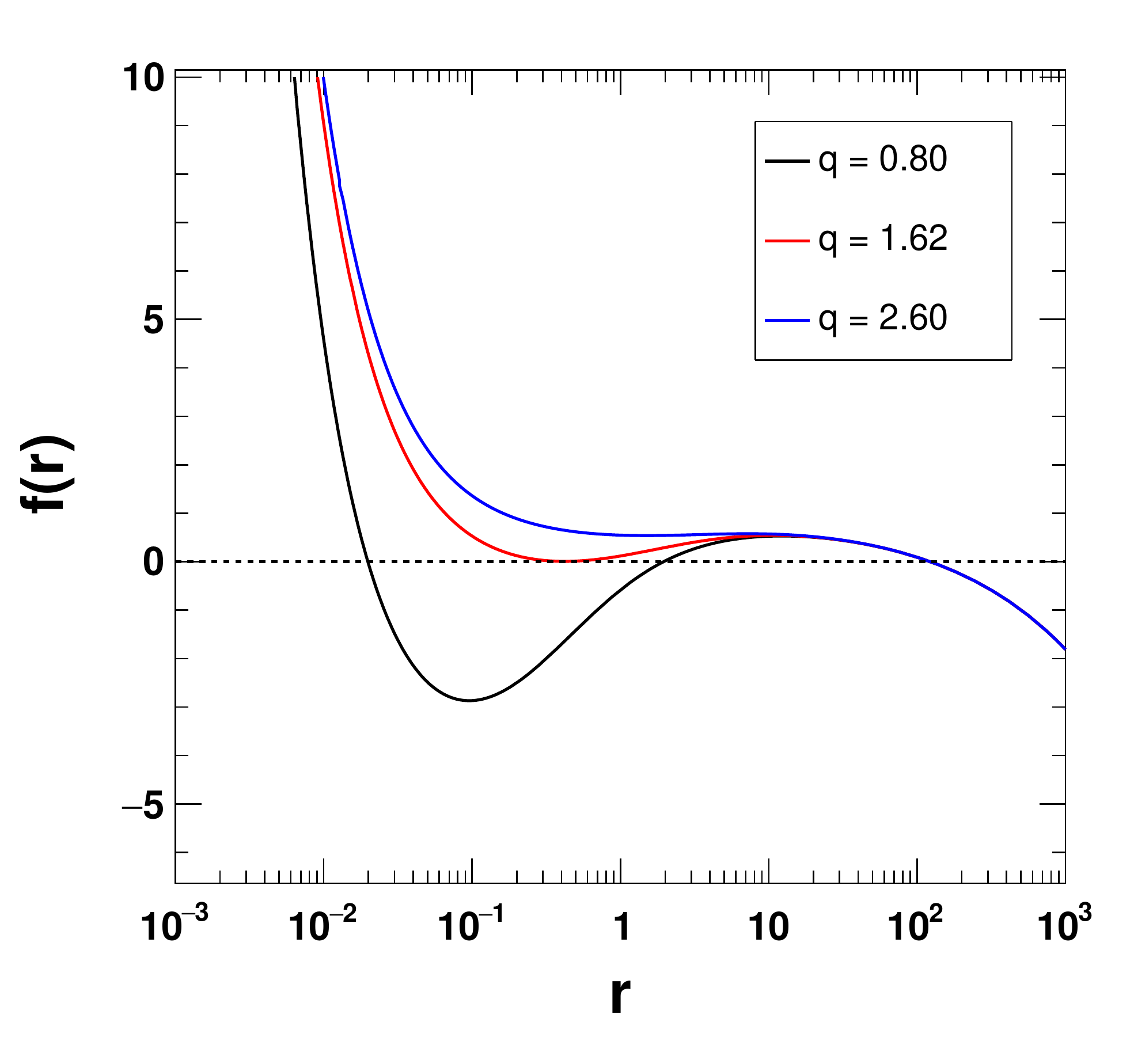}}
\vspace{-0.2cm}
\caption{Variation of metric function $f(r)$ with respect to $r$ for the black 
hole defined by metric \eqref{our_metric} with $\beta = 0.1$, $a=0.01$, 
$\zeta = 0.4$ and $M = 1$ (on left panel), and $\beta = 0.1$, $a=0.1$, 
$\zeta = -0.8$ and $M = 1$ (on right panel) for different $q$ values.}
\label{fig06}
\end{figure}

Similarly, the $f(r)$ function is plotted with respect to $r$ for different
values of the charge $q$ in Fig.\ \ref{fig06}. Here, on the left panel we have 
used $ \beta = 0.1$, $a = 0.01$, $\zeta = 0.4$ and $M = 1$, and on the right 
panel $ \beta = 0.1$, $a = 0.1$, $\zeta = -0.8$ and $M = 1$ are used. In both 
cases, we see that charge $q$ has negligible effect on the third horizon, i.e. 
the cloud of string horizon $r_c$. On the left panel, for $q$ around $0.953$, 
first and second horizons are combined to give a single horizon and 
for $q > 0.953$, first and the second horizons vanish completely. On the right 
panel, $q > 1.62$ values give a single horizon ($r_c$) of the black hole.

\begin{figure}[htb]
\centerline{
   \includegraphics[scale = 0.3]{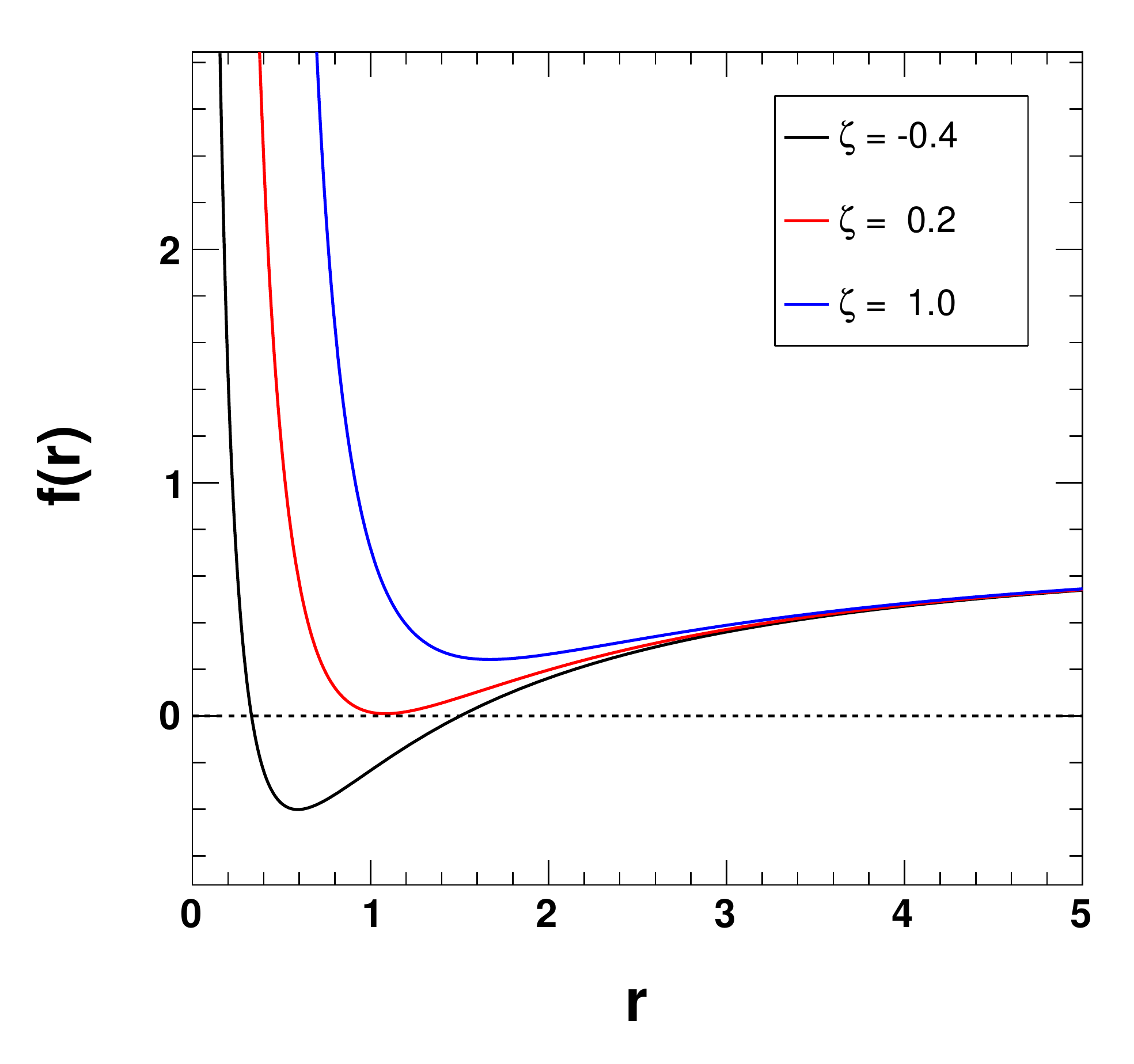}\hspace{1cm}
   \includegraphics[scale = 0.3]{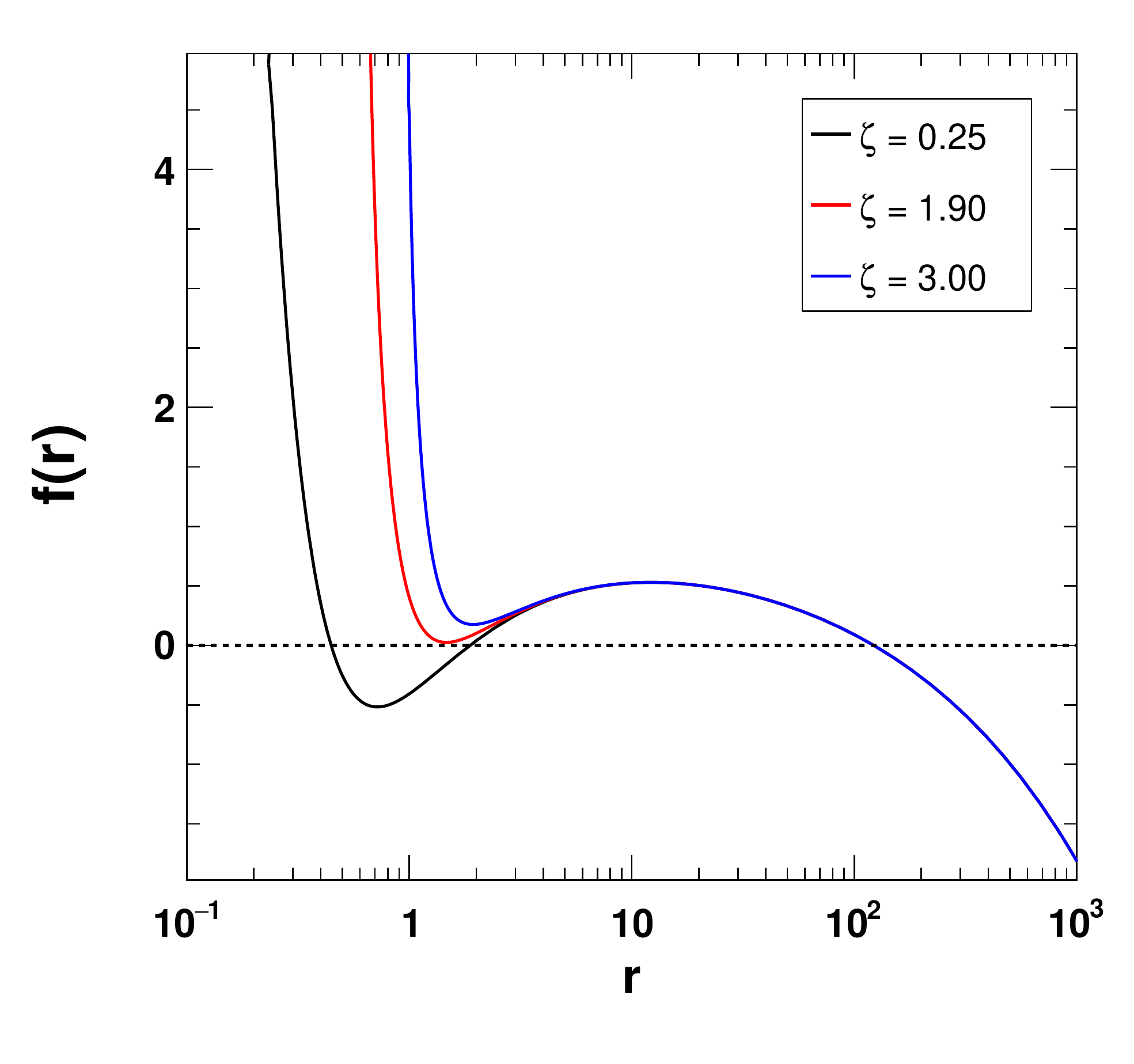}}
\vspace{-0.2cm}
\caption{Variation of metric function $f(r)$ with respect to $r$ for the black 
hole defined by metric \eqref{our_metric} with $\beta = 0.1$ $a=0.05$, $q = 1$ 
and $M = 1$ (on left panel), and $\beta = 0.1$, $a=0.1$, $q = 0.8$ and $M = 1$ 
(on right panel) for different $\zeta$ values.}
\label{fig07}
\end{figure}

Finally, in Fig.\ \ref{fig07} we have plotted the metric function with respect 
to $r$ for different values of the model parameter $\zeta$. We see that the 
impact of $\zeta$ on the metric function is similar to the charge $q$. On the 
left panel of this figure, we have used $\beta = 0.1$, $a = 0.05$, $q = 1$ and 
$M = 1$. We see that for $\zeta = 0.2$ ($\zeta_c$, the critical value), first 
and the second horizons of the black hole are coincided and for any other 
values greater than this, first and second horizons of the black hole vanish 
completely. On the right panel of the figure, we have increased $a$ to $0.1$ 
and decreased $q$ slightly to $0.8$. We observe that increase in the value of 
$a$, increases the critical value $\zeta_c$. For this case, it is $\zeta_c = 
1.90$ and for any other $\zeta > \zeta_c$, first and the second horizons 
vanish completely.

For the black hole metric \eqref{our_metric}, the expression for associated 
electric field is given by
 \begin{equation}
 E(r) =\frac{2 M^2 q \left(16 M^3 r^3-20 \zeta  M^2 q^2 r^2+\left(8 \zeta ^2+3\right) M q^4 r-\zeta  \left(\zeta ^2+1\right) q^6\right)}{\left(2 M r-\zeta  q^2\right)^3 \sqrt{\frac{1}{\left(\zeta -\frac{2 M r}{q^2}\right)^2}+1} \left(4 M^2 r^2-4 \zeta  M q^2 r+\left(\zeta ^2+1\right) q^4\right)}.
 \end{equation}
This expression shows that $\zeta$ has significant contributions to the 
electric field associated with the black hole. The electric field is regular 
everywhere and similar to the previous case, asymptotically it gives,
 \begin{equation}
 E(r) \approx \dfrac{q}{r^2}.
 \end{equation}
So, although the functional form is different, the charge distributions are 
equivalent in the asymptotic region. Both the black holes, therefore, can give 
Reissner-Nordstr\"om black hole in the limiting situation $q \rightarrow 0$,
i.e. for very small charge. To have a better view on the regularity of the 
black hole, we calculate Ricci scalar, Ricci squared and Kretschmann scalar 
for this black hole. Ricci scalar for this black hole metric \eqref{our_metric} is 
\begin{align}
R &= 8\,r^{-2} \left(\frac{a r^{\frac{4 \beta }{1-2 \beta }}}{16 \beta -4}+\frac{M^2 q^4 \left(4 \zeta  M^2 r^2-\left(4 \zeta ^2+1\right) M q^2 r+\zeta  \left(\zeta ^2+1\right) q^4\right) \sqrt{\frac{q^4}{\left(\zeta  q^2-2 M r\right)^2}+1}}{\left(2 M r-\zeta  q^2\right) \left(4 M^2 r^2-4 \zeta  M q^2 r+\left(\zeta ^2+1\right) q^4\right)^2}\right).
\end{align}
The Ricci squared is found as

\begin{align}
R_{\mu\nu}R^{\mu\nu} &= Z_5^{-1}\Big[Z_1 Z_2^2+2 (2-8 \beta )^2 r^2 Z_4^2+8 (1-4 \beta )^2 Z_6^2\Big],
\end{align}
where
$$Z_1 = (2-8 \beta )^2 \left(\zeta  q^2-2 M r\right)^2 Z_3^2,$$
$$Z_2 = a (1-2 \beta ) r^{\frac{4 \beta }{1-2 \beta }} \left(4 M^2 r^2-4 \zeta  M q^2 r+\left(\zeta ^2+1\right) q^4\right)+4 (1-4 \beta ) M^2 q^2 \sqrt{\frac{1}{\left(\zeta -\frac{2 M r}{q^2}\right)^2}+1},$$
$$ Z_3 = \left(4 M^2 r^2-4 \zeta  M q^2 r+\left(\zeta ^2+1\right) q^4\right),$$

$$Z_4 = \left(a \beta  r^{\frac{2}{1-2 \beta }-3} \left(2 M r-\zeta  q^2\right) Z_3^2+Z_7\right),$$

$$Z_5 = 2 (1-4 \beta )^4 r^4 \left(\zeta  q^2-2 M r\right)^2 Z_3^4,$$

$$Z_6 = \left(a \beta  r^{\frac{4 \beta }{1-2 \beta }} \left(2 M r-\zeta  q^2\right) Z_3^2+r Z_7\right)$$
and
$$Z_7 = 2 (4 \beta -1) M^3 q^2 \left(8 M^2 r^2-8 \zeta  M q^2 r+\left(2 \zeta ^2+1\right) q^4\right) \sqrt{\frac{1}{\left(\zeta -\frac{2 M r}{q^2}\right)^2}+1}.$$
The Kretschmann scalar is
\begin{align} \label{kret_GKG}
R_{\alpha\beta\mu\nu}R^{\alpha\beta\mu\nu} &= 4\, r^{-6} \left[4 A^2+4 (B+C+D)^2+\left\lbrace\frac{a (1-2 \beta )^2 r^{\frac{4 \beta }{1-2 \beta }+1}}{8 \beta ^2+2 \beta -1}+2 M \left(\text{csch}^{-1}\left(\frac{2 M r}{q^2}-\zeta \right)-1\right)\right\rbrace^2\right],
\end{align}
where
$$A = \frac{2 a \beta  (2 \beta -1) r^{\frac{4 \beta }{1-2 \beta }+1}}{8 \beta ^2+2 \beta -1}+D,$$
$$B = \frac{a \beta  (6 \beta -1) r^{\frac{4 \beta }{1-2 \beta }+1}}{8 \beta ^2+2 \beta -1}, $$
$$ C= \frac{4 M^3 q^2 r^2}{\left(2 M r-\zeta  q^2\right)^3 \sqrt{\frac{1}{\left(\zeta -\frac{2 M r}{q^2}\right)^2}+1}}+\frac{2 M^3 r^2}{q^4 \left(\zeta -\frac{2 M r}{q^2}\right)^5 \left(\frac{1}{\left(\zeta -\frac{2 M r}{q^2}\right)^2}+1\right)^{3/2}}$$
and
$$ D = \frac{2 M^2 q^2 r}{\left(\zeta  q^2-2 M r\right)^2 \sqrt{\frac{1}{\left(\zeta -\frac{2 M r}{q^2}\right)^2}+1}}+M \text{csch}^{-1}\left(\frac{2 M r}{q^2}-\zeta \right)-M. $$
From the above expressions it is clear that the Ricci scalar, Ricci square and 
the Kretschmann scalar depend on the charge $q$ of the black hole, model 
parameter $\zeta$ and the Rastall parameter $\beta$. One can see that for 
$\zeta = \arccsch(1) = 0.881374$ and $0.25 < \beta < 0.50$, the black hole is 
a regular one. But for  $\zeta \neq \arccsch(1)$, the black hole is singular 
one. Hence, the metric \eqref{our_metric} can represent both regular and 
singular black holes. However, the regular black hole represented by this 
metric is not identical to the ABG black hole represented by the metric 
\eqref{ABG_metric}.

\section{Quasinormal modes of charged black holes surrounded by a cloud of strings}\label{section3}

In this section, we shall compute the QNMs for scalar perturbation in case of 
the black holes defined earlier. To obtain QNMs for the black holes, we shall 
implement the higher order WKB approximation method which uses higher order 
derivatives of the black hole potential associated with scalar perturbation 
in tortoise coordinates. To obtain the expression for the associated black 
holes' potentials, at first we perturb the black hole with some probe coupled 
to a scalar field minimally, with the equation of motion,
\begin{equation}\label{scalar_eq01}
\dfrac{1}{\sqrt{-g}} \partial_\alpha(\sqrt{-g} g^{\alpha\beta} \partial_\beta) \mathit{\Phi} = \mu^2 \mathit{\Phi}.
\end{equation}
In this equation $\mu$ is the mass of the associated scalar field and it is possible to express $\mathit{\Phi}$ as
\begin{equation}\label{phi_expression}
\mathit{\Phi}(t, r, \theta, \phi) = e^{-i \omega t} \dfrac{\psi(r)}{r} Y^m_l(\theta, \phi),
\end{equation}
where $\psi(r)$ is the radial part and $Y^m_l(\theta, \phi)$ is the well known 
spherical harmonics. Using Eq.\ \eqref{phi_expression} we can transform Eq.\ 
\eqref{scalar_eq01} into a Schr\"{o}dinger like equation, given by
\begin{equation}\label{scalar_eq02}
\dfrac{d^2 \psi}{dx^2} + \Big(\omega^2 - V(x)\Big) \psi = 0,
\end{equation}
where $x$ is defined as
\begin{equation}
x = \int \dfrac{dr}{f(r)},
\end{equation}
which is known as the tortoise coordinate. In Eq.\ \eqref{scalar_eq02}, the 
effective potential is given by
\begin{equation}\label{gen_pot}
V(r) = f(r) \left\lbrace \dfrac{f'(r)}{r} + \dfrac{l(l+1)}{r^2} + \mu^2 \right \rbrace.
\end{equation}
To have a physically consistent system we need to consider boundary conditions to Eq.\ \eqref{scalar_eq02} both at the horizon of the black hole and 
infinity. For an asymptotically flat spacetime, the quasinormal conditions 
can be given by \cite{Ferrari, vishveshwara1970},
\begin{equation}
\psi(x) \rightarrow \begin{cases} A e^{+i \omega x} \;\;\; \text{if} \;\; x \rightarrow -\infty\\
B e^{-i \omega x} \;\;\; \text{if} \;\; x \rightarrow +\infty \end{cases},
\end{equation}
where $A$ and $B$ are the amplitudes of the wave. One can see that the purely ingoing wave physically implies 
that nothing can escape from the horizon of the black hole. On the other hand, 
the purely outgoing wave implies that no radiation comes from the infinity. 
These two expressions can effectively give the quasinormal requirement or condition 
which ensures existence of infinite set of discrete complex numbers, known as QNMs. For a normal Schwarzschild black hole, the QNMs basically depends on the 
mass $M$, overtone $n$ and multipole number $l$ only.

\begin{figure}[htb]
\centerline{
   \includegraphics[scale = 0.3]{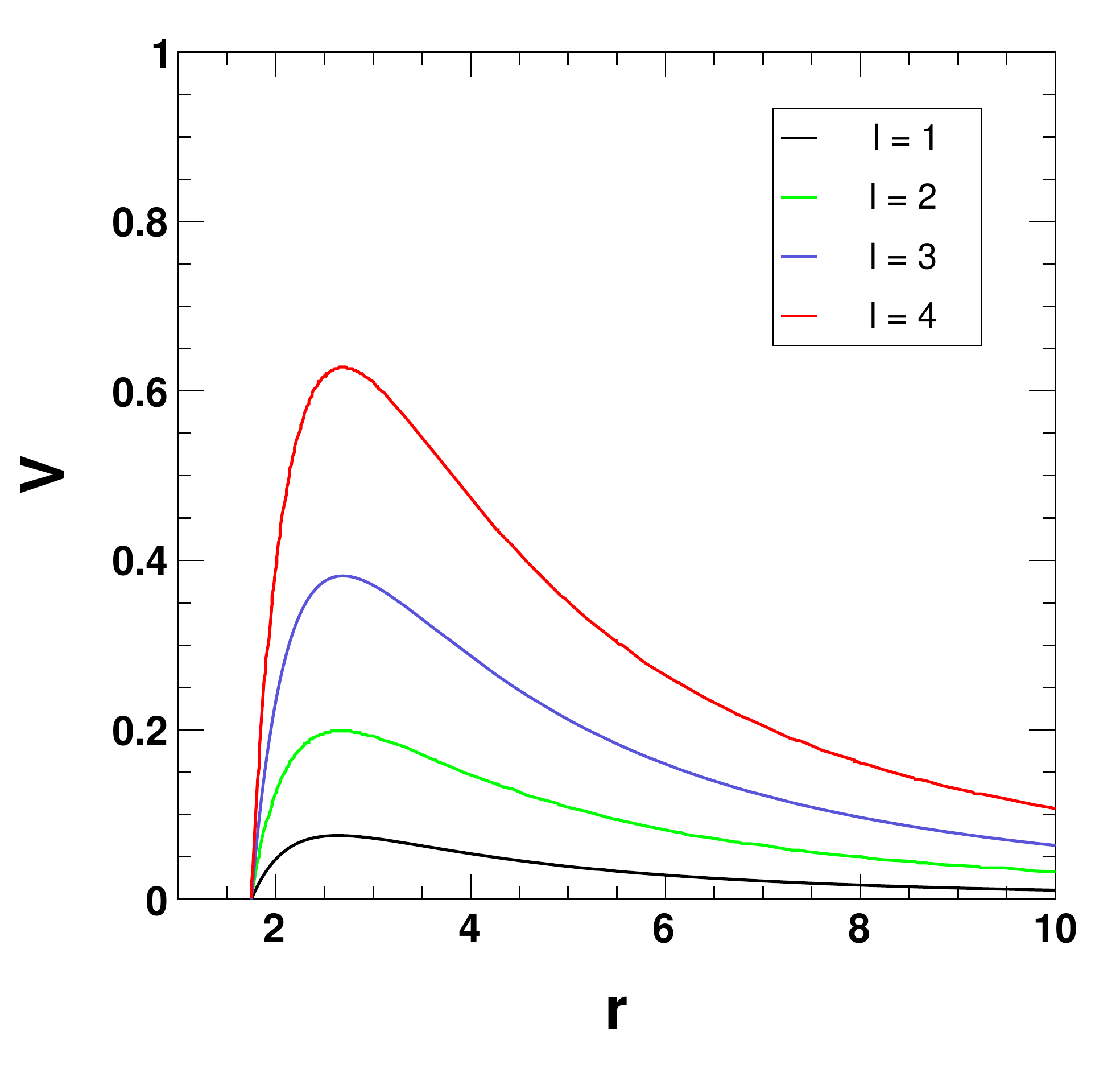}\hspace{1cm}
   \includegraphics[scale = 0.3]{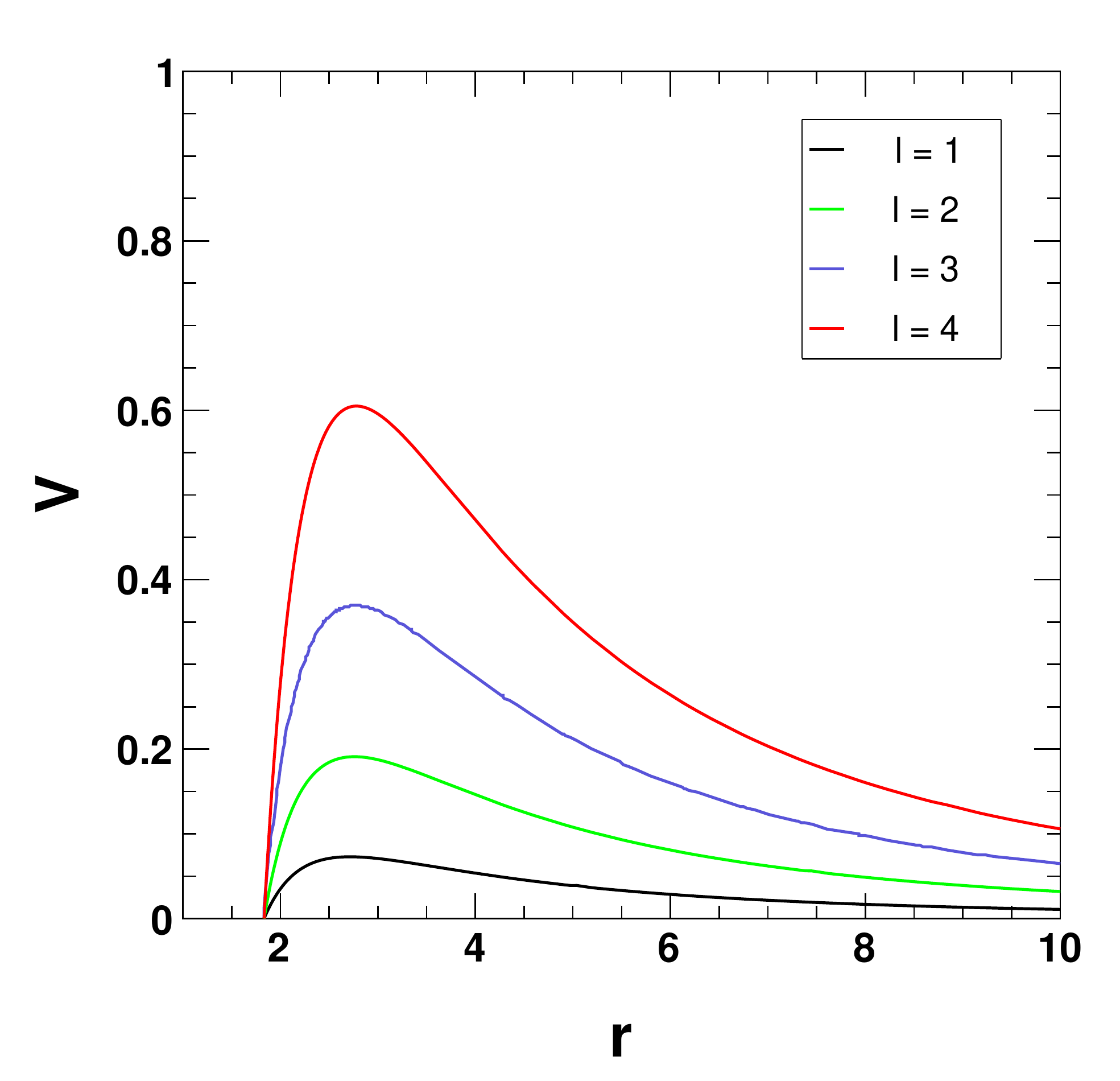}}
\vspace{-0.2cm}
\caption{Potential $V(r)$ versus $r$ for the ABG black hole (on left) and for 
the black hole defined by the metric \eqref{our_metric} (on right) with 
$a = 0.1$, $\beta = 0.1 $, $M = 1$, $q = 0.9$ and $\zeta = -0.6$.}
\label{fig_V_01}
\end{figure}

Before going to the QNMs of the black holes, we have compared the potentials 
associated with both of the black holes. In Fig. \ref{fig_V_01}, we have 
plotted the potential $V(r)$ versus $r$ for different values of multipole number $l$. The plot on 
the left panel shows the variation of the potential $V(r)$ with respect to $r$ 
for the ABG black hole and the plot on the right panel is for the metric 
\eqref{our_metric}. One can see that the pattern of variation of both the 
potentials is identical, but the maximum value of the potential for the ABG 
black hole is slightly greater than that of the other black hole given by 
the metric \eqref{our_metric}. However, this difference is significant only 
for higher values of the charge $q$ and for small values of the charge $q$, 
potentials of both the black holes are almost identical. This is due to the 
fact that the charge distribution functions of both the black holes behave 
identically in the asymptotic regime.

\begin{figure}[htb]
\centerline{
   \includegraphics[scale = 0.3]{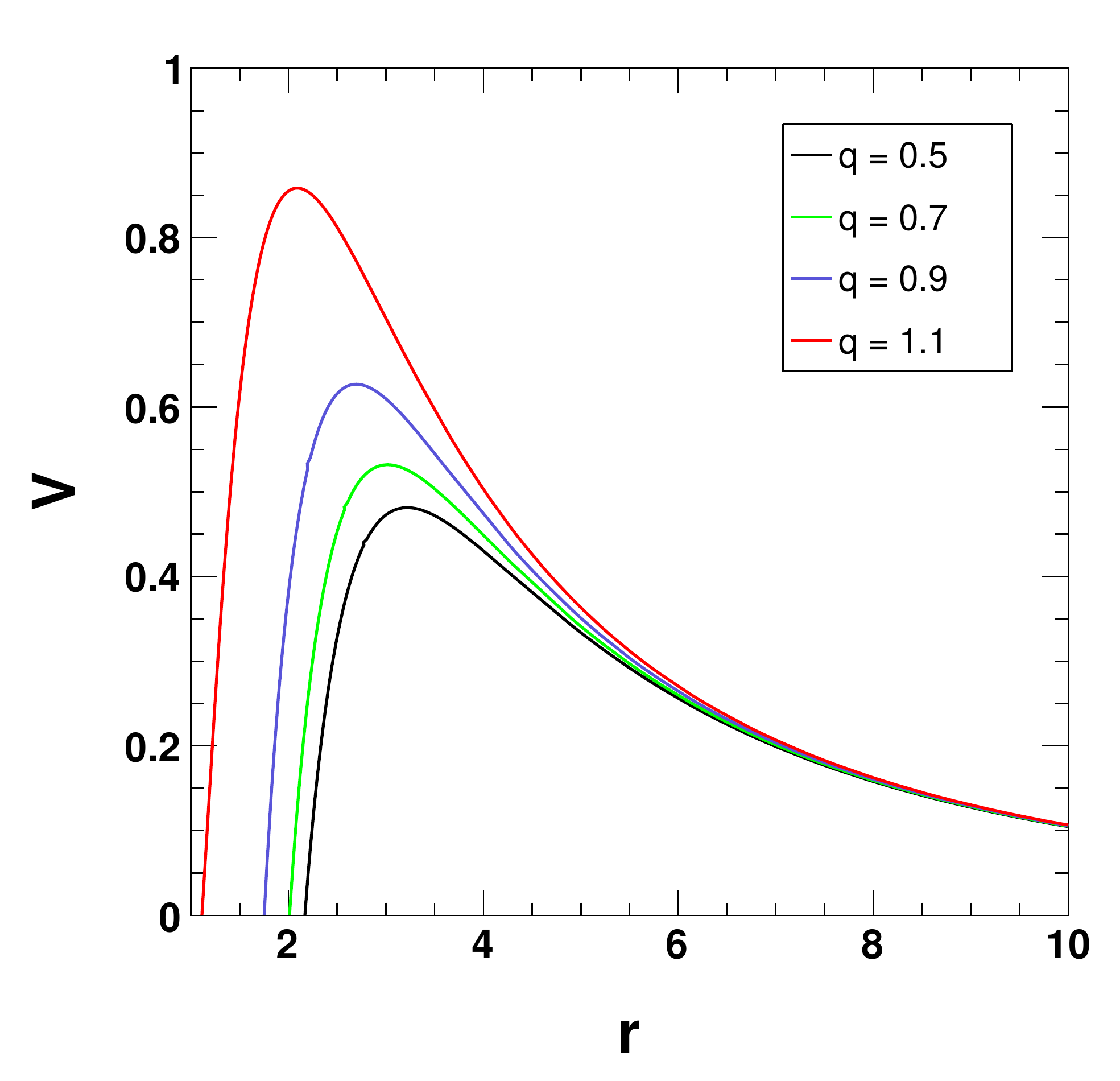}\hspace{1cm}
   \includegraphics[scale = 0.3]{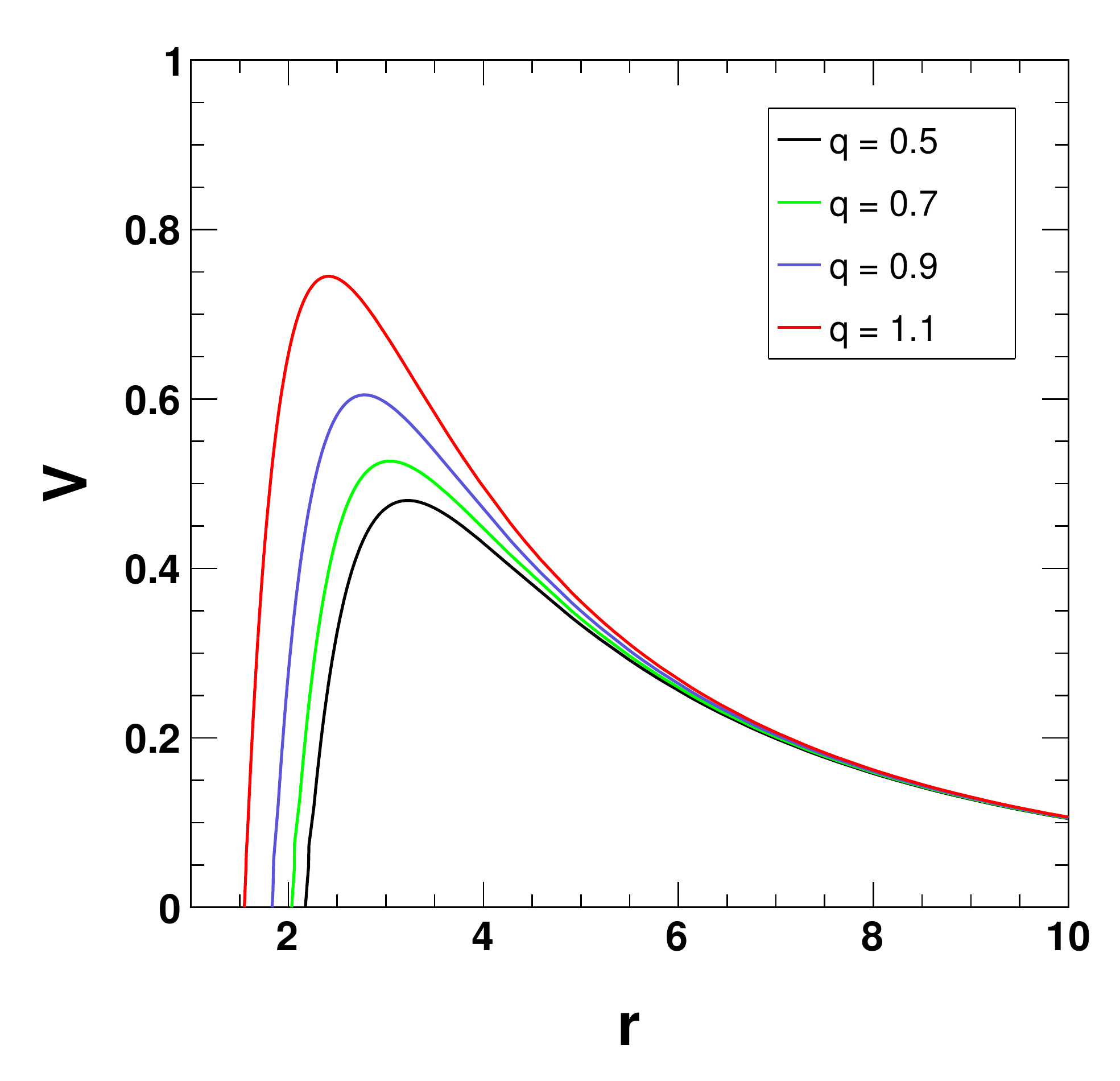}}
\vspace{-0.2cm}
\caption{Potential $V(r)$ versus $r$ for the ABG black hole (on left) and for 
the black hole defined by the metric \eqref{our_metric} (on right) with 
$a = 0.1$, $\beta = 0.1 $, $M = 1$, $l = 4$ and $\zeta = -0.6$.}
\label{fig_V_02}
\end{figure}

The potential $V(r)$ versus $r$ is plotted for different values of $q$ in 
Fig.\ \ref{fig_V_02}. Here, we have chosen $a = 0.1$, $\beta = 0.1$, $M = 1$, 
$l = 4$ and $\zeta = -0.6$. We see that in case of higher values of $q$, the  
maximum value of $V(r)$ i.e. $V_{max}$ for the ABG black hole (on left panel) 
is greater than the other black hole (on the right panel). Apart from this, 
value of $r$ corresponding to $V_{max}$ for the ABG black hole is smaller than 
the other black hole. However, this change is negligible for small values of 
the charge $q$.

\begin{figure}[htb]
\centerline{
   \includegraphics[scale = 0.3]{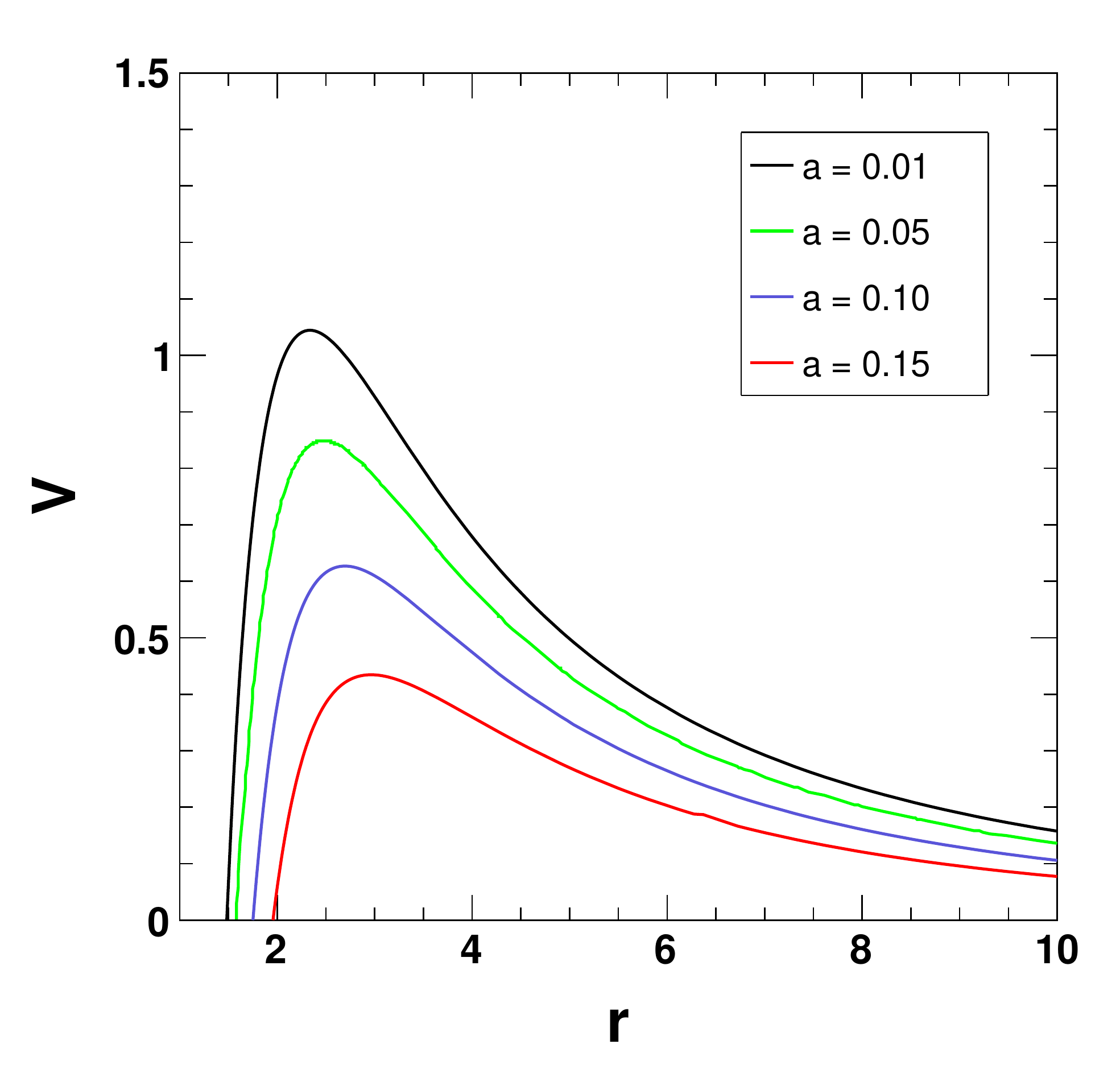}\hspace{1cm}
   \includegraphics[scale = 0.3]{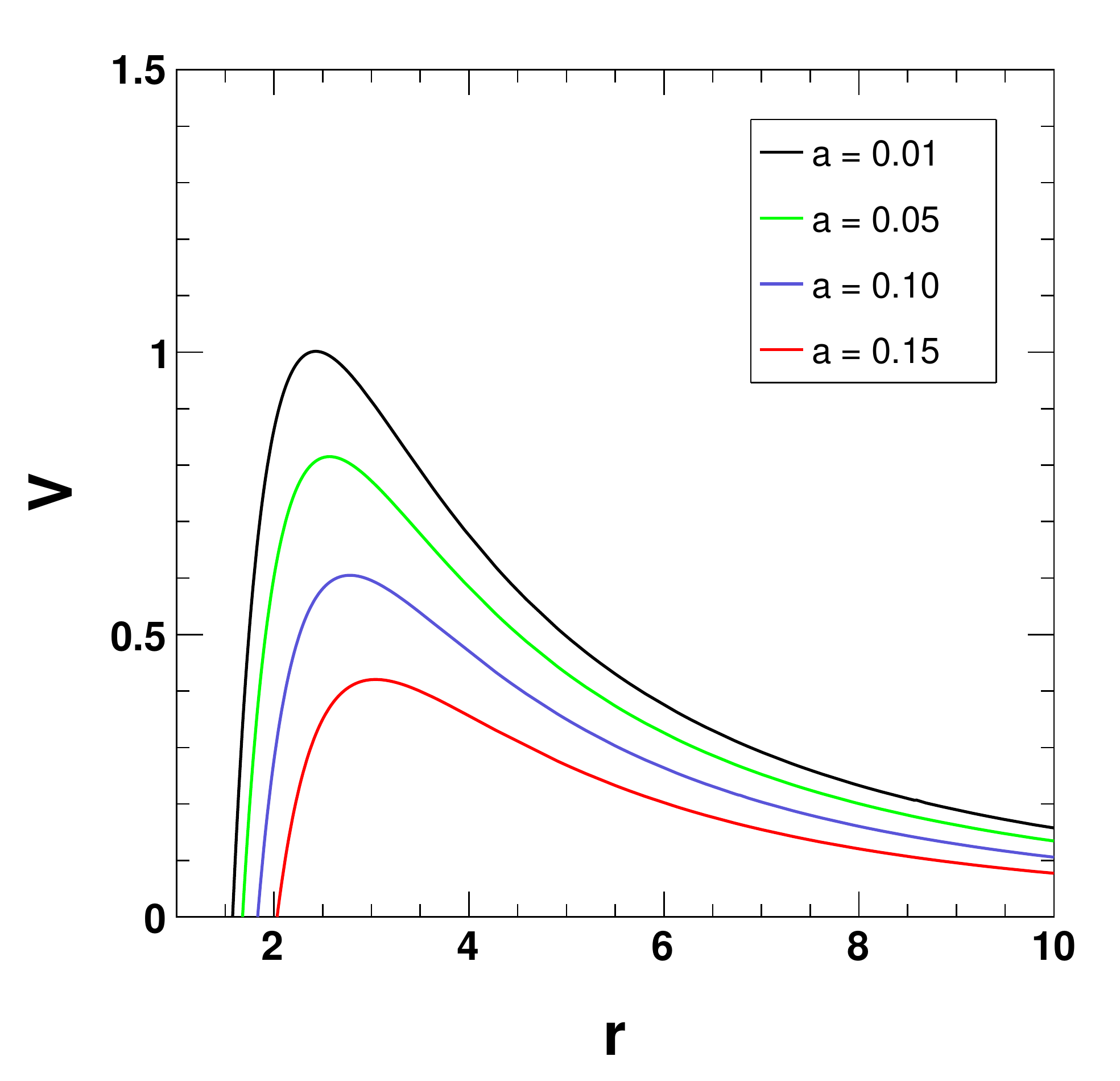}}
\vspace{-0.2cm}
\caption{Variation of potential $V(r)$ with respect to $r$ for the ABG black 
hole (on left) and for the black hole defined by the metric \eqref{our_metric} 
(on right) with $l = 4$, $\beta = 0.1 $, $M = 1$, $q = 0.9$ and $\zeta = -0.6$.}
\label{fig_V_03}
\end{figure}

We have shown the variation of potential $V(r)$ with respect to $r$ for both 
of the black holes in Fig.\ \ref{fig_V_03} for different values of cloud of 
string parameter $a$. It is seen from the figure that for higher values of 
$a$, potential $V(r)$ decreases following the same trend for both of the black 
holes. But for the ABG black hole, the potential is slightly higher than that 
of the other black hole given by the metric \eqref{our_metric}. This is again 
depends on the magnitude of the charge $q$ as seen from the previous results. 
For small magnitudes of charge $q$, the deviation is very very small.

\begin{figure}[htb]
\centerline{
   \includegraphics[scale = 0.3]{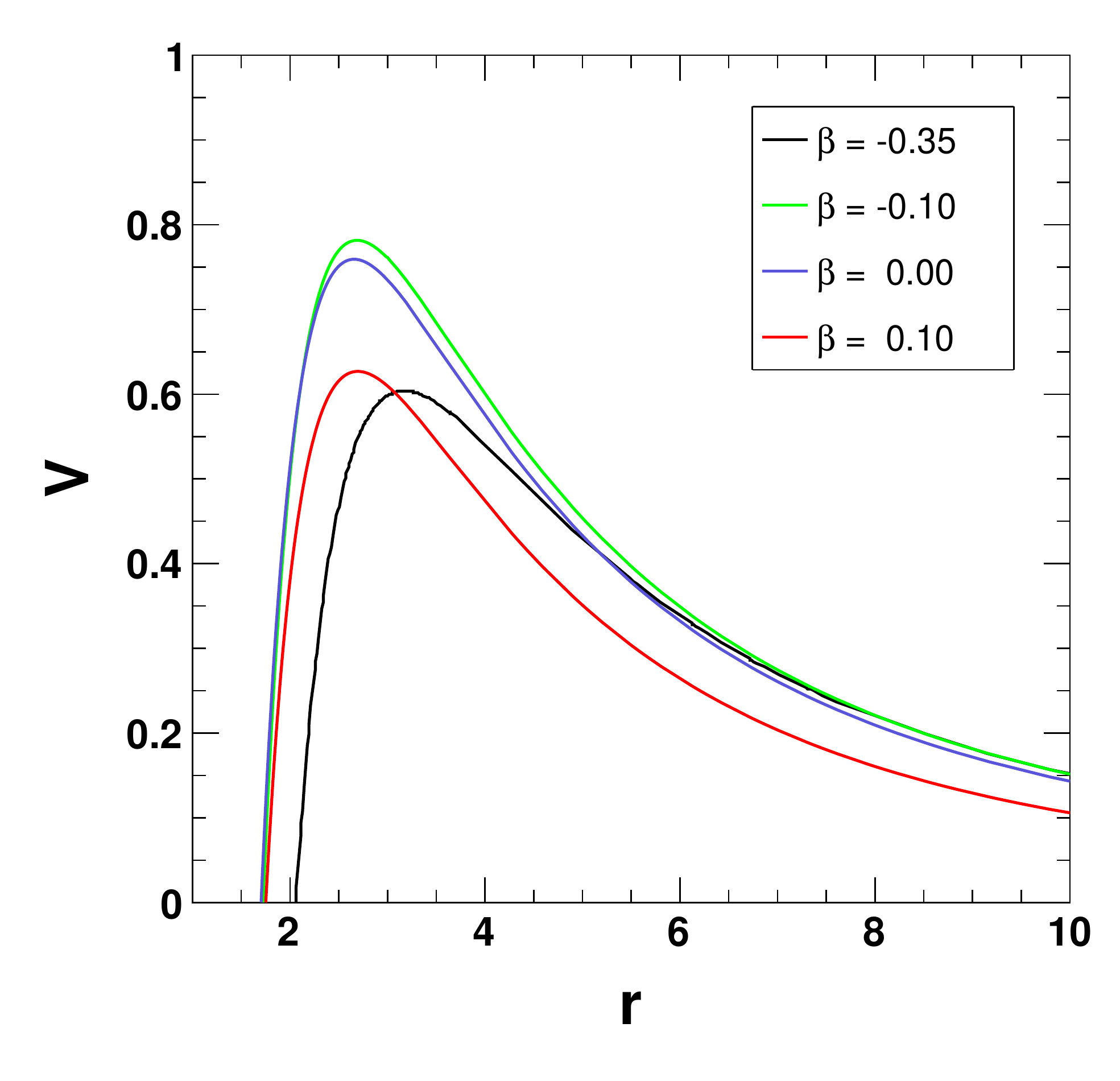}\hspace{1cm}
   \includegraphics[scale = 0.3]{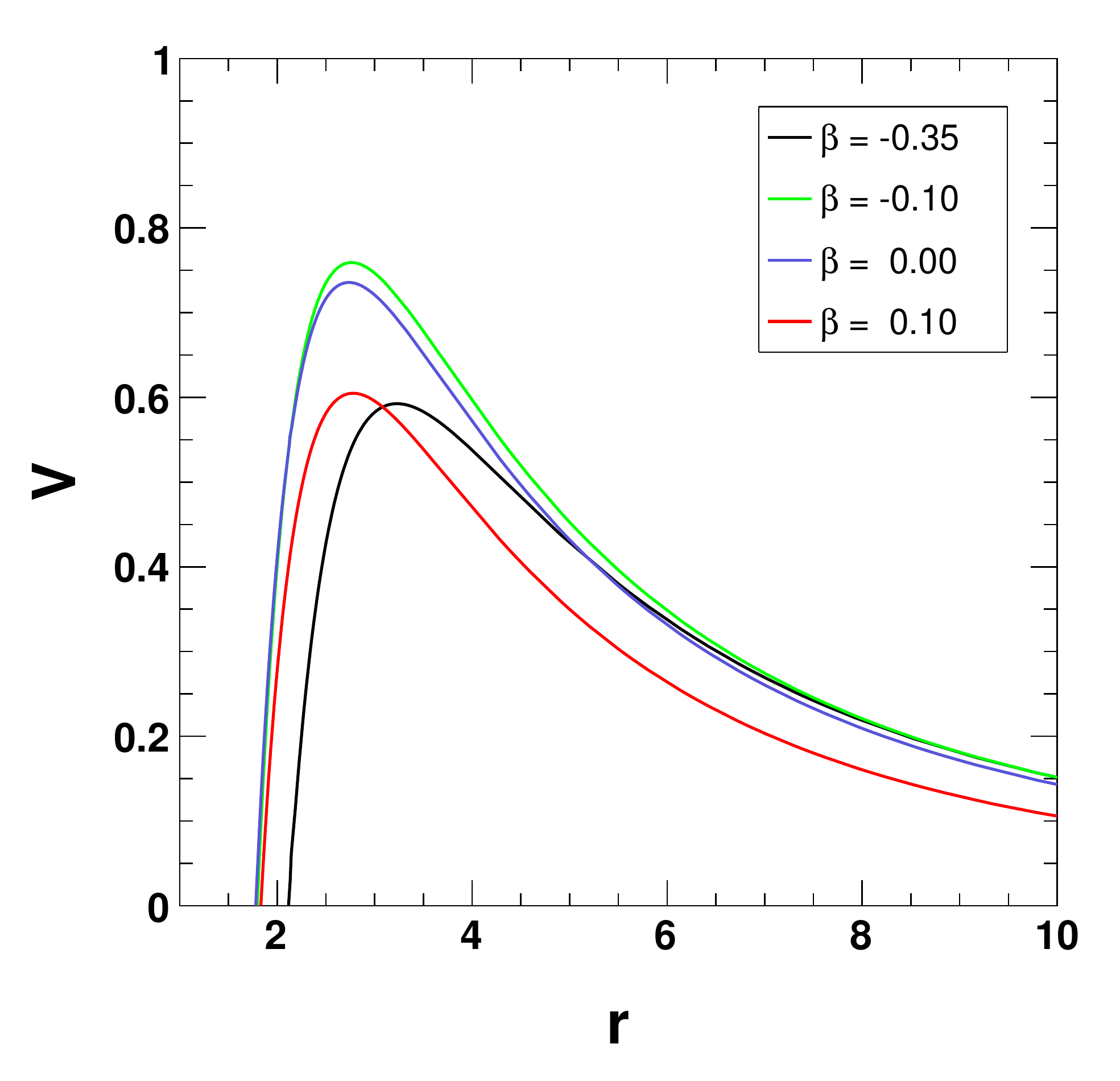}}
\vspace{-0.2cm}
\caption{Behaviour of $V(r)$ with respect to $r$ for the ABG black hole (on 
left) and for the black hole defined by the metric \eqref{our_metric} (on 
right) with $a = 0.1$, $l = 4$, $M = 1$, $q = 0.9$ and $\zeta = -0.6$.}
\label{fig_V_04}
\end{figure}

The dependency of the potential $V(r)$ with the Rastall parameter $\beta$ is 
shown in Fig.\ \ref{fig_V_04} for ABG black hole (on left panel) and for the 
black hole defined by the metric \eqref{our_metric} (on right panel). It is 
seen that for both the black holes, for $\beta = -0.10$, the potential has 
higher value. The GR limit corresponding to $\beta = 0$, gives slightly lower 
value of the potential and for $\beta > 0$ potential decreases accordingly. 
Similarly, for $\beta = -0.35$ the potential again decreases for both cases,
especially for the smaller $r$ values.

\begin{figure}[htb]
\begin{center}
   \includegraphics[scale = 0.3]{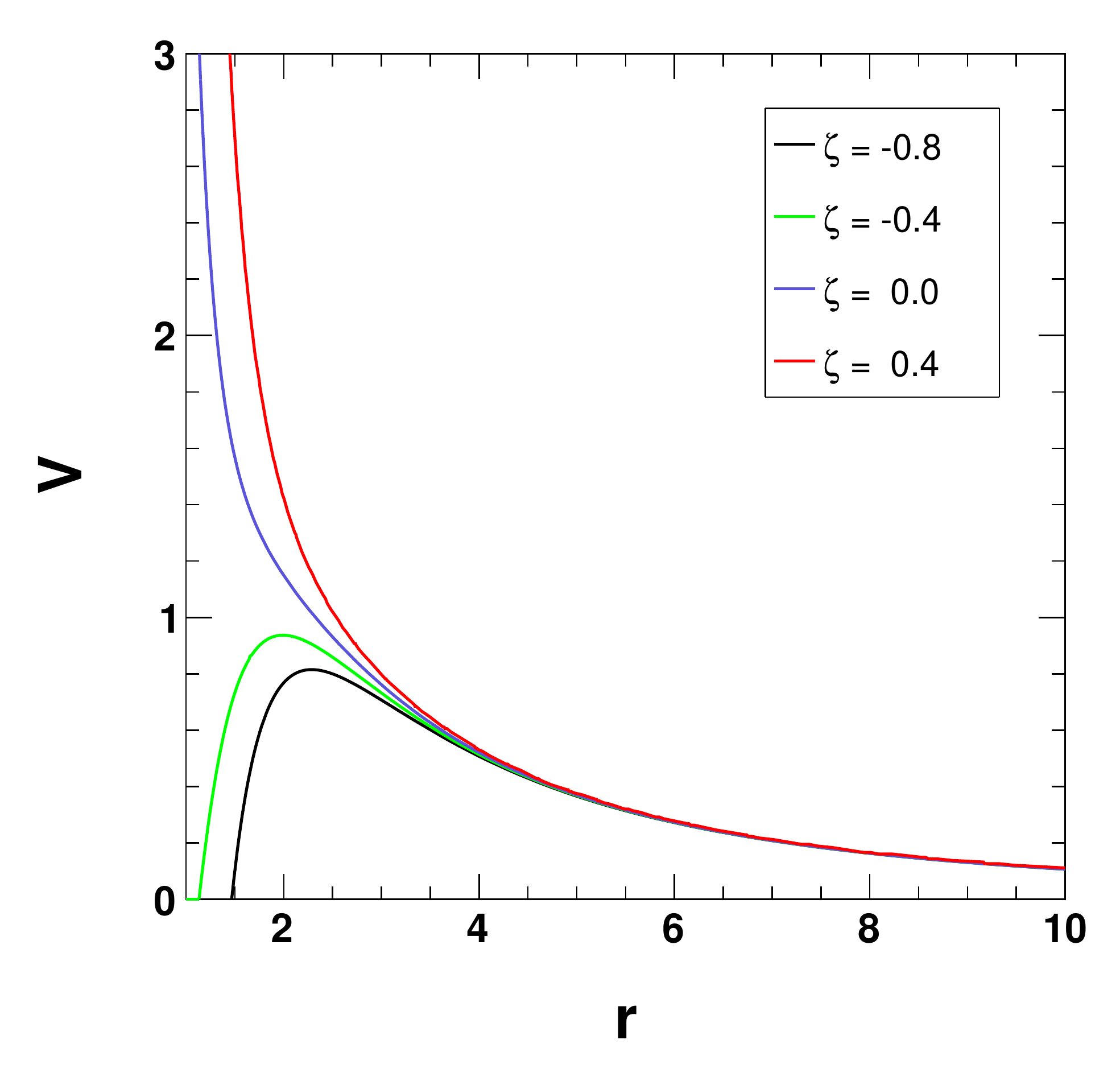}
\vspace{-0.2cm}
\caption{Behaviour of the potential $V(r)$ with respect to $r$ for the 
different values of the parameter $\zeta$ of the the black hole defined by the 
metric \eqref{our_metric} with $a = 0.1$, $\beta = 0.1 $, $M = 1$, $q = 1.2$ 
and $l = 4$.}
\label{fig_V_05}
\end{center}
\end{figure}

Another important parameter in the black hole represented by the metric 
\eqref{our_metric} is $\zeta$. We have plotted the potential corresponding to 
this black hole in Fig.\ \ref{fig_V_05} for different values of $\zeta$. It is 
observed that the parameter has very small impact on the potential for the 
small charges. However, if we increase the charge $q$ significantly, we 
observe noticeable changes in the pattern of the potential as shown in the 
figure. Here, we have used $q = 1.2$ in order to see the impacts of $\zeta$ on 
the potential significantly. It is clearly seen from the figure that higher 
positive values of $\zeta$ increases the potential near small $r$ values 
significantly. However, at far distance away from the black hole, the 
parameter $\zeta$ has no significant contributions. These impacts of $\zeta$ 
on both the metric and potential of the black hole suggest that the nature 
of this parameter is slightly similar to that of the charge $q$ of the black 
hole. 

The behaviour of the potential influences the QNMs significantly. From the 
above study of potentials associated with the black holes, we see that the 
black holes differ at higher magnitude of charges and asymptotically they 
behave identically. In this study of QNMs, we consider a massless scalar field 
with $\mu = 0$. To calculate the QNMs for these black holes, we have used 5th 
order WKB approximation method \cite{Kono2003, Konoplya2019} as in case of 6th 
order WKB approximation method, the errors associated with QNMs are 
comparatively higher. In the WKB approximation method, higher order 
approximations may not always give higher accuracy and in this case the 
non-linear charge distribution function also plays a crucial role 
\cite{Gogoi2021, Konoplya2019}. QNMs for both the black holes with different 
$n$ and $l$ values are shown in Table \ref{tab01}. In this table we have 
calculated a quantity $\vartriangle_k$ defined by \cite{Konoplya2019} 
$$ \vartriangle_k = |\dfrac{\omega_{k+1} - \omega_{k-1}}{2}|,$$ where $k$ 
denotes the order of WKB approximation. $\vartriangle_k$ gives a good 
measurement of the error order \cite{Konoplya2019} in QNMs. In this study, we 
have used the expression of $\vartriangle_k$ to calculate the error 
$\vartriangle_5$ associated with the 5th order WKB approximation for both of 
the black holes. Moreover, the table contains the $n<l$ cases only. It is 
because in WKB approximation method one can't have QNMs for $n \geq l$ with 
a satisfactory accuracy \cite{Konoplya2019}. It was also observed that with 
an increase in $(l-n)$ value higher accuracy can be obtained in the associated 
QNMs \cite{Gogoi2021}. From Table \ref{tab01}, one can see that for the chosen 
value of $\zeta$, the real quasinormal frequency for the black hole defined 
by metric \eqref{our_metric} i.e. for the second black hole is lower than that
for the ABG black hole or the first black hole. The magnitude of imaginary 
quasinormal frequency is comparatively higher for the second black hole. This 
suggests that the damping associated with the QNMs of the second black hole 
is higher. The absolute difference between the QNMs of the black holes 
increases with increase in $(l-n)$ values. Another observation from the table 
is that the errors associated the 5th order WKB approximation are much smaller
for the second black hole than that for first one and also higher $(l-n)$ 
value gives comparatively higher accuracy in the WKB approximation for both the
black holes. To have a better visualization of the variation of QNMs for 
both the black holes with respect to the black hole model parameters, we have 
plotted the real and imaginary quasinormal frequencies of the black holes. 

\begin{table}[ht]
\caption{QNMs of the ABG black hole and the other black hole defined by 
metric \eqref{our_metric} (BH02) for $a = 0.1$, $\beta = 0.1$, $q = 0.7$ and 
$\zeta = -0.8$ obtained using the 5th order WKB approximation method.}
\label{tab01}
\begin{center}
\begin{scriptsize}{
\begin{tabular}{c c c c c c c}
    \hline
    \hline
    $n$ & $l$  & ABG & $\Delta_5$(ABG) & BH02 & $\Delta_5$(BH02) & $\mid$ ABG$-$BH02 $\mid$ \\
    \hline
    \hline
   \multirow{5}{*}{\;\;\;$n=0$\;\;\;} & \multirow{1}{*}{\;\;\;$l=1$\;\;\;} & $0.240478 -0.0695003 i$ & $0.174093$ & $0.238837 -0.0698842 i$ & $0.0000737562$ & $0.0016856$\\
    & \multirow{1}{*}{$l=2$} & $0.40208 -0.068745 i$ & $0.00185342$ & $0.399421 -0.069109 i$ & $0.0000115004$ & $0.00268394$\\
    & \multirow{1}{*}{$l=3$} & $0.563475 -0.0685916 i$ & $0.00132465$ & $0.559769 -0.068895 i$ & $2.94799\times 10^{-6}$ & $0.00371822$\\
    & \multirow{1}{*}{$l=4$} & $0.724769 -0.0685155 i$ & $0.00210424$ & $0.720018 -0.0688071 i$ & $1.06331\times10^{-6}$ & $0.00475931$\\
    & \multirow{1}{*}{$l=5$} & $0.886016 -0.0684734 i$ & $0.0000182862$ & $0.880219 -0.0687625 i$ & $4.72176\times10^{-7}$ & $0.00580376$\\
    \hline
    \multirow{4}{*}{$n=1$} & \multirow{1}{*}{$l=2$} & $0.390301 -0.208608 i$ & $0.0233237$ & $0.387631 -0.21019 i$ & $0.0000447605$ & $0.00310411$\\
    & \multirow{1}{*}{$l=3$} & $0.55497 -0.207148 i$ & $0.0162886$ & $0.551067 -0.208172 i$  & $0.0000123964$ & $0.0040349$\\
    & \multirow{1}{*}{$l=4$} & $0.718092 -0.20642 i$ & $0.0274191$ & $0.713161 -0.207326 i$ & $4.62062\times10^{-6}$ & $0.00501355$\\
    & \multirow{1}{*}{$l=5$} & $0.88052 -0.206017 i$ & $0.000242116$ & $0.874572 -0.206895 i$ & $2.08662\times10^{-6}$ & $0.00601286$\\
    \hline
     \multirow{3}{*}{$n=2$} & \multirow{1}{*}{$l=3$} & $0.538845 -0.349602 i$ & $0.0931067$ & $0.53476 -0.351735 i$ & $0.000160981$ & $0.00460834$\\
    & \multirow{1}{*}{$l=4$} & $0.705234 -0.346878 i$ & $0.17811$ & $0.699966 -0.3485 i$ & $0.0000614924$ & $0.00551146$\\
    & \multirow{1}{*}{$l=5$} & $0.869821 -0.345336 i$ & $0.00156762$ & $0.863563 -0.346825 i$ & $0.0000281748$ & $0.00643289$\\
    \hline
\end{tabular}
}\end{scriptsize}
\end{center}
\end{table}

\begin{figure}[htb]
\centerline{
   \includegraphics[scale = 0.3]{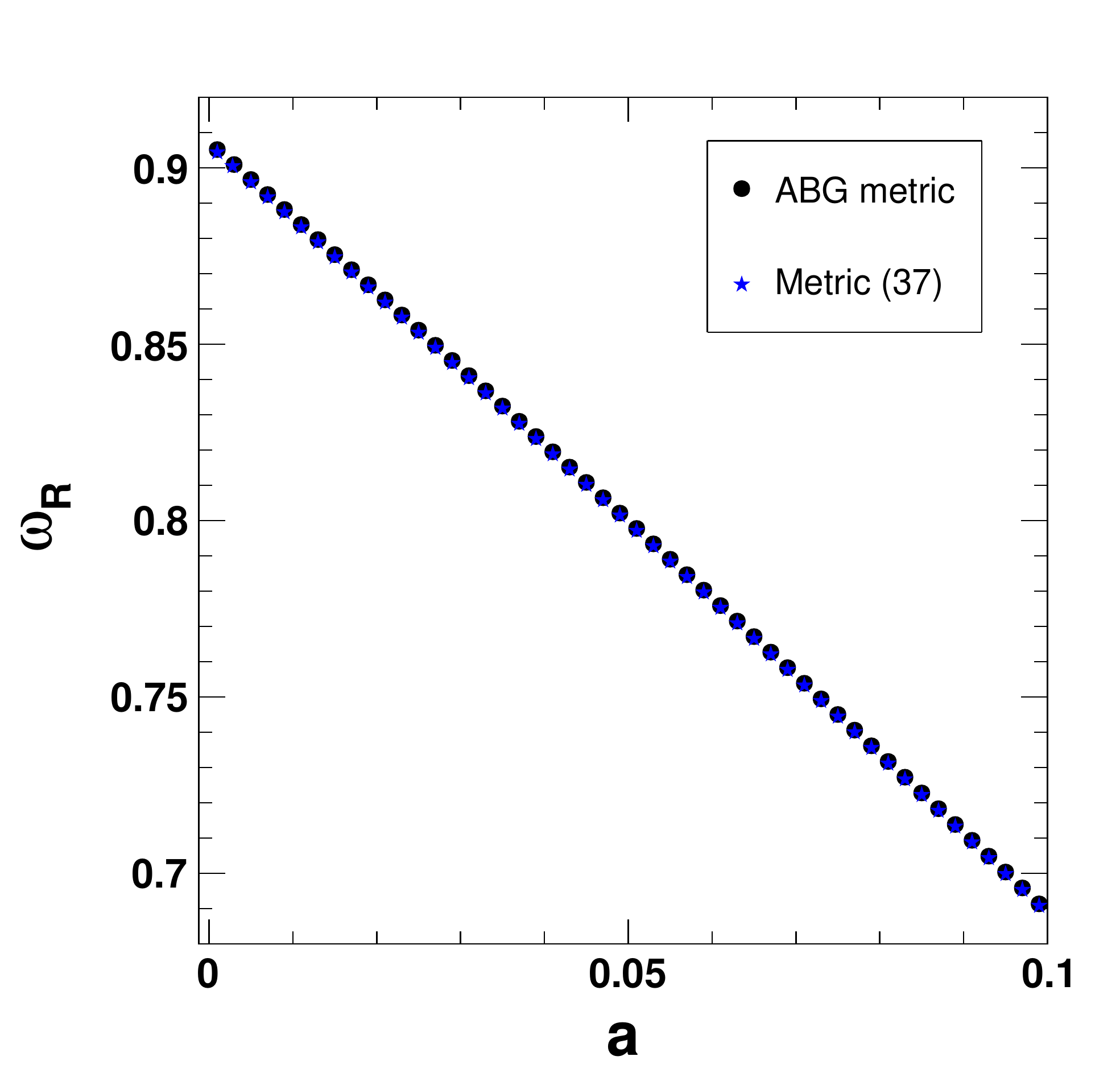}\hspace{1cm}
   \includegraphics[scale = 0.3]{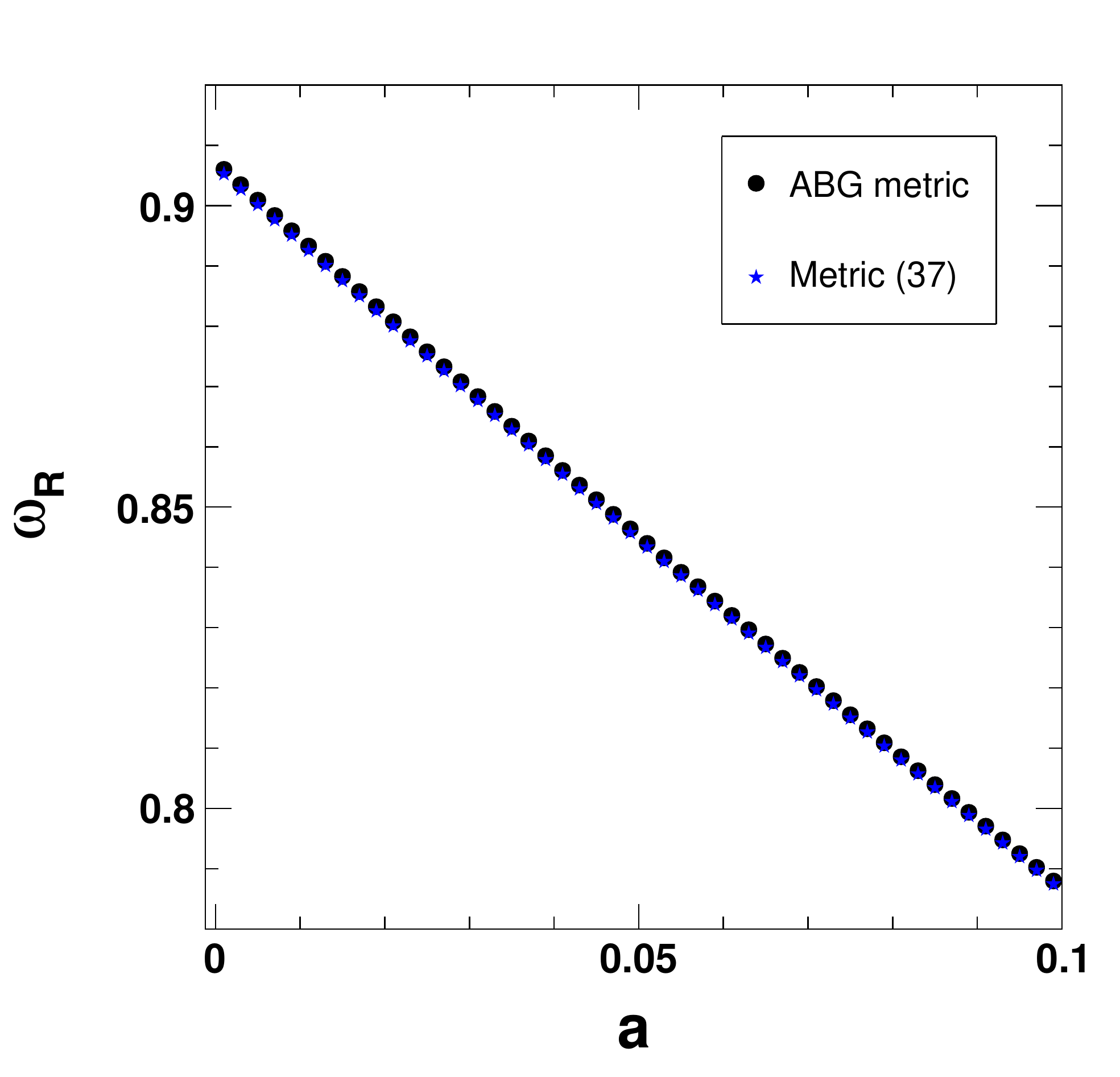}}
\vspace{-0.2cm}
\caption{Real QNMs versus cloud of strings parameter $a$ for the ABG black hole 
and the black hole defined in Eq.\ \eqref{our_metric} with $n=0$, $l=4$, $q = 0.5$, $M = 1$, 
$\zeta = -0.4$ and $\beta = 0.1$ (on left panel), and $\beta = -0.1$ (on right 
panel). }
\label{fig13}
\end{figure}

In Fig. \ref{fig13} we have plotted real QNMs with respect to the cloud of 
strings parameter $a$ for the ABG black hole and the black hole defined 
by the metric \eqref{our_metric} with $q = 0.5$, $M = 1$, $\zeta = -0.4$ and 
$\beta = 0.1$ (on left panel), and $\beta = -0.1$ (on right panel). We can see 
that for the charge $q = 0.5$, QNMs for both the black holes are very close to 
each other. However, the difference is expected to be more for higher values 
of the charge $q$ from the previous study of the black holes. For both the 
black holes, real quasinormal frequencies, $\omega_R$ decrease with 
increase in $a$ identically. But for $\beta = -0.1$, the frequencies decrease 
slowly in comparison to $\beta = 0.1$ case, where for $a=0.1$, $\omega_R$ 
decreases below $0.7$. This suggests that $\beta$ might have some significant 
impacts over the QNMs.

\begin{figure}[htb]
\centerline{
   \includegraphics[scale = 0.3]{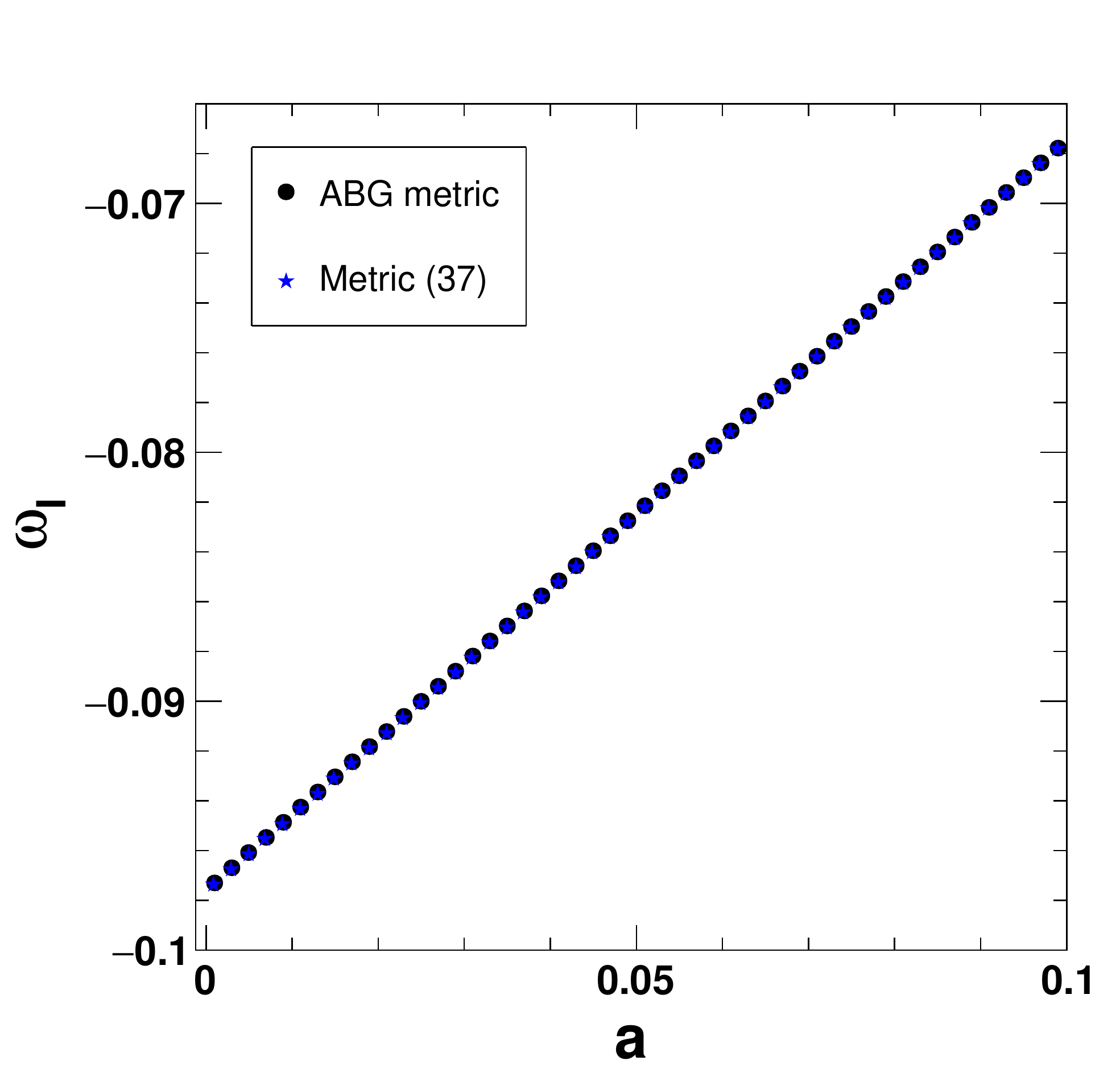}\hspace{1cm}
   \includegraphics[scale = 0.3]{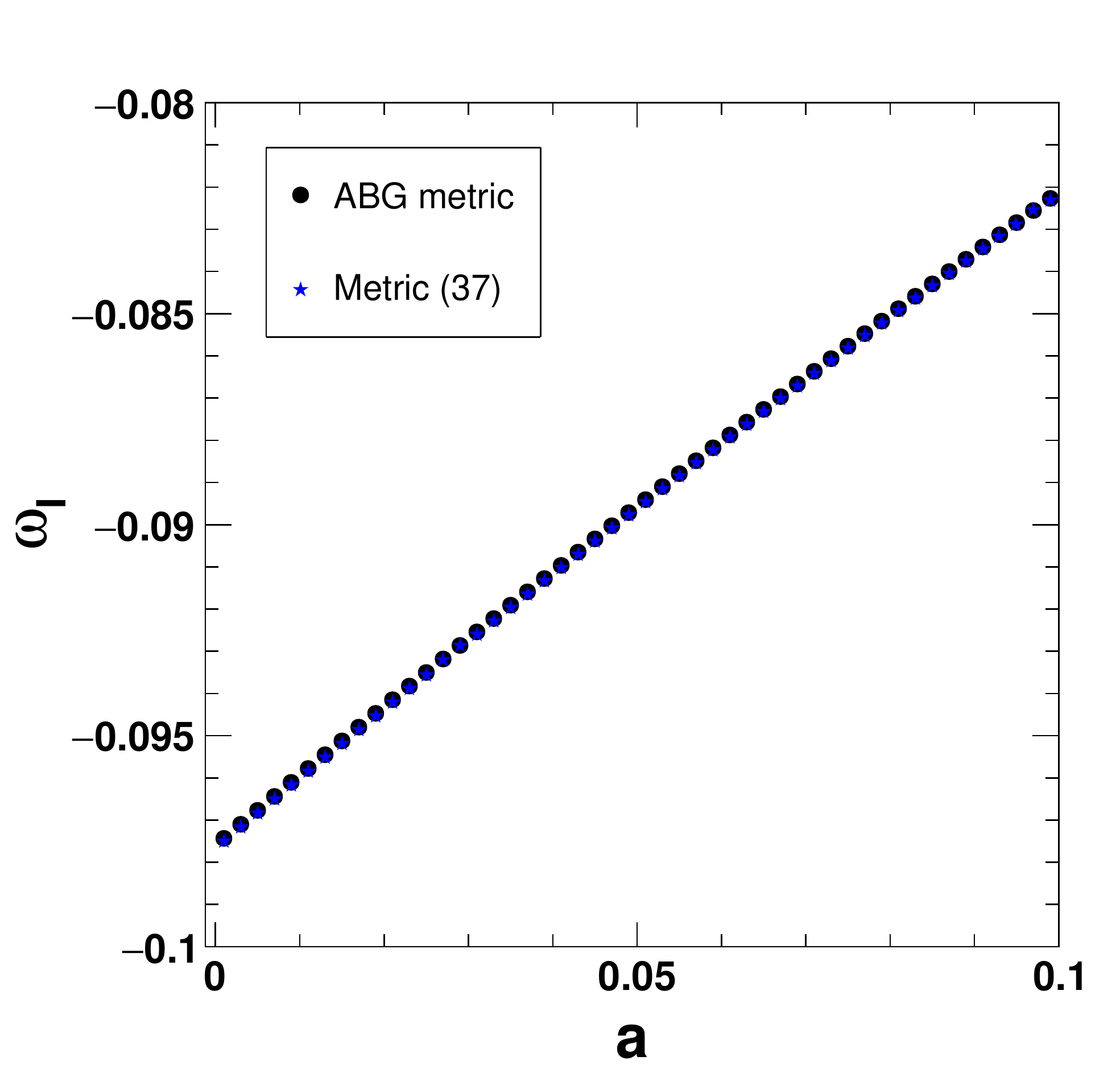}}
\vspace{-0.2cm}
\caption{Behaviour of Imaginary QNMs with respect to the cloud of strings 
parameter $a$ for the ABG black hole and the black hole defined in 
Eq. \eqref{our_metric} with $n=0$, $l=4$, $q = 0.5$, $M = 1$, $\zeta = -0.4$ and 
$\beta = 0.1$ (on left panel), and $\beta = -0.1$ (on right panel).}
\label{fig14}
\end{figure}

In Fig.\ \ref{fig14}, imaginary QNMs are plotted with respect to $a$ for both
the black holes with $q = 0.5$, $M = 1$, $\zeta = -0.4$ and $\beta = 0.1$ (on 
left panel), and $\beta = -0.1$ (on right panel). Here, the magnitude of the 
imaginary quasinormal frequencies, $\omega_I$ decreases more rapidly with $a$ 
for $\beta = 0.1$ showing the less damping. As in case of the real quasinormal 
frequencies, here also the variation of frequencies with respect to $a$ is 
almost linear. These results show that the GWs decay slowly with increase in 
the cloud of strings parameter $a$. This result is consistent with a recent 
study, where the authors have considered a Schwarzschild black hole in Rastall 
gravity surrounded by a cloud of strings \cite{Cai_2020}.

\begin{figure}[htb]
\centerline{
   \includegraphics[scale = 0.3]{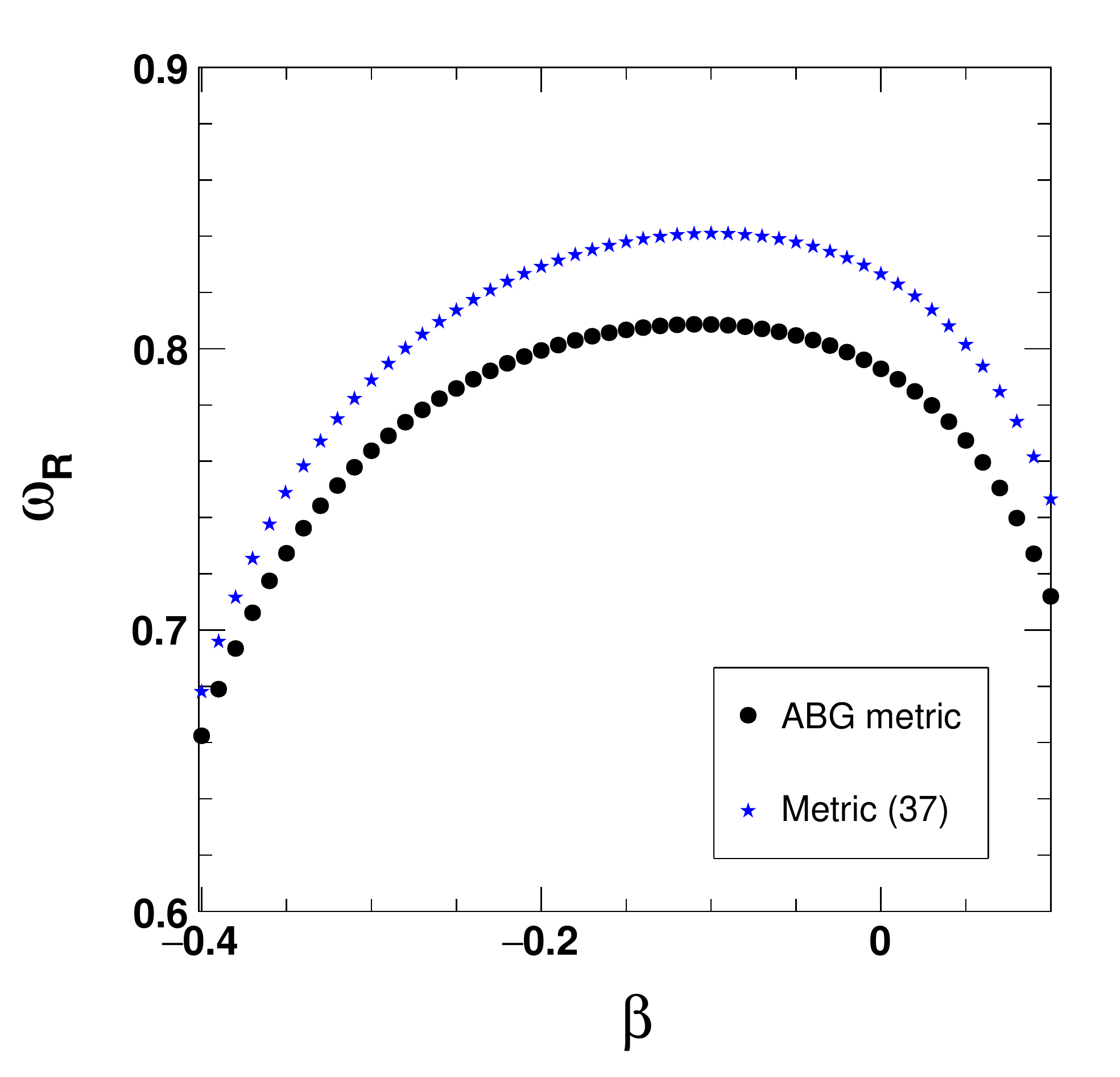}\hspace{1cm}
   \includegraphics[scale = 0.3]{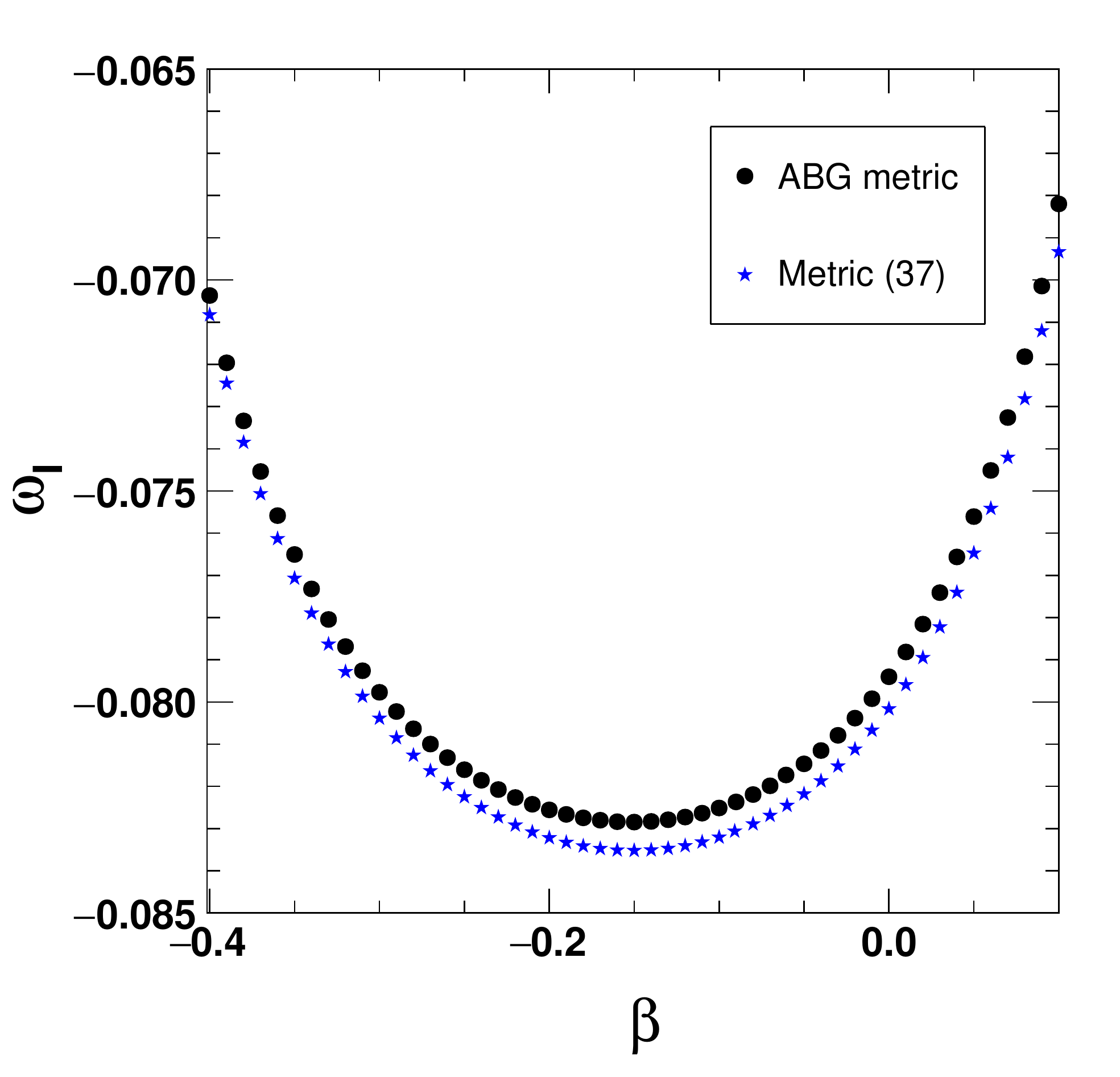}}
\vspace{-0.2cm}
\caption{Variation of QNMs in terms of $\beta$ for the ABG black hole and the 
black hole defined in Eq. \eqref{our_metric} with $n=0$, $l=4$, $q = 0.8$, $M = 1$, 
$\zeta = -0.4$ and $a = 0.1$.}
\label{fig15}
\end{figure}

\begin{figure}[htb]
\centerline{
   \includegraphics[scale = 0.3]{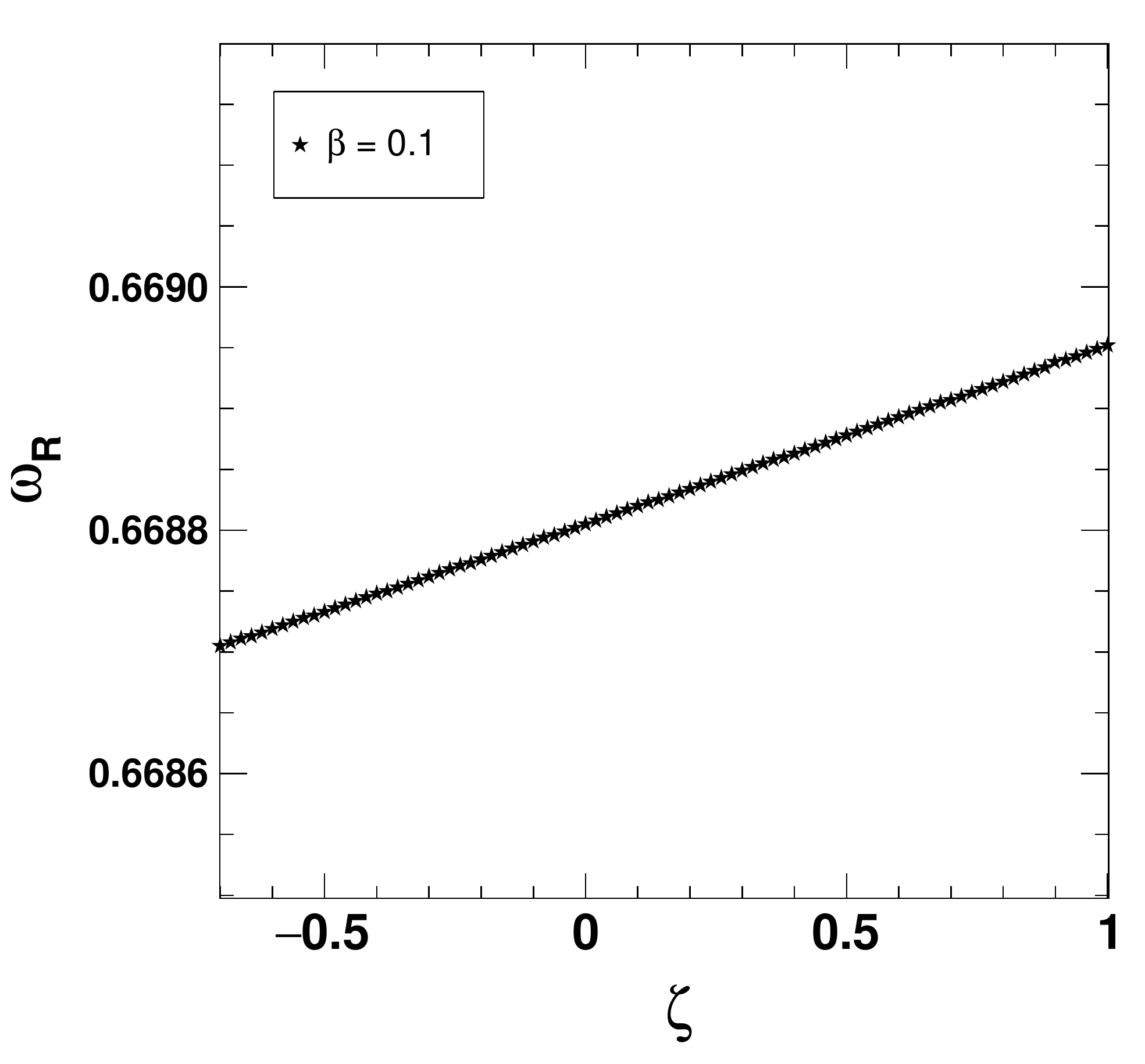}\hspace{1cm}
   \includegraphics[scale = 0.3]{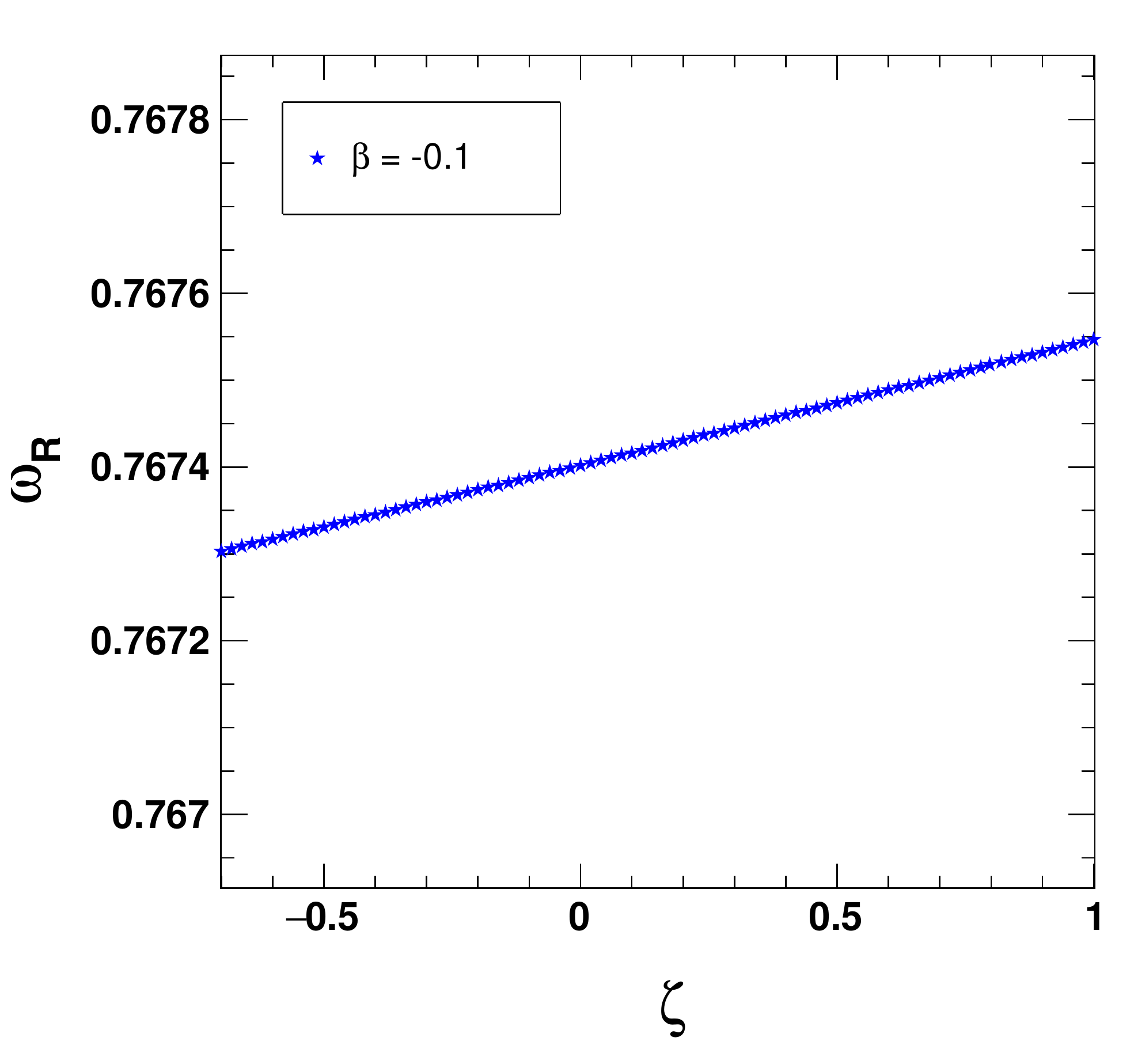}}
\vspace{-0.2cm}
\caption{Behaviour of real QNMs with respect to the model parameter $\zeta$ 
for the black hole defined in Eq.\ \eqref{our_metric} with $n=0$, $l=4$, $q = 0.3$, $M = 1$, 
$a = 0.1$ and $\beta = 0.1$ (on left panel), and $\beta = -0.1$ (on right 
panel).}
\label{fig16}
\end{figure}

\begin{figure}[htb]
\centerline{
   \includegraphics[scale = 0.3]{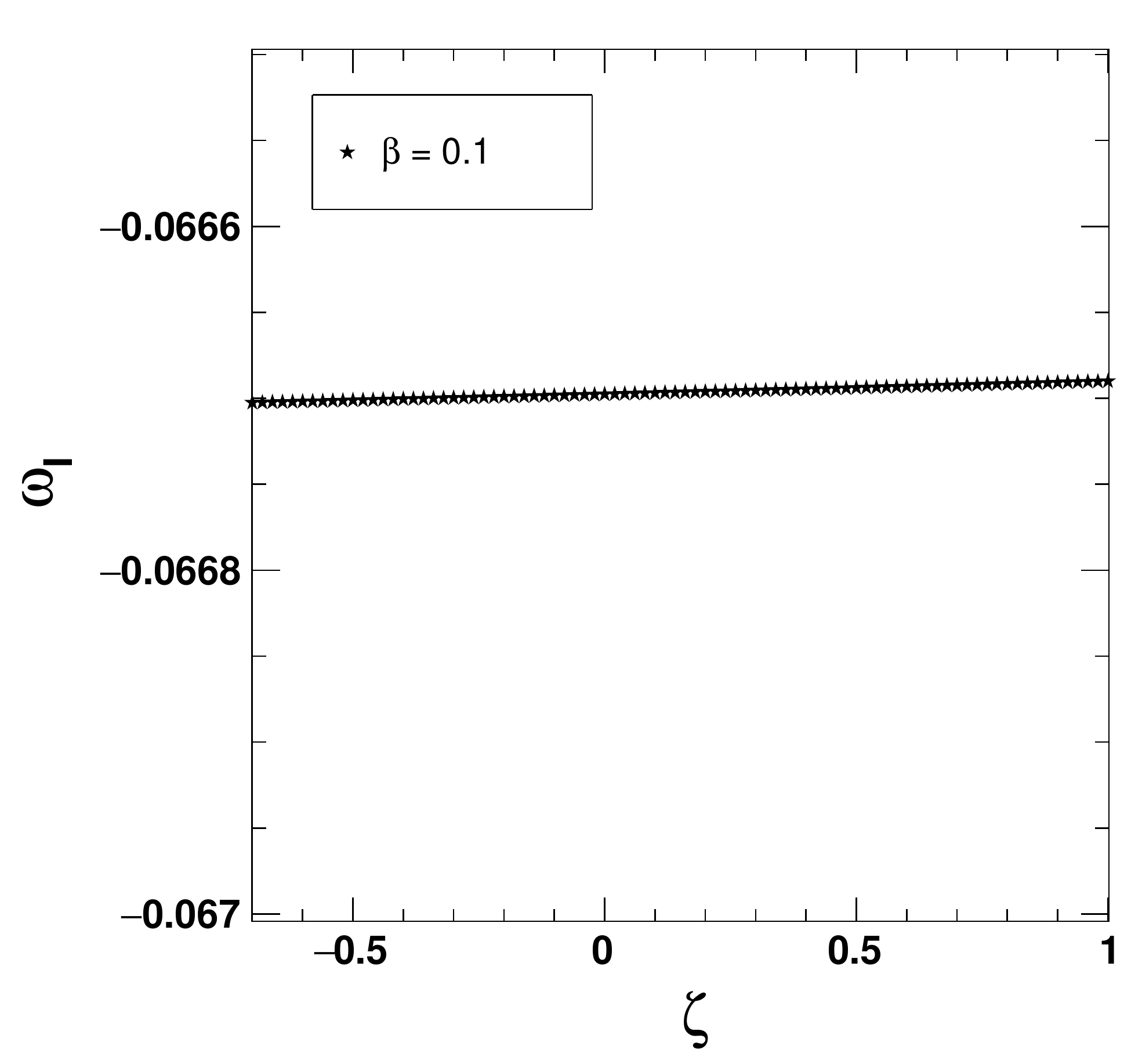}\hspace{1cm}
   \includegraphics[scale = 0.3]{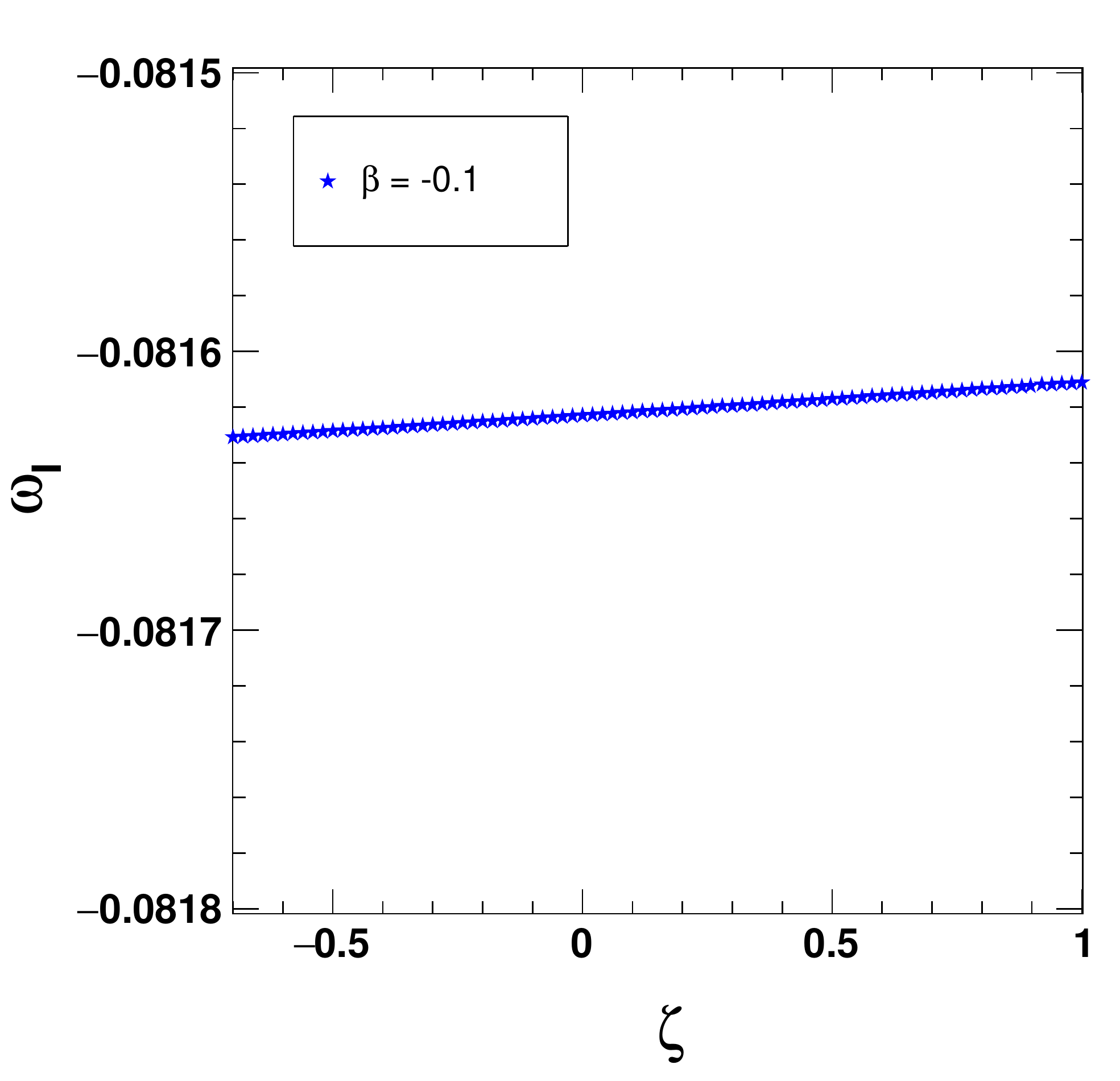}}
\vspace{-0.2cm}
\caption{Behaviour of imaginary QNMs with respect to the model parameter 
$\zeta$ for the black hole defined in Eq.\ \eqref{our_metric} with $n=0$, $l=4$, $q = 0.3$, 
$M = 1$, $a = 0.1$ and $\beta = 0.1$ (on left panel), and $\beta = -0.1$ (on 
right panel).}
\label{fig17}
\end{figure}

In Fig.\ \ref{fig15} we have plotted $\omega_R$ versus $\beta$ (on left panel) 
and $\omega_I$ versus $\beta$ (on right panel) with $q = 0.8$, $M = 1$, 
$\zeta = -0.4$ and $a = 0.1$ for both of the black holes. Here in case of real 
frequencies, with increase in $\beta$, $\omega_R$ increases and reaches a 
maximum value at $\beta$ around $-0.1$ and then again starts to decrease for 
both of the black holes. Similarly, in case of $\omega_I$, the magnitude 
increases with increase in $\beta$ and reaches a maximum value at around 
$-0.15$. The second black hole seems to be more sensitive in terms of 
both $\omega_R$ and $\omega_I$ for the considered set of parameters. In both 
the plots, $\beta = 0$ corresponds to GR limit and it can be seen that in 
Rastall gravity, depending on the value of Rastall parameter $\beta$ QNMs can 
be greater or less than the corresponding GR values with an upper 
bound on the magnitude of quasinormal frequencies. However, this upper bound 
depends on the other parameters also. Another observation in this case is that 
for the second black hole i.e. black hole defined by our metric 
\eqref{our_metric}, the magnitude of quasinormal frequencies are greater than 
the ABG black hole. But in Table \ref{tab01}, we observed an opposite 
scenario. It is due to the fact that here we have considered a higher value of 
$q$ and $\zeta$. From the previous part, we have seen that $\zeta$ can be 
more sensitive to potential and hence QNMs for higher values of $q$. So, 
these results apparently suggest that an increase in $\zeta$ might increase 
the real quasinormal frequencies. To see this clearly, we have plotted real 
QNMs versus $\zeta$ for a positive and negative $\beta$ value in 
Fig.\ \ref{fig16}. The figure shows that with increase in $\zeta$, $\omega_R$ 
increases linearly. Aslo we found that the slope of the curve increases with 
increase in the value of $q$. The magnitude of $\omega_I$ decreases with 
increase in $\zeta$ for both positive and negative values of $\beta$ (see 
Fig.\ \ref{fig17}). However, $\zeta$ is less sensitive to $\omega_I$ in 
comparison to $\omega_R$. These results suggest that with increase in 
$\zeta$, gravitational waves decay slowly. Again in case of $\beta$, GWs decay 
rapidly up to $\beta \sim -0.15$ and beyond this value, GWs start to decay 
slowly (see Fig.\ \ref{fig15}).

\begin{figure}[htb]
\centerline{
   \includegraphics[scale = 0.3]{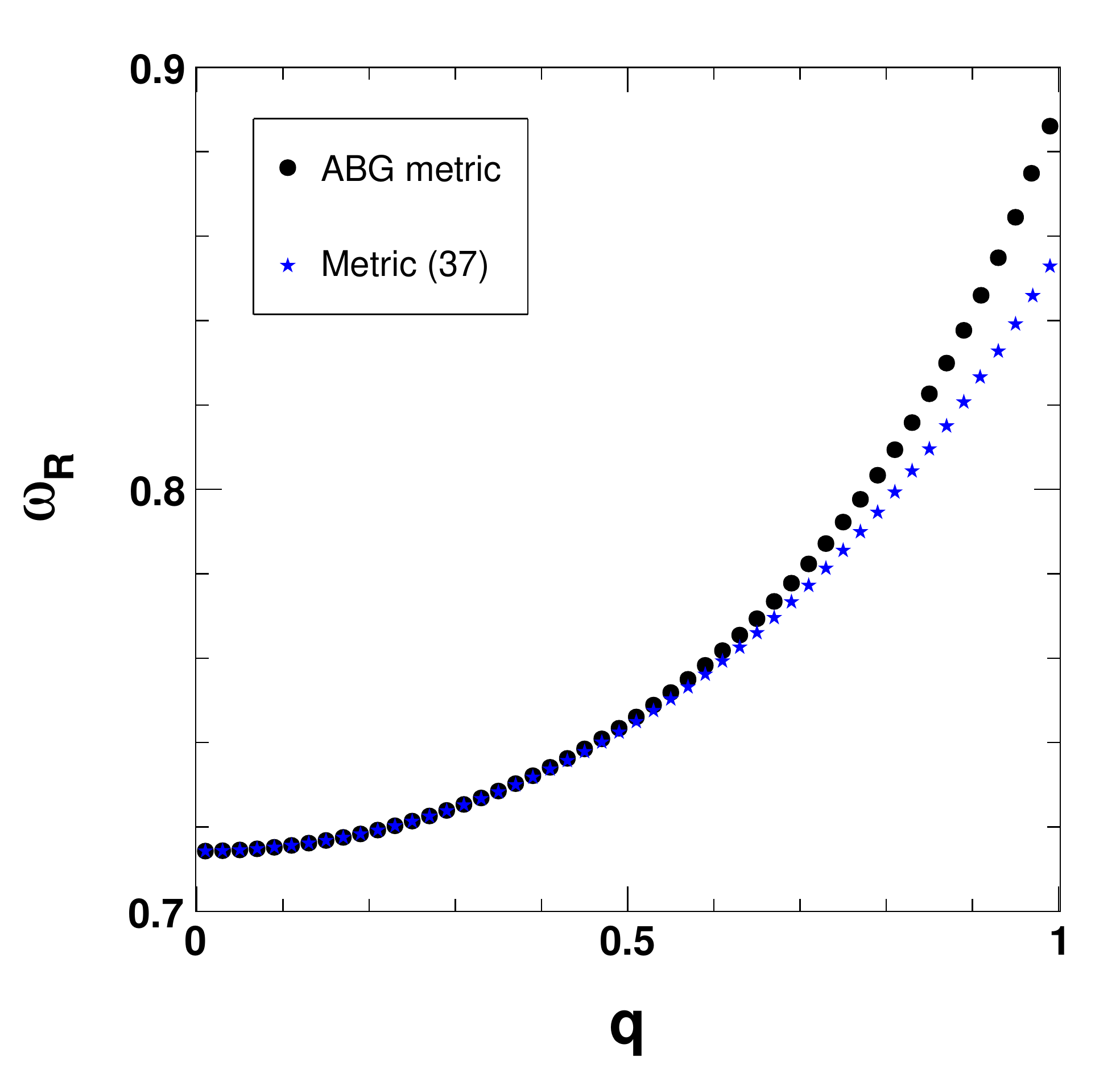}\hspace{1cm}
   \includegraphics[scale = 0.3]{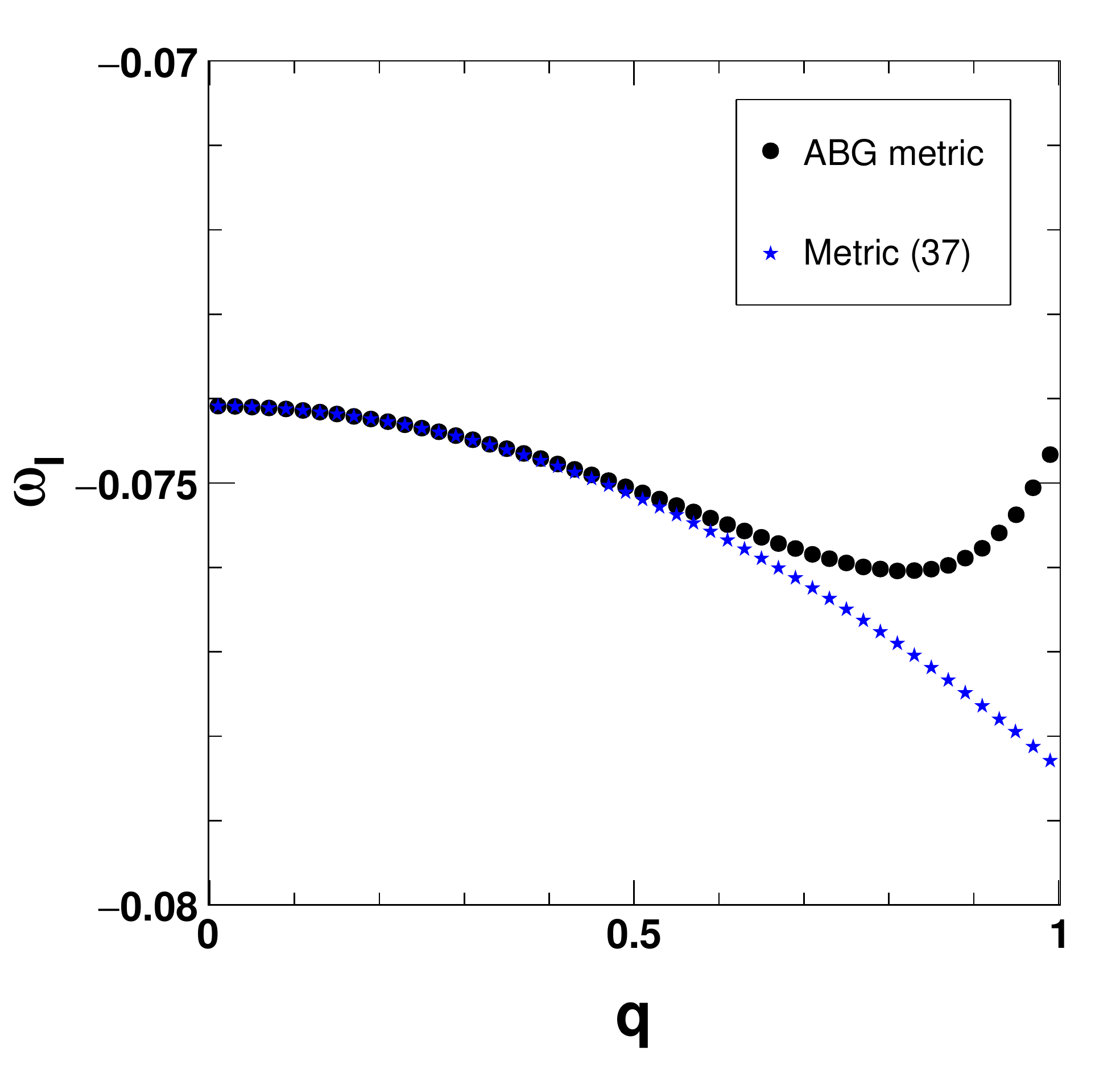}}
\vspace{-0.2cm}
\caption{Variation of QNMs with respect to $q$ for the ABG black hole and the 
black hole defined in Eq.\ \eqref{our_metric} with $n=0$, $l=4$, $\beta = 0.05$, $M = 1$, 
$\zeta = -0.8$ and $a = 0.1$.}
\label{fig18}
\end{figure}

Finally, we observed the variation of $\omega_R$ and $\omega_I$ with respect to 
$q$ for both the black holes in Fig.\ \ref{fig18} with $\beta = 0.05$, $M = 1$, 
$\zeta = -0.8$ and $a = 0.1$. QNMs for both the black holes are very close to 
each other in the asymptotic region i.e. for very small $q$. However, with 
increase in $q$, depending on the value of $\zeta$, $\omega_R$ as well as 
$\omega_I$ for the second black hole starts deviating from the first one. In 
this case, for $\zeta = -0.8$, with increase in $q$, $\omega_R$ for the second 
black hole starts decreasing from the first one. In case of the imaginary 
quasinormal frequency, the magnitude of $\omega_I$ increases for higher values 
of $q$ for the second black hole. But for the first black hole, magnitude of 
$\omega_I$ increases to a certain value and then starts to decrease towards 
$q=1$. These results suggest that in case of the second black hole, GWs decay 
more rapidly than the ABG black hole when the charge $q$ increases towards $1$.

\section{Characteristics of Hawking Temperature of the Black hole solutions}\label{section4}
\begin{figure}[htb]
\centerline{
   \includegraphics[scale = 0.3]{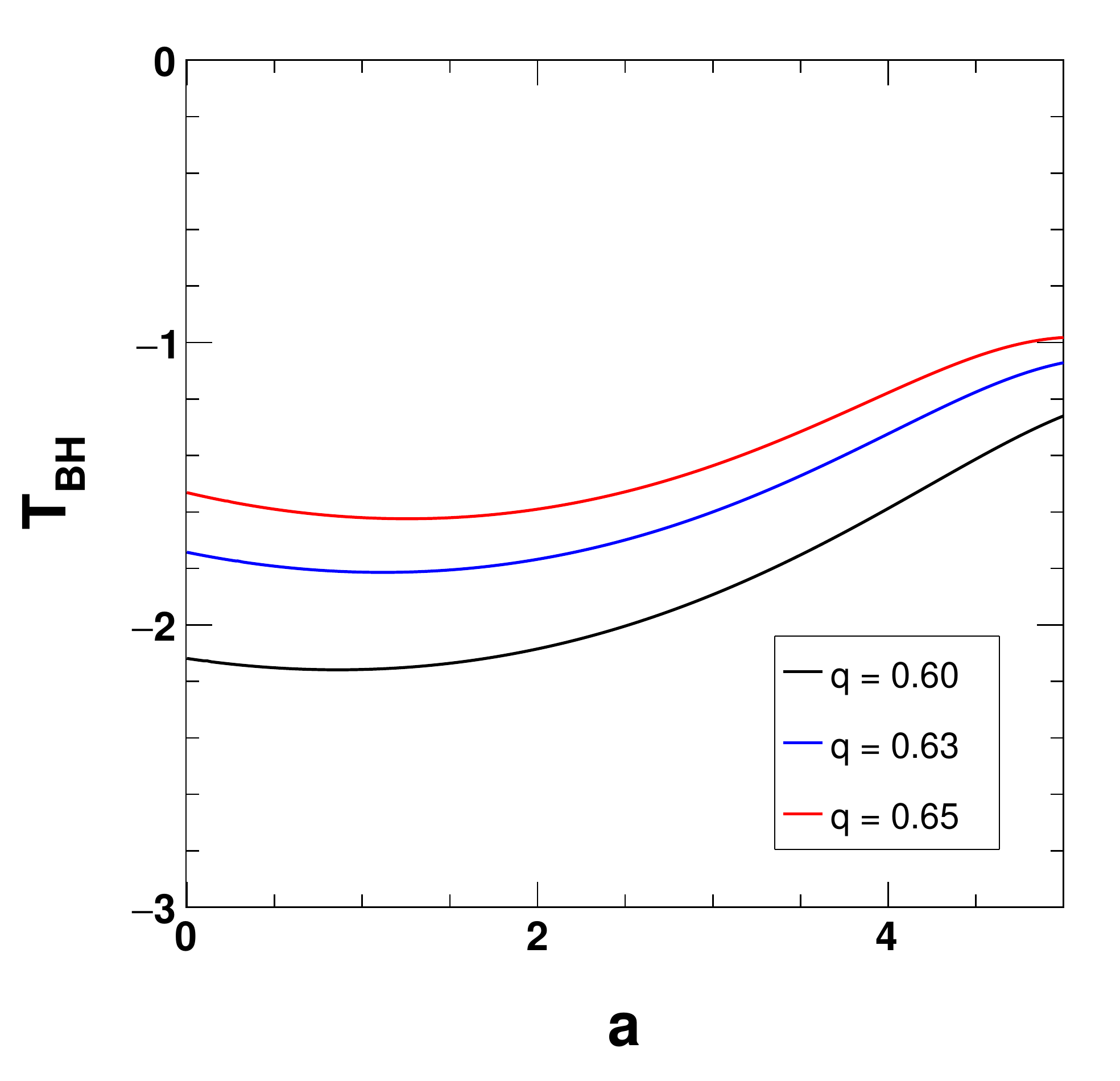}\hspace{1cm}
   \includegraphics[scale = 0.3]{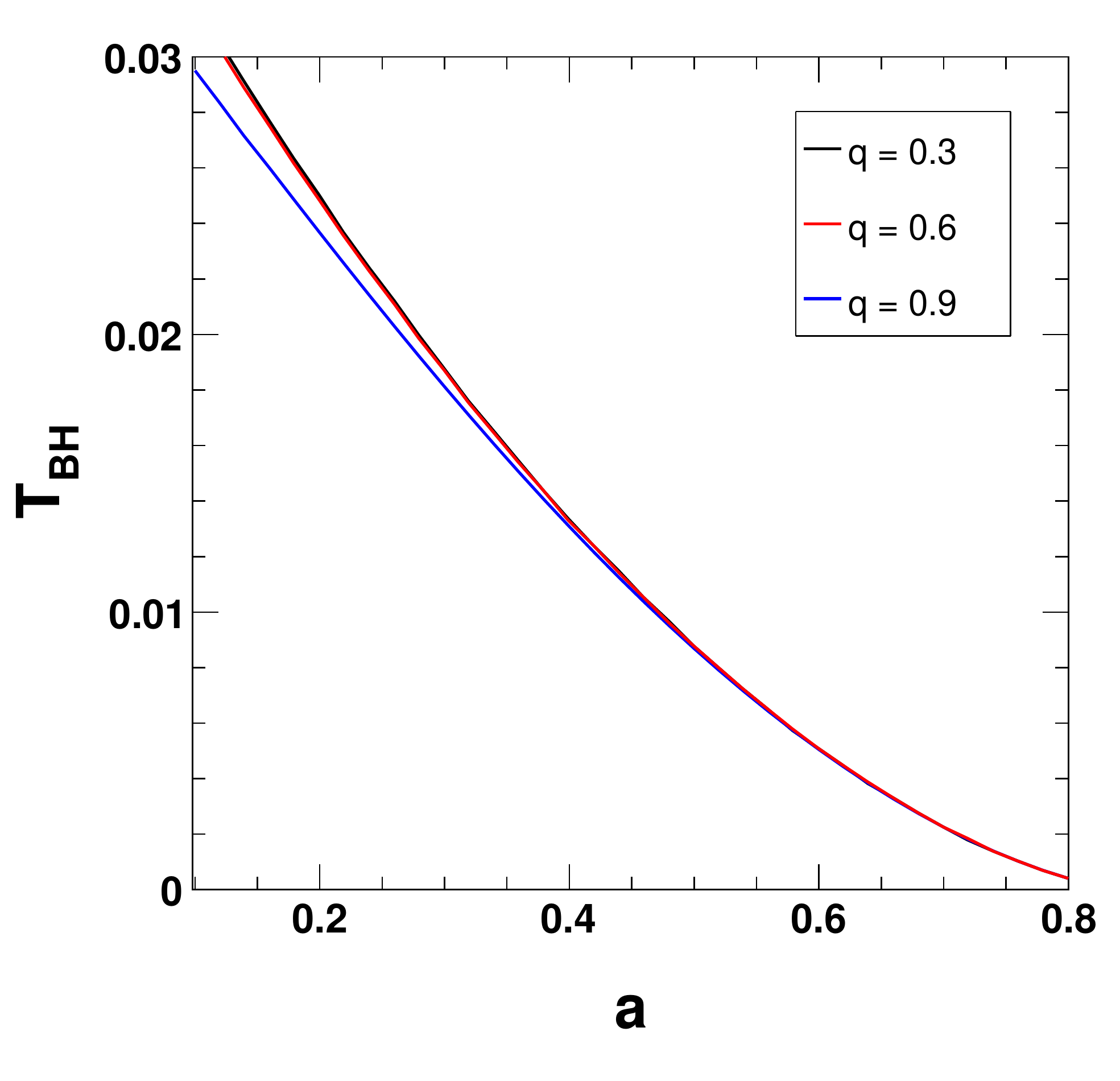}}
\vspace{-0.2cm}
\caption{Variation of $T_{BH}$ with respect to $a$ for different values of
charge $q$ for the first horizon $r_-$ (on left panel) with $M = 1$ 
and $\beta = 0.1$, and for the second horizon $r_+$ (on right panel) with 
$M = 1$ and $\beta = 0.01$ in case of the ABG black hole.}
\label{fig19}
\end{figure}

\begin{figure}[htb]
\centerline{
   \includegraphics[scale = 0.3]{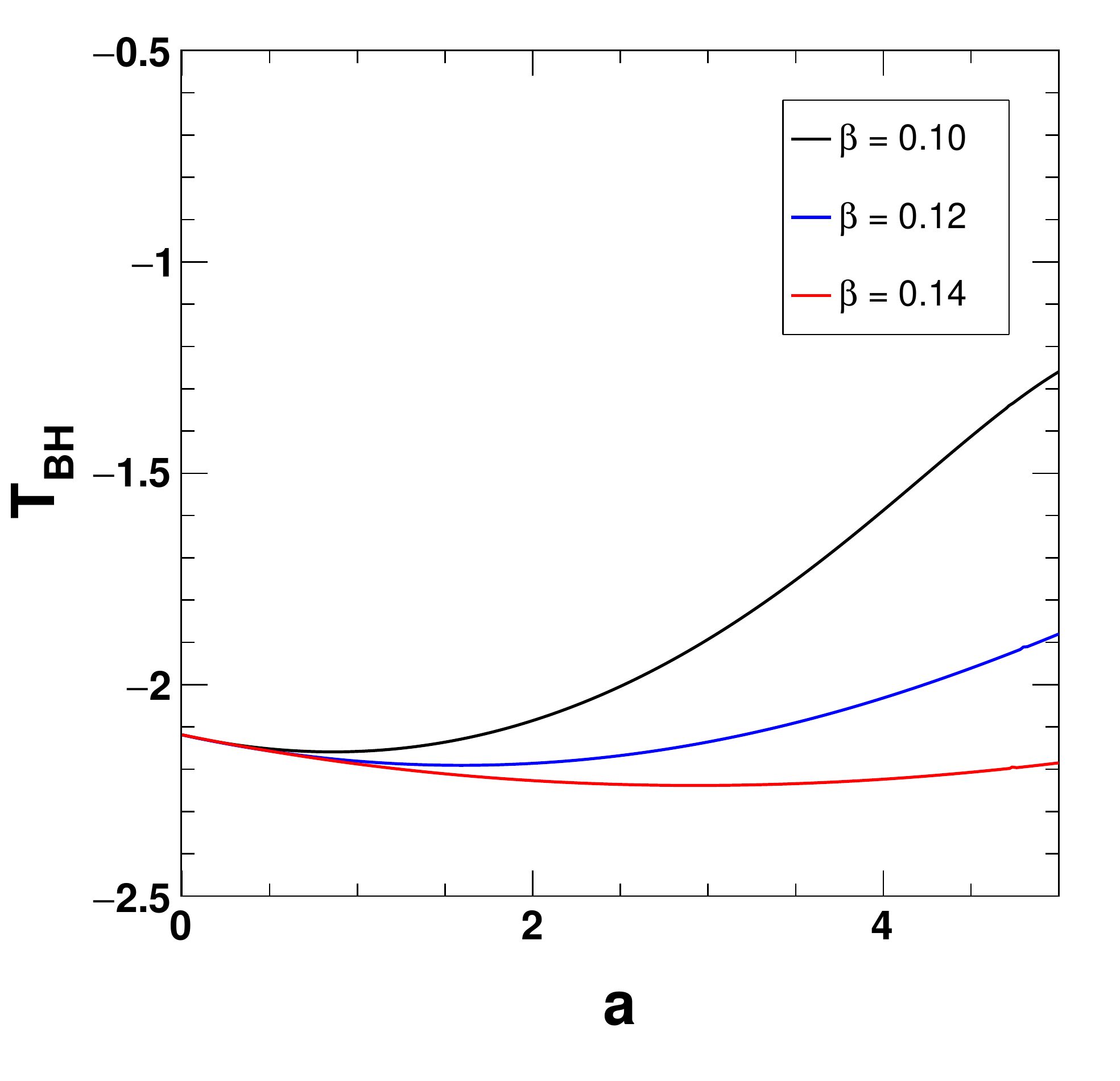}\hspace{1cm}
   \includegraphics[scale = 0.3]{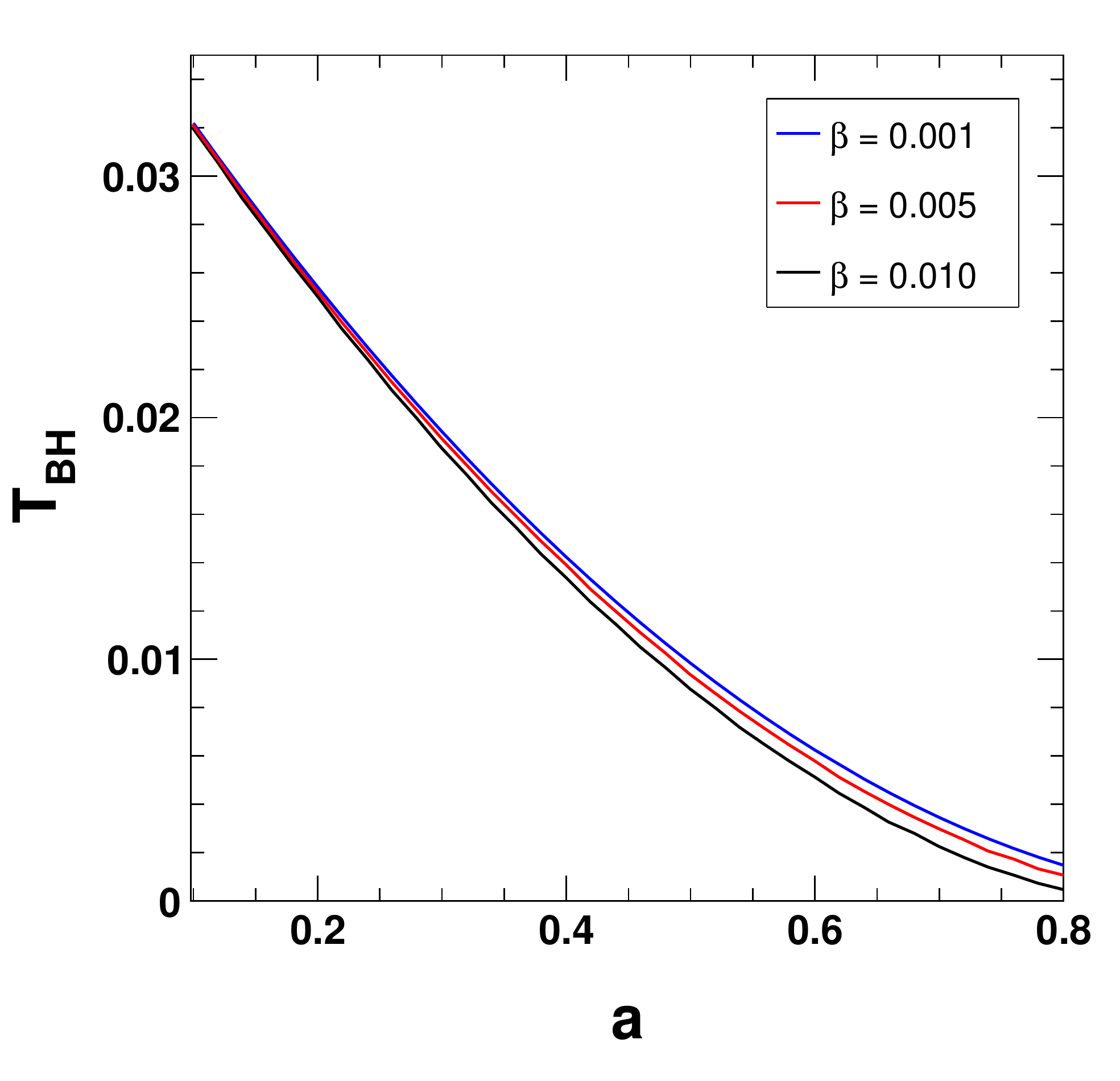}}
\vspace{-0.2cm}
\caption{Variation of $T_{BH}$ versus $a$ for different values of $\beta$ for 
the first horizon $r_-$ (on left panel) with $M = 1$ and $q = 0.6$,
and for the second horizon $r_+$ (on right panel) with $M = 1$ and $q = 0.3$ 
in case of the ABG black hole.}
\label{fig20}
\end{figure}

\begin{figure}[h!]
\centerline{
   \includegraphics[scale = 0.3]{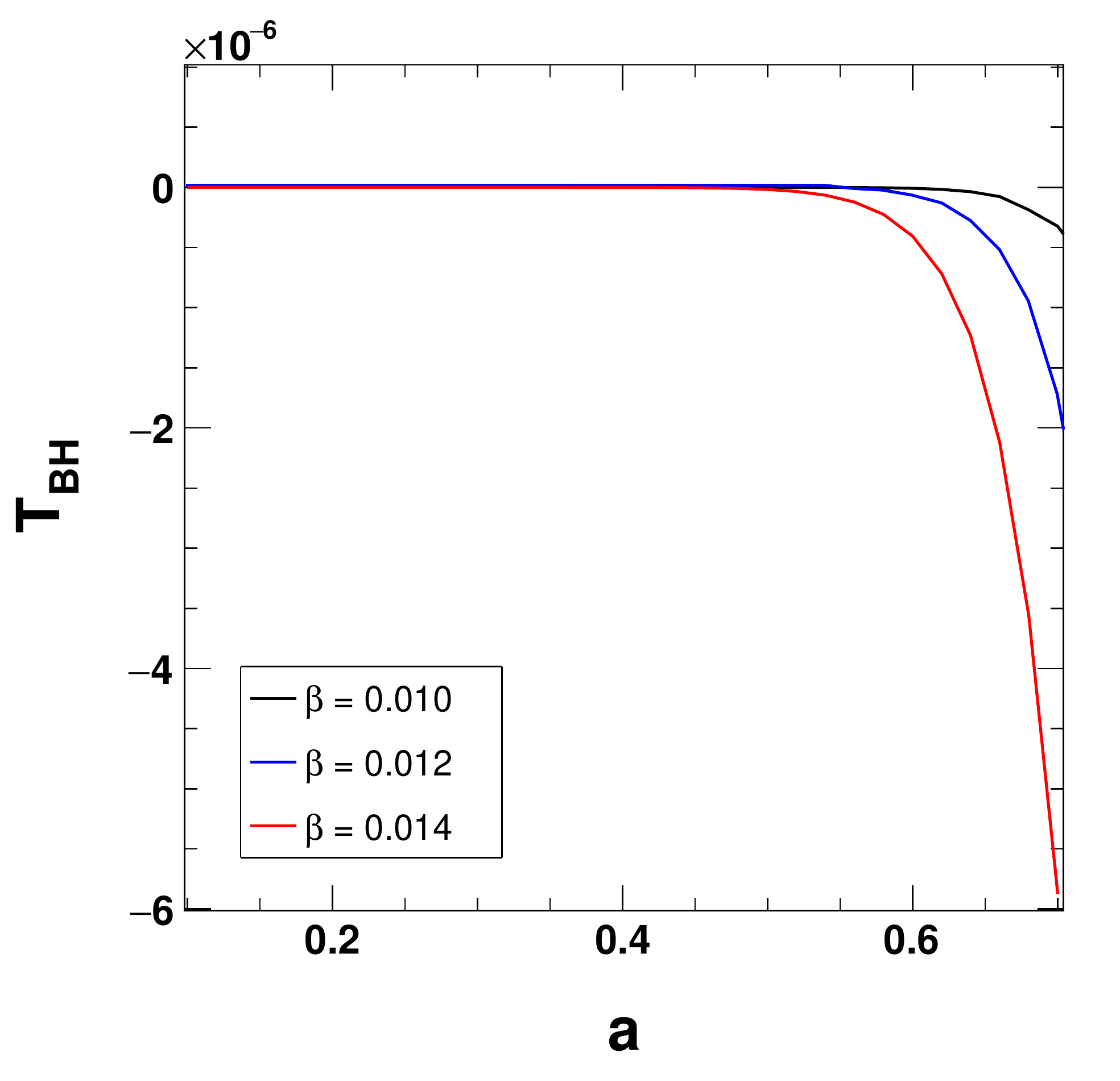}}
\vspace{-0.2cm}
\caption{Variation of $T_{BH}$ with respect to $a$ for different values of 
$\beta$ for the third horizon or cloud of string horizon $r_c$ with $M = 1$ 
and $q = 0.6$ in case of the ABG black hole.}
\label{fig21}
\end{figure}

For both the black holes, from the metrics \eqref{ABG_metric} and 
\eqref{our_metric}, we can calculate the event horizon radius using the 
condition: $f(r_H) = 0$, where $r_H$ is the event horizon radius in general. 
For a general Schwarzschild black hole, this condition gives a simple relation 
between the mass and radius of the black hole. In our cases, the relation will 
be different from that of Schwarzschild and it will depend on the other metric 
function parameters also. In this section, we shall study the Hawking 
temperature of the black holes for different cases to have a visualization of 
the black hole characteristics. To do so, we obtain the surface gravity of the 
black holes by using definition:
\begin{equation}
\kappa_{r_H} = \dfrac{1}{2} \dfrac{d f(r)}{dr} \Big|_{r\,=\,r_H}.
\end{equation}
Here $r_H = r_-, r_+$ and $r_c$ for inner horizon or Cauchy horizon, event 
horizon and cloud of string horizon respectively. Using the surface gravity 
of a black hole one can easily obtain the Hawking temperature as
\begin{equation}
T_{BH} = \dfrac{\hbar \kappa_{r_H}}{2 \pi}.
\end{equation}
In Fig.\ \ref{fig19} we have plotted the Hawking temperature $T_{BH}$ with 
respect to cloud of string parameter $a$ for the first horizon $r_-$ (on the 
left panel) and for the second horizon $r_+$ (on the right panel) for the ABG 
black hole with different values of black hole charge $q$. It is seen that for 
the $r_-$, the black hole temperature $T_{BH}$ is negative and increases 
towards higher values of $a$. For the smaller charge $q$, the black hole 
temperature decreases drastically. But in case of second horizon $r_+$, 
the temperature $T_{BH}$ is positive and it increases slowly with decrease in 
charge $q$. In this case, $T_{BH}$ is a very small positive number and it 
decreases and moves towards zero with increase in surrounding field $a$. For 
large values of $a$, one can see that a variation in charge $q$ does not show 
a significant changes in $T_{BH}$. Here, we have not plotted $T_{BH}$ for the 
third horizon $r_c$ as from our previous study of the black hole 
characteristics, it is clear that $q$ has not any significant impact over the 
third horizon. However, one can see Fig.\ \ref{fig21}, which is plotted for 
$r_c$ and 
shows that $T_{BH}$ is again negative. In Fig.\ \ref{fig20}, we have plotted 
the Hawking temperature $T_{BH}$ with respect to the cloud of string parameter 
$a$ for the first horizon $r_-$ (on the left panel) and for second horizon 
$r_+$ (on the right panel) for the ABG black hole with different values of 
Rastall parameter $\beta$. In case of the first horizon $r_-$, we see that for 
smaller values of $a$, variation of $\beta$ does not have any significant 
impacts on $T_{BH}$. However, with increase in $a$, $T_{BH}$ becomes more 
$\beta$ dependent. This due to the fact that $\beta$ is coupled with the cloud 
of string parameter $a$ in the third term of the metric function. One can see 
that for the first horizon $r_-$, decrease in $\beta$ increases $T_{BH}$ for 
large $a$. A closer look shows that in the small $a$ regime, $T_{BH}$ slightly 
decreases with increase in $a$ and for higher values of $\beta$ this pattern 
continues. But in case of smaller values of $\beta$, $T_{BH}$ increases 
drastically with increase in  $a$. In this case also, $T_{BH}$ is negative.  
For the second horizon $r_+$, we observe an opposite scenario in comparison 
to the previous case. Here, for the small $a$ regime, $T_{BH}$ is almost 
indistinguishable for different values of $\beta$. However, for large $a$, 
smaller $\beta$ has higher temperature $T_{BH}$. Finally, we plot $T_{BH}$ 
with $a$ for the third horizon $r_c$ with different values of $\beta$ in 
Fig.\ \ref{fig21} as mentioned above. It is seen that $T_{BH}$ is again 
negative here but with a very small magnitude. An increase in $a$, decreases 
$T_{BH}$ further and for higher values of $\beta$, $T_{BH}$ decreases more 
drastically. 

\begin{figure}[htb]
\centerline{
   \includegraphics[scale = 0.3]{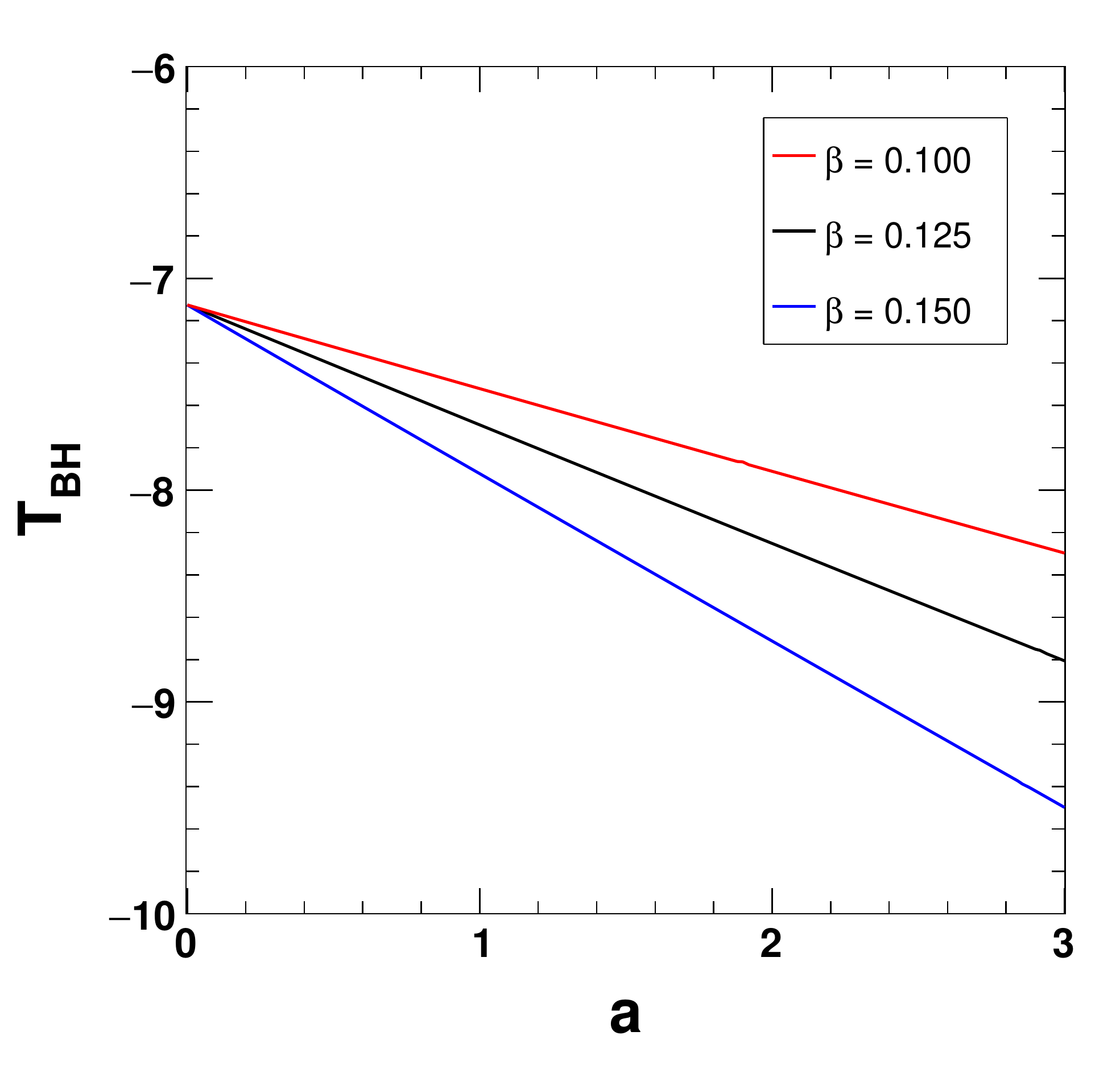}\hspace{1cm}
   \includegraphics[scale = 0.3]{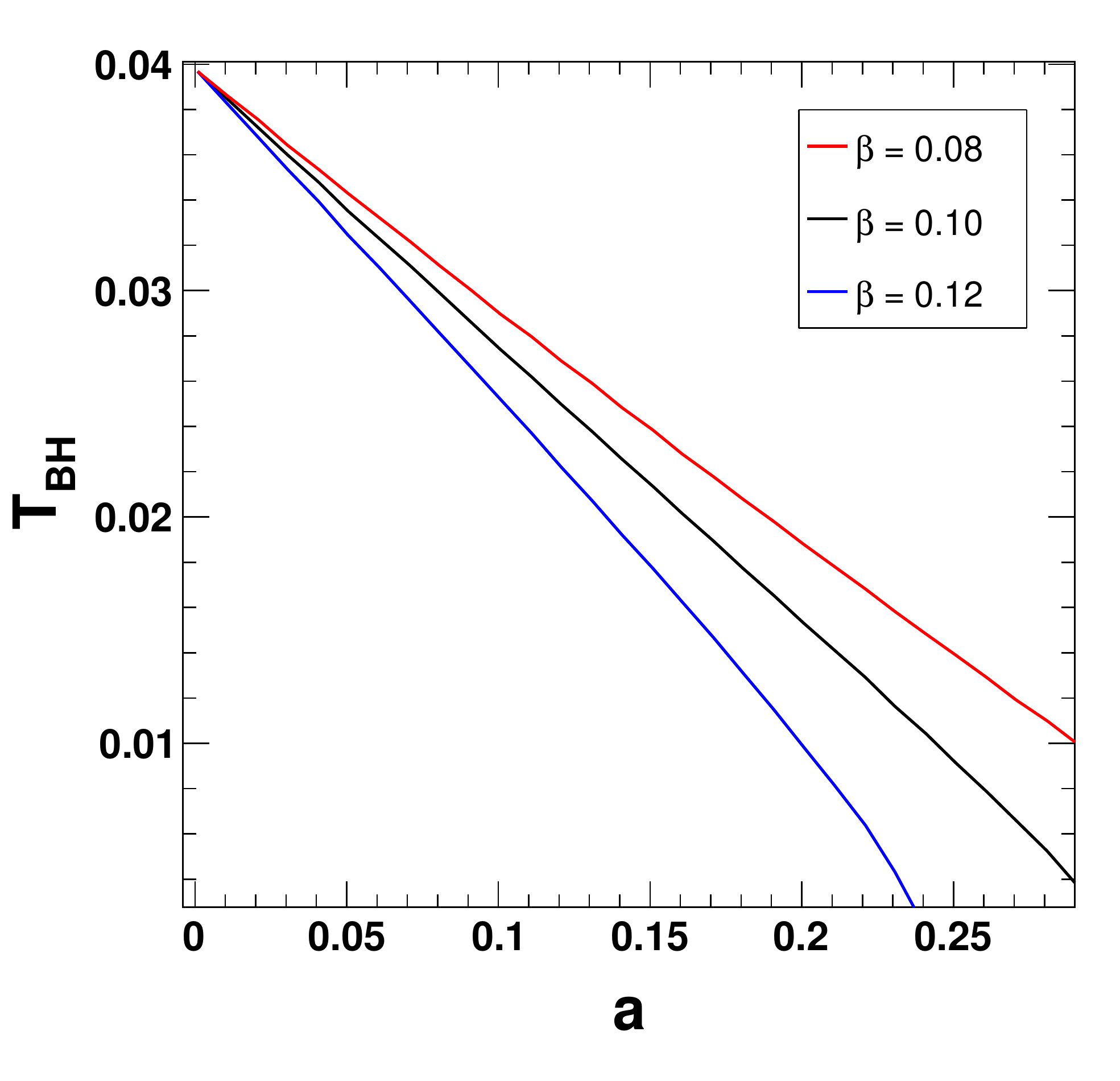}}
\vspace{-0.2cm}
\caption{$T_{BH}$ versus $a$ for the first horizon $r_-$ (on left panel) with $M = 1$, $q = 0.6$ and $\zeta = -0.4$, and for the second horizon $r_+$ (on right panel) with $M = 1$, $q = 0.3$ and $\zeta = -0.8$ for the black hole defined by the metric \eqref{our_metric}.}
\label{fig22}
\end{figure}

\begin{figure}[htb]
\centerline{
   \includegraphics[scale = 0.3]{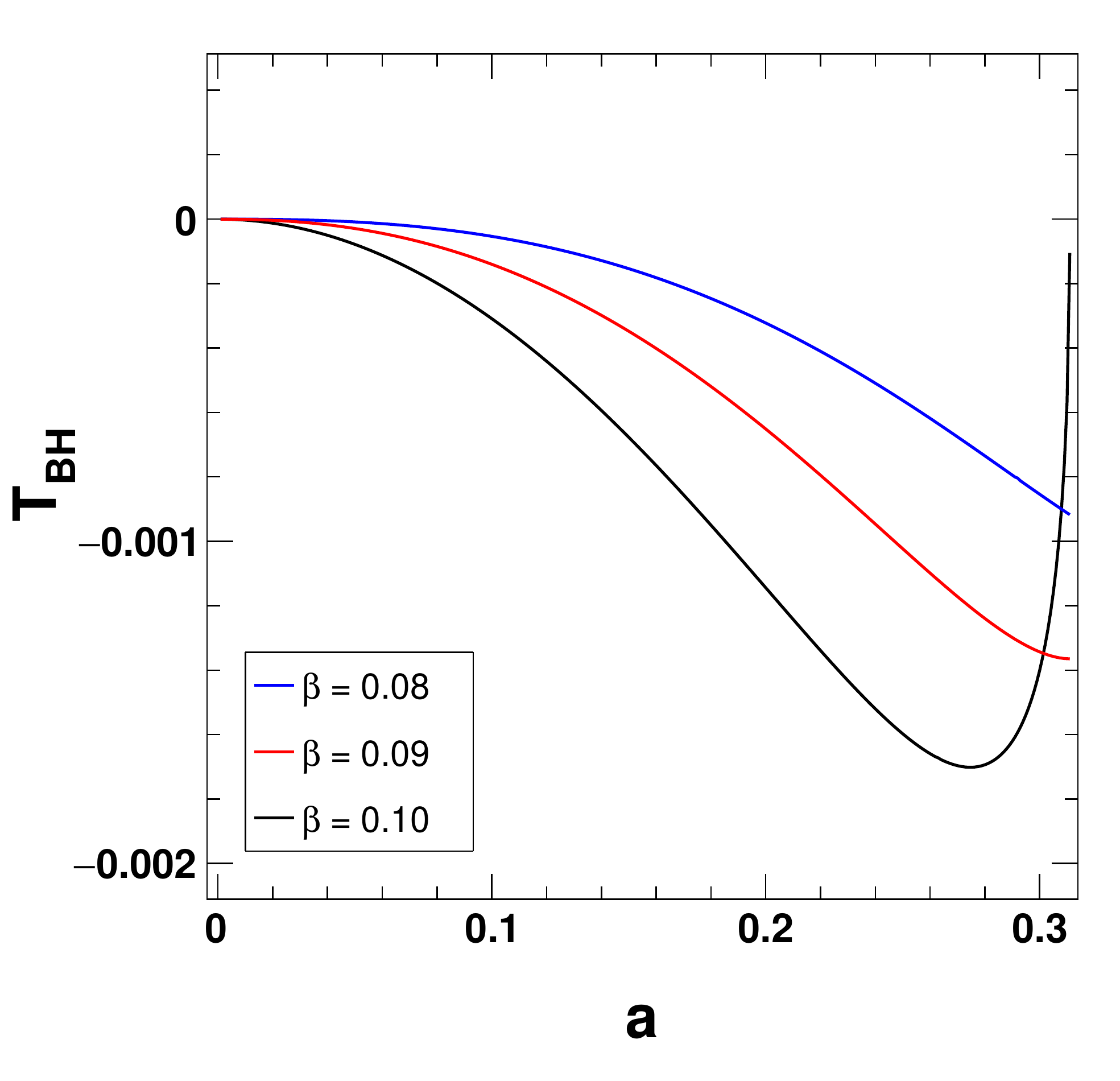}}
\vspace{-0.2cm}
\caption{$T_{BH}$ with respect to $a$ for the third horizon or cloud of string 
horizon $r_c$ with parameters $M = 1$, $q = 0.6$ and $\zeta = -0.4$ for the 
black hole defined by the metric \eqref{our_metric}. }
\label{fig23}
\end{figure}

\begin{figure}[htb]
\centerline{
   \includegraphics[scale = 0.3]{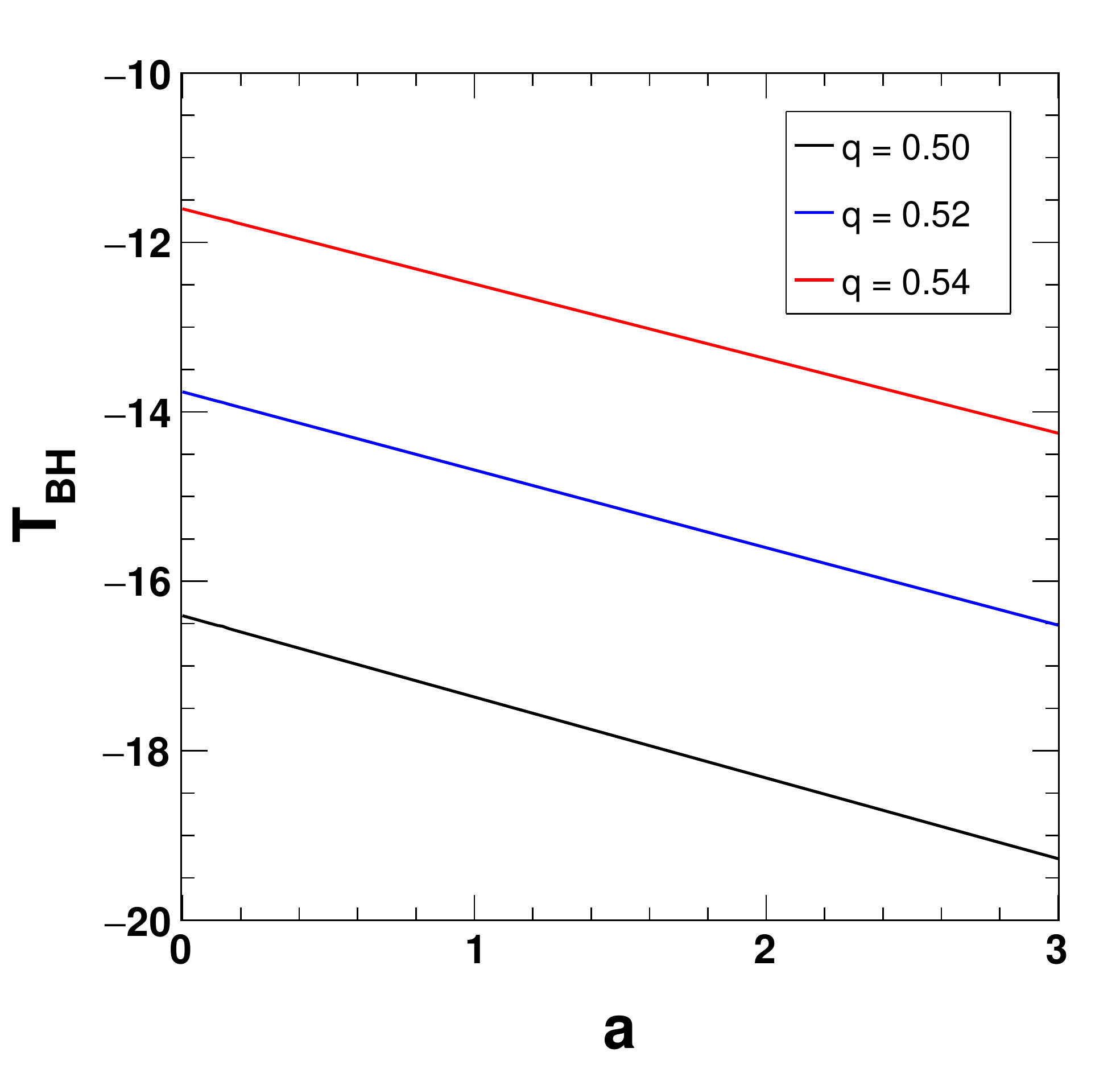}\hspace{1cm}
   \includegraphics[scale = 0.3]{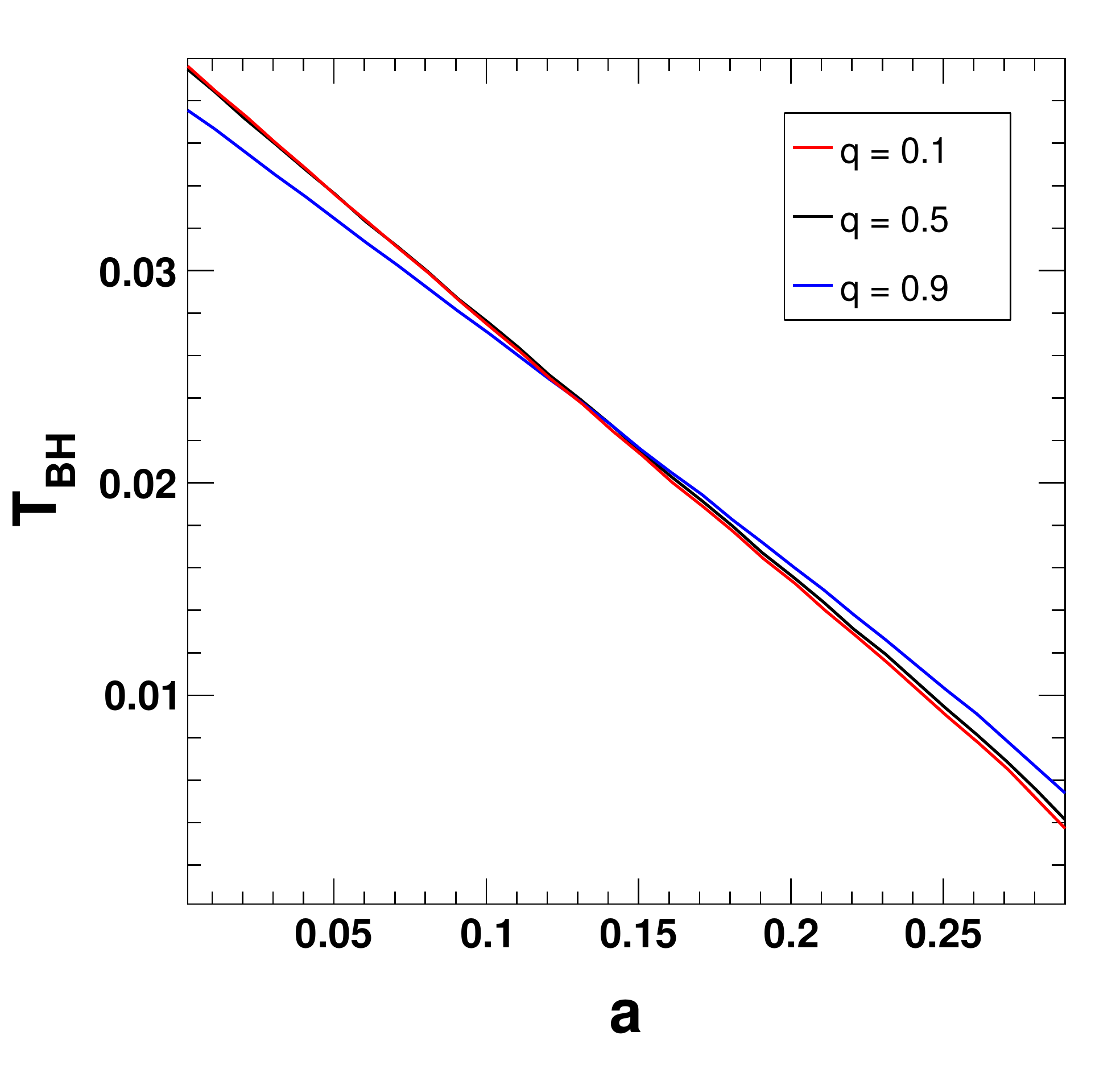}}
\vspace{-0.2cm}
\caption{Behaviour of $T_{BH}$ with respect to $a$ for the first horizon $r_-$ 
(on left panel), and for the second horizon $r_+$ (on right panel) with 
parameters $M = 1$, $\beta = 0.1$ and $\zeta = -0.4$ for the black hole 
defined by the metric \eqref{our_metric}.}
\label{fig24}
\end{figure}

\begin{figure}[htb]
\centerline{
   \includegraphics[scale = 0.3]{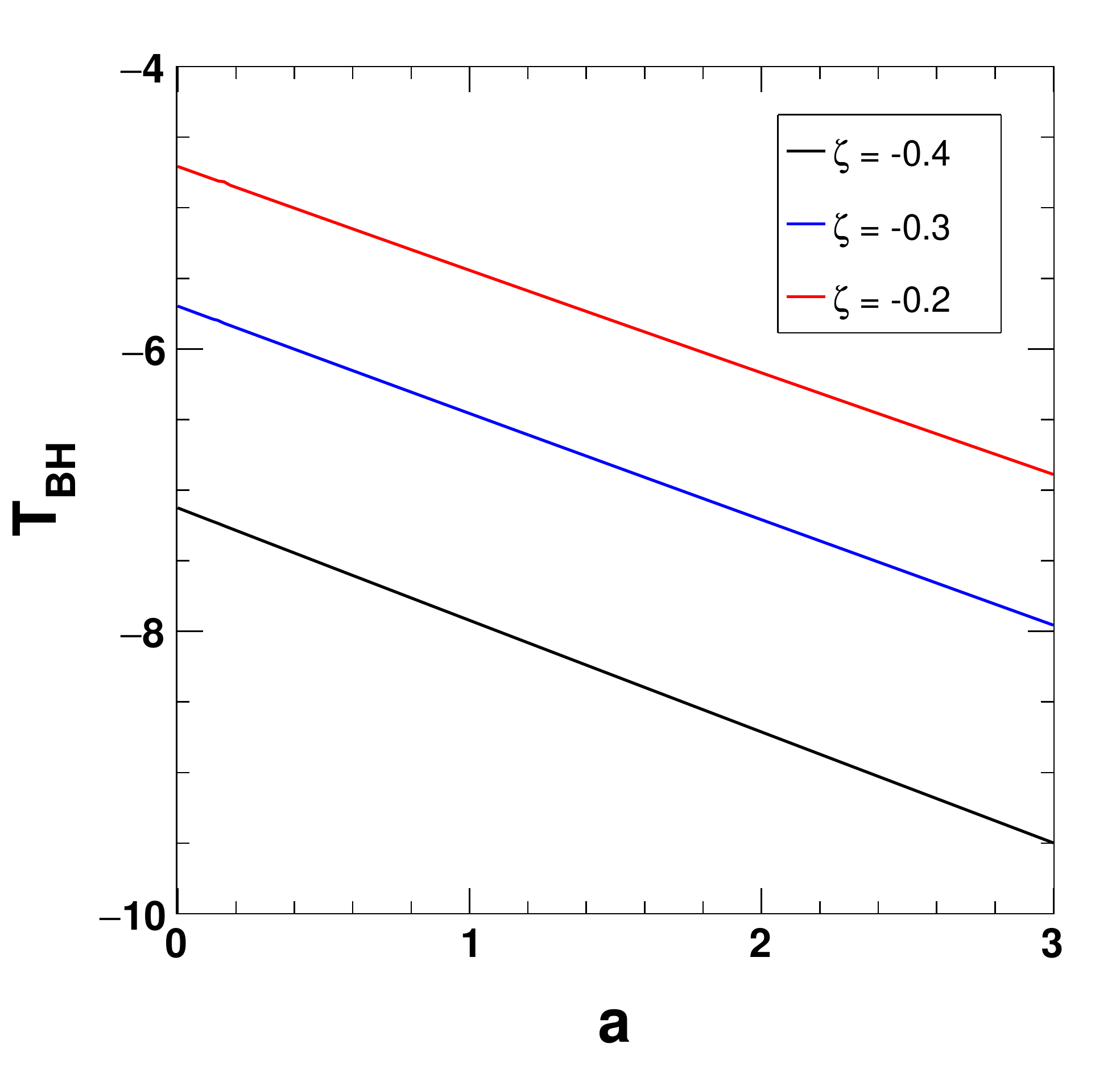}\hspace{1cm}
   \includegraphics[scale = 0.3]{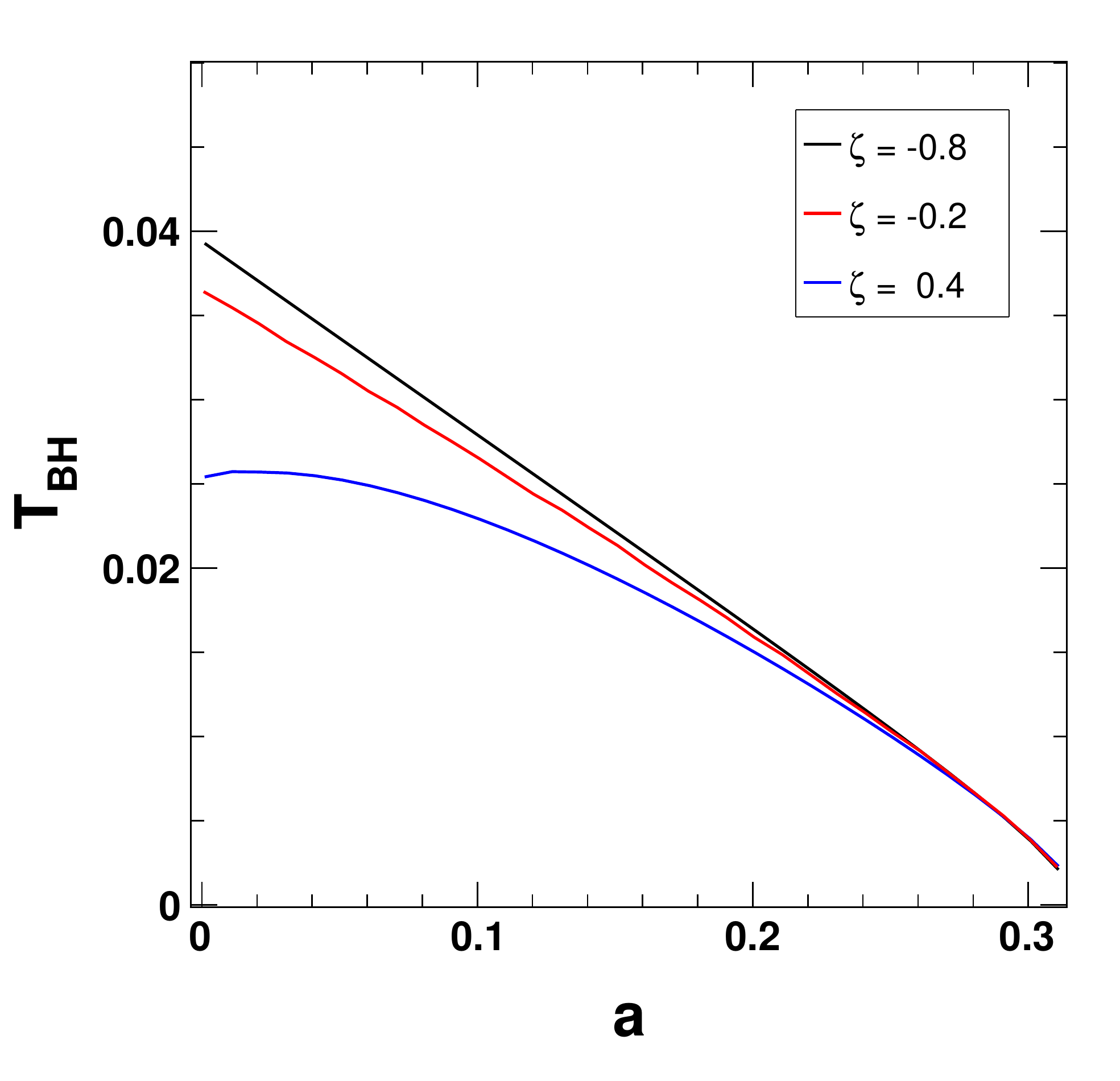}}
\vspace{-0.2cm}
\caption{Variation of $T_{BH}$ in terms of $a$ for the first horizon $r_-$ (on 
left panel) with parameters $M = 1$, $\beta = 0.1$ and $q = 0.6$, and for the 
second horizon $r_+$ (on right panel) with parameters $M = 1$, $\beta = 0.1$ 
and $q = 0.9$ for the black hole defined by the metric \eqref{our_metric}.}
\label{fig25}
\end{figure}

Similarly, we have studied the variation of Hawking temperature $T_{BH}$ with
respect to the cloud of strings parameter $a$ for the black hole defined by 
the metric \eqref{our_metric}. In Fig.\ \ref{fig22} we have plotted the Hawking
temperature $T_{BH}$ versus the cloud of strings parameter $a$ for the first 
horizon $r_-$ (on the left panel) and for the second horizon $r_+$ (on the 
right panel) for different values of $\beta$. In both cases, $T_{BH}$ decreases
with increase in cloud of strings parameter $a$. For higher values of $\beta$, 
$T_{BH}$ has smaller values corresponding to non-zero $a$ values. For the 
first case, $T_{BH}$ is negative and for the second case, it is a small 
positive number close to zero. From the previous study of the metric function, 
it is clear that the Rastall parameter $\beta$ has a significant impact over 
the third horizon $r_c$ of the black hole. This suggests that $T_{BH}$ 
corresponding to the third horizon will show noticeable changes for different 
values of $\beta$. This variation can be seen in Fig.\ \ref{fig23}. For 
$\beta = 0.10$, $T_{BH}$ decreases upto $a = 0.28$ and then suddenly increases 
and approaches to zero at around $a = 0.311029183$. For the considered 
parameter 
set, i.e. for $M = 1$, $q = 0.6$, $\zeta = -0.4$ and $\beta = 0.10$, when $a$ 
approaches $0.311029183$, the third horizon $r_c$ and the second horizon $r_+$ 
coincide to give a single horizon. Beyond this critical point, both $r_+$ and 
$r_c$ start to vanish. From the figure, it is also clearly seen that this 
critical $a$ value increases with decrease in $\beta$. In Fig.\ \ref{fig24}, 
we have plotted $T_{BH}$ with respect to $a$ for different values of $q$ in 
case of $r_-$ and $r_+$. We see that for $r_-$, $T_{BH}$ decreases with 
increase in $a$ and higher charge $q$ has higher $T_{BH}$ values. In case of 
$r_+$, it is seen that $T_{BH}$ decreases and approaches zero with increase in 
$a$. But here we observe a junction point for some $a$ at which $T_{BH}$ for 
all $q$ coincides and flips the variation pattern. This is clearly seen from 
the right panel of this figure, where for $q = 0.9$, $T_{BH}$ is smaller than 
the other two cases with $q = 0.5$ and $0.1$ in the small $a$ regime. But as 
$a$ increases, we observe an opposite pattern. Finally, in Fig.\ \ref{fig25}, 
we have plotted $T_{BH}$ with respect to $a$ for different values of $\zeta$ 
for $r_-$ and $r_+$ respectively. For $r_-$, the behaviour is almost identical 
to the previous case with different values of $q$. But in case of $r_+$ we 
observe a slightly different pattern for small $a$ regime with higher 
$\zeta$ value. But towards higher values of $a$, $\zeta$ dependency of 
$T_{BH}$ ceases and the graphs become almost indistinguishable.

Thus for both the black holes we see that for the first and the third horizon, 
$T_{BH}$ is negative which is an anomalous behaviour. This is actually not a 
new outcome for a black hole in modified gravity as it was encountered earlier 
also \cite{Park_2008, Toledo_2020, Singh_2020, Wei_2011}. In Ref.\ \cite{Park_2008}, the possibility of negative Hawking temperatures of black holes was shown 
with some added matter-energy content, for example quintessence, higher 
dimensional spacetimes and MTGs. In a recent study \cite{Toledo_2020}, it was 
mentioned that the Hawking temperature can be negative for quintessence and 
cloud of strings fields. Therefore our result resonates these earlier 
observations in case of Hawking temperature. 

\section{Conclusion} \label{conclusion}
In this work, we have studied the QNMs and properties of the ABG black hole 
and a new black hole surrounded by a cloud of strings field in Rastall gravity.
The main motivation of this study is to show the dependency of QNMs and black 
hole characteristics on the non-linear charge distribution surrounded by a 
cloud of string field in Rastall gravity. Recently a study has been done in 
Ref.\ \cite{Cai_2020}, where the authors studied the QNMs and spectroscopy of 
a Schwarzschild black hole in Rastall gravity. But in our work we have 
considered the Reissner-Nordstr\"om black hole in Rastall gravity surrounded 
by a cloud of strings in Rastall gravity at first and then we have extended 
the black hole metric to non-linear charge distribution by considering the ABG
black hole and a new black hole. The main characteristic of ABG black hole is 
that it gives a regular black hole in GR surrounded by a cloud of strings. But 
in case of Rastall gravity, the cloud of strings parameter is coupled with the 
Rastall parameter in the black hole metric function giving a possibility of 
singularity from the last term of the metric function. However, we have found 
that this issue can be resolved by setting $0.25 < \beta < 0.50$. Thus, we have 
shown that by controlling the Rastall parameter it is still possible to 
have a regular black hole in this case also. Inspired from the ABG black hole, 
we have introduced a new black hole with a different charge distribution 
function and with an extra model parameter $\zeta$. This parameter can control 
the singularity arising from the charge and mass coupled term in the metric 
function. We found found that $\zeta = \arccsch(1) = 0.881374$ and 
$0.25 < \beta < 0.50$ can make the second black hole a regular one. The main 
advantage of using this new parameter is that for the ABG black hole, in GR, 
we always get a regular black hole, but in case of the second black hole, in 
GR, only $\zeta = \arccsch(1) = 0.881374$ can give a regular black hole. For 
other cases, the black hole has a physical singularity.

We have also shown that both the black holes behave identically in the 
asymptotic regime. But with increase in charge $q$, the black holes start to 
deviate from each other. In case of QNMs also we have observed the same 
results. However, the free parameter $\zeta$ can be controlled to adjust the 
black hole properties and QNMs up to some extent. Due to this property, such 
a model can be easily fitted with experimental results. Thus there is a good 
possibility of physical consistency with the experimental constraints on the 
QNMs in case of the second black hole.

In a recent work, QNMs of black holes with a non-linear charge distribution 
in Rastall gravity surrounded by dark energy fields have been studied 
\cite{Gogoi2021}. We observe that the dependency of QNMs with the Rastall 
parameter studied in that work is completely different from this work. Hence, 
the impact of the cloud of strings parameter and the dark energy fields on the 
QNMs is not identical.

We believe that the impacts of the cloud of strings on the black holes in 
Rastall gravity can be studied more explicitly in the near future. Specially 
an anisotropic cloud of strings field in extended Rastall gravity can shed 
more light on the dependency of different properties of the black holes 
including QNMs with these situations.



\end{document}